\documentclass[fleqn,usenatbib, onecolumn]{mnras}
\usepackage[utf8]{inputenc}
\usepackage{amsmath,amssymb, graphicx, multicol, array, upgreek, natbib, booktabs, multirow, bm, dsfont} %amsthm
\usepackage{simplewick}
\usepackage{scalerel}
\usepackage[mathscr]{euscript}
\usepackage{extarrows}
\usepackage{caption}
\usepackage{subcaption}

\usepackage[dvipsnames]{xcolor}
\definecolor{color26}{rgb}{0.9, 0.9, 0.9}
\definecolor{color27}{rgb}{0.2, 0.6, 0.8}
\definecolor{color38}{rgb}{0.9,.17,0.31}
\newcommand{\jiamin}[1]{\textcolor{color27}{#1}}

\definecolor{myblue}{rgb}{0.1, 0.4, 0.5}
\definecolor{color1}{rgb}{0.5, 0.5, 0.5}

\newcommand{\mathcolorbox}[2]{\colorbox{#1}{$\displaystyle #2$}}

\usepackage{tikz}
\usetikzlibrary{tikzmark}

\def\begeqar{\begin{eqnarray}}
\def\endeqar{\end{eqnarray}}
\def\begeq{\begin{equation}}
\def\endeq{\end{equation}}
\def\non{\nonumber\\}

\def\quijote{{\textsc{Quijote }}}
\def\logn{{{lognormal }}}
\def\mdpatchy{{\textsc{MultiDark-Patchy }}}
\def\patchy{{\textsc{Patchy }}}
\def\nbk{{\textsc{Nbodykit }}}
\def\mpch{{h^{-1}\rm{Mpc}}}

\def\bfzero{{\bf 0}}
\def\hr{{\hat{r}}}
\renewcommand{\vec}[1]{\boldsymbol{#1}}
\def\hs{{\hat{s}}}
\def\bfshat{{\hat\bfs}}
\def\bfk{{\bf k}}
\def\bfK{{\bf K}}
\def\hk{{\hat{k}}}

\def\bfkhat{{\hat\bfk}}
\def\bfKhat{{\hat\bfK}}
\def\bfr{{\bf r}}
\def\bfR{{\bf R}}
\def\bfrhat{{\hat\bfr}}
\def\bfRhat{{\hat\bfR}}
\def\calB{{\cal B}}
\def\calC{{\cal C}}
\def\calD{{\cal D}}
\def\calE{{\cal E}}
\def\calL{{\cal L}}

\def\calG{{\cal G}}
\def\calH{{\cal H}}
\def\calM{{\cal M}}
\def\calN{{\cal N}}
\def\calO{{\cal O}}
\def\calQ{{\cal Q}}
\def\calY{{\cal P}}
\def\calR{{\cal R}}
\def\calS{{\cal S}}
\def\bfx{{\bf x}}

\def\calZ{{\zeta}}
\def\cov{{\rm{Cov}}}

\def\rmM{{\rm M}}

\newcommand{\av}[1]{\left\langle{#1}\right\rangle} 
\newcommand{\vk}{\vec k}

\newcommand{\vx}{\vec x}
\newcommand{\vs}{\vec s}

\newcommand{\vr}{\vec r}
\newcommand{\bfs}{\mathbf s}

\newcommand{\hn}{\hat{\vec n}}

\renewcommand{\L}{\Lambda}
\renewcommand{\P}{\mathcal{P}}

\newcommand{\tj}[6]{\begin{pmatrix} {#1} & {#2} & {#3}\\ {#4} & {#5} & {#6}\end{pmatrix}}

\newcommand{\six}[6]{\left(\begin{array}{ccc}
									{#1}& {#2}& {#3}\\
									{#4}& {#5}& {#6} \\
\end{array}\right)}

\newcommand{\nine}[9]{\left\{\begin{array}{ccc}
									{#1}& {#2}& {#3}\\
									{#4}& {#5}& {#6} \\
							    	{#7}& {#8}& {#9} \\
\end{array}\right\}}

\begin{document}
\title{Analytic Gaussian Covariance Matrices for Galaxy $N$-Point Correlation Functions}
\author[J. Hou et al.]{
Jiamin Hou$^{1}$\thanks{Email: \href{mailto:jiamin.hou@ufl.edu}{jiamin.hou@ufl.edu}},
Robert N. Cahn$^{2}$, Oliver H.\,E. Philcox$^{3,4}$, and Zachary Slepian$^{1,2}$\vspace*{4pt} \\ 
$^{1}$Department of Astronomy, University of Florida, 211 Bryant Space Science Center, Gainesville, FL 32611, USA\vspace*{-2pt}\\
$^{2}$Lawrence Berkeley National Laboratory, 1 Cyclotron Road, Berkeley, CA 94720, USA\vspace*{-2pt}\\
$^{3}$Department of Astrophysical Sciences, Princeton University, Princeton, NJ 08540, USA\vspace*{-2pt}\\
$^{4}$School of Natural Sciences, Institute for Advanced Study, 1 Einstein Drive, Princeton, NJ 08540, USA\\}

% These dates will be filled out by the publisher
\date{Accepted XXX. Received YYY; in original form ZZZ}

% Enter the current year, for the copyright statements etc.
\pubyear{2021}
\pagerange{\pageref{firstpage}--\pageref{lastpage}}
\maketitle
\label{firstpage}

\begin{abstract}
    We derive analytic covariance matrices for the $N$-Point Correlation Functions (NPCFs) of galaxies in the Gaussian limit. Our results are given for arbitrary $N$ and projected onto the isotropic basis functions of \citet{Cahn202010}, recently shown to facilitate efficient NPCF estimation.
    %This is the first time a covariance matrix formalism in the isotropic basis as proposed in~\citet{Cahn202010} with an arbitrary $N$ is being presented, and recently the isotropic basis has been shown to be an appropriate basis for NPCF computation. 
    A numerical implementation of the 4PCF covariance is compared to the sample covariance obtained from a set of \logn simulations, \quijote dark matter halo catalogues, and \textsc{MultiDark-Patchy} galaxy mocks, with the latter including realistic survey geometry.
    %Based on this formalism, we implement numerically the covariance matrix for the real-space four-point Correlation Function and compare it to a set of \logn mock simulations, \quijote dark matter halo catalogues, and \textsc{MultiDark-Patchy} galaxy mock catalogues with survey geometry. 
    The analytic formalism gives reasonable predictions for the covariances estimated from mock simulations with a periodic-box geometry. Furthermore, fitting for an effective volume and number density by maximizing a likelihood based on Kullback-Leibler divergence is shown to partially compensate for the effects of a non-uniform window function.
\end{abstract}

\section{Introduction}

Large-scale structure (LSS) is a powerful observable with which to elucidate cosmic evolution. To characterize its spatial distribution, various summary statistics have been proposed, of which the most prominent are 
%Its spatial distribution can be well characterized by various summary statistics. Much of the effort in the LSS analysis has been dedicated into 
the two-point statistics, \textit{i.e.} the 2-Point Correlation Function (2PCF) and its Fourier-space counterpart, the power spectrum~\citep[e.g.,][]{BOSSCollaboration2017, eBOSSCollaboration2021}.

Although two-point statistics fully capture information in the early Universe, assuming a standard inflationary model with adiabatic perturbations, gravitational evolution induces non-linearities in the LSS at late times, spreading information into higher-order statistics. Furthermore, different mechanisms during inflation can generate distinctive non-Gaussian signatures ~\citep{Kofman199101,Linde199707,Komatsu200309,Chen200701,Chen201002}. These two effects justify pushing beyond the power spectrum or 2PCF. Examples include the 3-Point Correlation Function~\citep[3PCF;][]{Peebles1978,Fry1993, Slepian201506,Slepian201510,Slepian201702,Portillo201808}, the bispectrum~\citep{Scoccimarro199803,Scoccimarro200012,Pearson201808},
skew spectra~\citep{Dizgah202004,Schmittfull202103}, the marked density field \citep{marked2,marked1}, and the integrated bispectrum and trispectrum~\citep{Sefusatti200503,Chiang201405}. Methods such as BAO reconstruction~\citep{Eisenstein200708,Padmanabhan200903,White201504,Schmittfull201512,Schmittfull201707}, forward-modeling of the galaxy density field~\citep{Jasche201304,Seljak201712,Jasche201905,Schmidt201901}, and machine learning techniques have also been proposed as alternative but complementary approaches to summary statistics. Previous work has demonstrated that combining two- and higher-point statistics can break the degeneracy between linear bias and the amplitude of matter fluctuations, tighten constraints on standard $\Lambda$CDM parameters~\citep{Agarwal202103,Gil-Marin201702,Sugiyama202012,Gualdi202104}, and provide further insights into the neutrino mass~\citep{Ruggeri201803,deBelsunce201902,euclid_forecast,Hahn202003, Kamalinejad202011,Aviles202106} and modified gravity~\citep{Bartolo2013J03,Alam202011}. Gravitational evolution imprints a useful shape on the $N$-point statistics;~\citet{Samushia202102} showed that for $N=3$ this shape can potentially provide complementary information to BAO reconstruction when it is used as standard ruler.

To infer cosmological parameters from the $N$-Point Correlation Functions (NPCFs) using Bayes theorem with a Gaussian likelihood, %\jiamin{does this have to be a Gaussian likelihood? oliver: only a Gaussian likelihood is fully determined by its covariance. Jiamin: what else are needed if likelihoods differ from Gaussian when using Bayes' theorem for parameter inference?}\textcolor{red}{OP: In order to perform parameter inference, we must be able to write down the likelihood. The Gaussian distribution is fully characterized by its covariance matrix. If it were non-Gaussian, the covariance isn't enough to write down the likelihood (e.g. there could be non-zero skew, kurtosis etc...), or we might not even have an analytic likelihood.}, 
a covariance matrix is required. Usually, this is obtained by sampling independent
realizations of the statistic from simulations. However this approach introduces sampling variance, which then propagates into the parameter estimates~\citep{Dodelson201309,Percival201404,Taylor201406,Sellentin201602}. 
% Although these paper \bob{this paper?  Dodelson?  or our paper?} also provides normalization of the parameter covariance up to second order \bob{second order in what?} by assuming Gaussian distributed data and likelihood, we still need at least $N_{\rm{S}}$ realizations to meet the minimum requirement $N_{\rm S}>N_{\rm D}-N_{\rm P}+2 / \nu^{2}+1$, with $N_{\rm{D}}$ the dimension of the covariance \bob{what is the dimension of the covariance?} \jiamin{number of data entries used to estimate the covariance}, $N_{\rm P}$ the number of parameters, and $\nu$ the fractional error of the covariance~\citep{Taylor201406}. 
To reduce this variance, the number of mock catalogs must be much larger than the dimension of the NPCFs; if the statistic contains many bins, the computational cost of this poses a significant challenge.
%thus mock-based sample covariance poses a challenge for computational cost. 

An alternative approach is to compute the covariances analytically. This has been intensively studied especially for two- and three-point statistics~\citep{Grieb2016,Li201901,Scoccimarro199912,Slepian201506,Slepian201804,rascal1,Barreira201903,rascal2,Philcox201910,rascal3,Wadekar202012,Sugiyama202009}. Recent work in~\citet{Philcox2021encore} developed an efficient algorithm to measure the NPCF for arbitrary $N$; given the high dimensionality of the NPCFS for large $N$, this poses a further challenge for covariance estimation. % given its high dimensionality. 
Thus far, few studies have considered the covariance of the NPCFs with $N>3$. To address this, we here derive an analytic expression for the NPCF covariance at arbitrary $N$.
In order to efficiently characterise the NPCF we work with the isotropic basis functions developed in~\citet{Cahn202010}; these have rotational symmetry in 3D, and may be related to the quantum-mechanical angular momentum basis states.% and are well suited for describing our Universe under the assumption of homogeneity and isotropy on the large scales.

An important assumption in our modeling is that the two-point statistics are the dominant contribution to the covariance, \textit{i.e.} we ignore contributions from three- and higher-point statistics. To test this assumption, we will use simulations that include non-Gaussian effects. For the majority of this paper, we will assume the two-point statistics to be isotropic, such that the spatial distribution of the galaxy pairs is independent of the line of sight (l.o.s). In practice, a galaxy's peculiar velocity, induced by its local gravitational environment, can give rise to redshift space distortions (RSD) and thus break isotropy. Although the main tests in this paper will be focused on the isotropic case, we will show in the Appendix an analytic expression that includes the effects of RSD, by expanding the anisotropic two-point statistics in multipoles with respect to the l.o.s. Finally, we will compare the results of our formalism to the covariance estimated from mock catalogues with a realistic survey geometry.

In \S\ref{sec:review_iso_basis} we briefly review the isotropic basis and its properties, before the NPCF estimator is defined in \S\ref{sec:npcf}. 
In \S\ref{sec:cov_gaussian} we present our formalism for the theoretical covariance in the Gaussian Random Field (GRF) limit, starting with the basic elements as building blocks for constructing the Gaussian covariance, then presenting the general formalism for the NPCF covariance, and ending with explicit expressions for the case of $N=4$. 
In \S\ref{sec:numerics_and_simulations} we compare our numerical implementation of the Gaussian NPCF covariance to a set of \logn mocks, a set of halo catalogues from N-body simulations using \quijote simulations and \patchy mocks, where the latter include realistic survey geometry. We summarize our results in \S\ref{sec:summary}. Appendices \ref{appendix:generalized_gaunt_N234}, ~\ref{appendix:basic_elements}, and~\ref{appendix:fabrikant_rbin} provide intermediate derivation steps as well as consistency checks, Appendix \ref{appendix:cov_pc} discusses the covariance contribution from the disconnected piece of the NPCF estimators, and Appendix \ref{appendix:cov_fc_rsd_iso} presents the derivation of the covariance including RSD. The code for computing the covariance of the connected 4PCFs is publicly available.\footnote{See ~\url{https://github.com/Moctobers/npcf_cov.git}}

\section{Review of the Isotropic Basis Functions}
\label{sec:review_iso_basis}
In this section we will provide a summary of the isotropic basis functions, including a number of important properties that will be needed later for the derivation of the theoretical covariance. Further details are presented in~\citet{Cahn202010}.
\subsection{Construction of the basis function $\calY_{\Lambda}$}
In our notation, the isotropic functions $\mathcal{P}_{\Lambda}$ are sums of products of $n$ spherical harmonics $Y_{\ell m}$ multiplied by a product of Clebsch-Gordan coefficients, denoted by $\calC^\Lambda_{\rm M}$. 
%\bob{why N-1 rather than N?} \jiamin{hope this is now clearer with Eq (1) in the introduction and its surrounding text/} 
They are constructed so as to be invariant under simultaneous rotation of all $n$ coordinates:
\begeqar
\calY_\Lambda(\bfRhat)&=&\sum_{m_1...m_{n}}\calC^{\Lambda}_{\rm M} Y_{\ell_1 m_1}(\bfrhat_1)\cdots Y_{\ell_{n} m_{n}}(\bfrhat_{n}),
%\calC^\Lambda_{\rm M}&=&\sum_{m_{12},m_{123},\cdots}\left<\ell_1 m_1\ell_2 m_2|\ell_{12} m_{12}\right>\cdots\left<\ell_{12..   N-2}m_{12.. N-2}\ell_{N-1} m_{N-1}|\ell_{12..N-1} m_{12..N-1}\right>\non
%&&\qquad\qquad\qquad\times\left< \ell_{12..N-1} m_{12..N-1}\ell_N m_N|00\right>,\non
\endeqar
where $\bfRhat$ stands for a collection of unit vectors $\bfrhat_1,...,\bfrhat_{n}$. Each unit vector $\bfrhat_i$ is associated with a rotation generator $\mathbf{L}_i$, \textit{i.e.} the angular momentum operator. 
The isotropic $\calY_{\Lambda}$ function is an eigenfunction of each operator $\mathbf{L}^2_i$ with eigenvalue $\ell_i(\ell_i+1)$ and of the operator $(\sum_{i=1}^{n} \mathbf{L}_i)^2$ with eigenvalue zero (see also the discussion in~\citealt{Philcox202106ND} for a generalization of this to $D$ dimensions). We denote the orbital angular momenta by $\ell_i$, with $m_i$ being its projection onto the $z$-axis.\footnote{We adopt the language of quantum mechanics to describe the spherical harmonics.} 
%Since for NPCF we have in effect only $(N-1)$ coordinates. 
For $n>3$ the combination of a given set of orbital angular momenta, $\ell_1,\ldots,\ell_n$, is not unique: we need to specify intermediate orbital angular moment. These  are constructed from the primary orbital angular momenta, for example, $(\mathbf{L}_1+\mathbf{L}_2)^2$ with eigenvalue $\ell_{12}(\ell_{12}+1)$, and analogously for $(\mathbf{L}_1+\mathbf{L}_2+\mathbf{L}_3)^2$, \textit{et cetera}.
%$\sum_{i=1}^{K}\mathbf{L}_i$ ($K\leq n-3$)
%RNC: can't use M here because you use it below.  K leq N-4 not N-1
%whose square has the eigenvalue $\ell_{12\ldots n-2}(\ell_{12\ldots n-2}+1)$, to define the state uniquely. 
For brevity, we will hereafter call the $\ell_i$ `primary' angular momenta and the $\ell_{12}$, $\ell_{123}$, \ldots, `intermediate' angular momenta. Further, we use $\Lambda$ to indicate  the collection of angular momenta $\{\ell_1, \ell_2, (\ell_{12}), \ell_3, (\ell_{123}),\ldots, \ell_{n}\}$, with intermediate angular momenta in the brackets, and $\rmM$ to represent the collection of azimuthal angular momentum components $\{m_1,m_2,...,m_{n}\}$, with each $m_i=\{-\ell_i,\ldots,\ell_i\}$, $m_{12}=\{-\ell_{12..},\ldots,\ell_{12..}\}$ and $\sum_i^{N-1} m_i=0$. In our convention, the primary angular momenta $\ell_1,\ell_2,\ldots$ follow the ordering of the unit vectors: $\ell_1$ corresponds to $\bfrhat_1$, $\ell_2 $ corresponds to $\bfrhat_2$, etc. %$|\bfr_i|=r_i$:  $\ell_1$ corresponds to $r_1$, $\ell_2 $ corresponds to $r_2$, the second smallest, etc.
%pairs with the order of the amplitude of directional vectors $r_1<r_2...<r_N$, where $r_i=|\bfr_i|$. 

The $\calC^{\Lambda}_{\rm M}$ coefficient can be expressed using Wigner 3-$j$ symbols:
\begeqar 
\label{eqn:C_threej}
\calC^\Lambda_\rmM&=&\calE(\Lambda)\sqrt{2\ell_{12}+1}\times\cdots\times\sqrt{2\ell_{12\ldots n-2}+1}\non
&&\times\sum_{m_{12} \ldots }(-1)^\kappa
\six{\ell_1}{\ell_2}{\ell_{12}}{m_1}{m_2}{-m_{12}}
\six{\ell_{12}}{\ell_3}{\ell_{123}}{m_{12}}{m_3}{-m_{123}}\cdots\non
&&\qquad\times\six{\ell_{12\ldots n-2}}{\ell_{n-1}}{\ell_{n}}{m_{12\ldots n-2}}{m_{n-1}}{m_{n}}\label{eq:calC2}
\endeqar 
where $\calE(\Lambda)=(-1)^{\sum_i\ell_i}$ and $\kappa=\ell_{12}-m_{12}+\ell_{123}-m_{123}+\cdots+\ell_{12\ldots n-2}-m_{12... n-2}$.  If the sum of the angular momenta is even, then $\calE(\Lambda)=1$ and $\calY_\Lambda$ is real.  Otherwise, $\calE(\Lambda)=-1$ and $\calY_\Lambda$ is imaginary. 
For $n=2$ and $n=3$, $\calC^\Lambda_{\rm M}$ becomes:
\begeqar\label{eqn:clebsch_gorden_N2N3}
\calC^{\ell\ell'}_{mm'}&=&\frac{(-1)^{\ell-m}}{\sqrt{2\ell+1}}\delta^{\rm K}_{\ell,\ell'}\delta^{\rm K}_{m,-m'},\label{eqn:clebsch_gorden_N3}\\
\calC^{\ell_1\ell_2\ell_3}_{m_1m_2m_3}&=&(-1)^{\ell_1+\ell_2+\ell_3}\six{\ell_1}{\ell_2}{\ell_3}{m_1}{m_2}{m_3}\label{eqn:clebsch_gorden_N3},
\endeqar
with $\delta^{\rm K}_{\ell_i \ell'_i}$ being the Kronecker delta. The result in the second line is non-zero only when $\ell_1,\ell_2$, and $\ell_3$  satisfy the triangular inequality, $|\ell_1-\ell_2|\leq\ell_3\leq\ell_1+\ell_2$. Furthermore, if any of the angular momenta are zero, the second line reduces to the first~\cite[eq.~34.3.1]{NIST:DLMF}.

The form of the $C_{\rm M}^\Lambda$ coefficient is chosen to ensure orthonormality of the isotropic basis functions. 
%isotropic function is constructed from the 3-$j$ symbols and the spherical harmonics to guarantee orthonormalilty and completeness. 
The orthonormality relation is:
\begeqar\label{eqn:orthogonal_iso_function}
\int  \,d\bfRhat\, \calY_{\Lambda}(\bfRhat)\calY^*_{\Lambda'}(\bfRhat)=\delta^{\rm K}_{\ell_1\ell'_1}\delta^{\rm K}_{\ell_2\ell'_2}\times\cdots\times\delta^{\rm K}_{\ell_{12}\ell'_{12}}\times\cdots\times\delta^{\rm K}_{\ell_{n}\ell'_{n}}.
\endeqar
Using this, we can expand an arbitrary isotropic function in this basis
\begeqar\label{eqn:f_iso_expand}
\zeta(\bfR)&=&\sum_\Lambda \calZ_\Lambda(R)\calY_\Lambda(\bfRhat),
\endeqar
with $R \equiv \{r_1, r_2, \ldots, r_{n}\}$ and $\bfR \equiv \{\bfr_1, \bfr_2, \ldots, \bfr_{n}\}$.
By invoking the orthonormality relation Eq.~\eqref{eqn:orthogonal_iso_function} we can obtain the expansion coefficient
\begeqar\label{eqn:f_iso_expand}
\calZ_\Lambda(R)&=&\int d\bfRhat\, \zeta(\bfR)\calY^*_\Lambda(\bfRhat).
\endeqar
In our context, $\zeta(\bfR)$ is the $N$-point correlation function. If we expand the function in the basis $\calY_{\Lambda}$, parity-even correlators will have real coefficients, but parity-odd correlators will have purely imaginary coefficients.
%defined as $\delta(\mathbf{x})=\rho(\mathbf{x})/\bar{\rho}-1$.

\subsection{Useful properties}% of the isotropic $\calY_\Lambda$ functions}
\label{subsec:calY-from-avg}
We define some useful quantities derived from the isotropic basis that will be of use later.
Consider a product of $n$ spherical harmonics. If we represent integration over the rotations, $\calR$, by $d \calR$ with $\int d \calR=1$ then, as shown in the previous work~\citep{Cahn202010}, averaging over the rotation group projects out the isotropic components:
\begeqar\label{eq:all-var}
\int d \calR\, \prod_{j=1}^{n}Y_{\ell_jm_j}(\calR\bfrhat_j)=\sum_\Lambda \calC^\Lambda_\rmM\calY_\Lambda(\bfRhat).
\endeqar
The result is non-zero only if $\sum_j m_j=0$ and the $\ell_i$ satisfy a generalized triangular inequality, namely that they can be combined to make a state of zero total angular momentum. The sum over $\Lambda$ includes all possibilities that can be constructed from the given primary $\ell_j$. 

A useful consequence is the identity

\begeqar\label{eq:all-varplus}
\int d\calR \prod_{j=1}^{n}\sum_{m_j}Y_{\ell_jm_j}(\calR\bfrhat_j)Y^*_{\ell_jm_j}(\bfkhat_j)=\sum_\Lambda\sum_{m_j} \calC^\Lambda_\rmM Y^*_{\ell_jm_j}(\bfkhat_j)\calY_\Lambda(\bfRhat)=\sum_\Lambda \calY_\Lambda(\bfRhat)\calY^*_\Lambda(\bfKhat),
\endeqar
where $\bfKhat=\{\bfkhat_1,...,\bfkhat_{n}\}$, and the sum is over all $\Lambda$ that can be constructed from the primary $\ell_j$.

The rotational average of a product of $n$ spherical harmonics with a common argument is determined in a similar fashion:
\begeqar\label{eq:one-var}
\int d\calR \prod_{j=1}^n Y_{\ell_jm_j}(R\bfrhat)&=&(4\pi)^{-n/2}\prod_{j=1}^n\sqrt{2\ell_j+1} \sum_\Lambda\calC^\Lambda_\rmM\calC^\Lambda_\bfzero\non
&=&(4\pi)^{-n/2}\sum_\Lambda\calD^{\rm P}_\Lambda\calC^\Lambda_\bfzero\calC^\Lambda_\rmM,
\endeqar
where ${\rm M}$ stands for all the $m_j$ and the subscript $\bfzero \equiv \{0,0,...\}$, and the sum is over all $\Lambda$ consistent with the given $\ell_i$ (by the introduction of intermediate $\ell_{12}$, etc.).  We have defined the following coefficient involving the primary angular momenta:%a coefficient of the products of square root of primary angular momenta
\begeqar\label{eqn:defcalD}
\calD^{\rm P}_\Lambda=\prod_{j=1}^n\sqrt{2\ell_j+1}.
\endeqar
The superscript ${\rm P}$ stands for ``primary". Since we will use it often, we write out $\calD^{\rm P}_\Lambda$ for $n=3$ explicitly:
\begeqar\label{eqn:defcalD_var}
\calD^{\rm P}_{\ell_i\ell'_i\ell''_i} = \sqrt{(2\ell_i+1)(2\ell'_i+1)(2\ell''_i+1)}.
\endeqar

When calculating the covariance matrix, we will encounter pairs of galaxy $n$-tuplets involving directional vectors $\{\bfr_i\}$ and $\{\bfr'_i\}$ with origins separated by a vector $\bfs$. Hence it is practical to consider a product of $n$ isotropic $\calY$ functions of three arguments
\begeqar
    \prod_{i=1}^n \calY_{\ell_i\ell'_i\ell''_i}(\bfrhat_i,\bfrhat'_i,\bfshat)=\prod_{i=1}^n\sum_{m_i,m'_im''_i}
    \calC^{\ell_i\ell'_i\ell''_i}_{m_im'_im''_i}Y_{\ell_im_i}(\bfrhat_i)Y_{\ell'_im'_i}(\bfrhat'_i)Y_{\ell''_im''_i}(\bfshat).
\endeqar
Since the isotropic basis does not encode the absolute orientation of each galaxy $n$-tuplet, we can average over orientation of the $\bfrhat_i$, $\bfrhat'_i$, and $\bfshat$ via Eq.~\eqref{eq:all-var}  with the relative orientations of directional vectors within each galaxy $n$-tuplets fixed. Following this, we find
\begeqar\label{eq:averaging}
    &&\int d\calR\, d\calR'\, d\calS\, \prod_{i=1}^n \calY_{\ell_i\ell'_i\ell''_i}(\bfrhat_i,\bfrhat'_i,\bfshat)\non
    &&\qquad\quad = (4\uppi)^{-n/2}\sum_{\Lambda\Lambda'\Lambda''}\calQ^{\Lambda\Lambda'\Lambda''}\calD^{\rm P}_{\Lambda''}\calC^{\Lambda''}_\bfzero\calY_\Lambda(\bfRhat)\calY_{\Lambda'}(\bfRhat').
\endeqar
where the $\Lambda$, $\Lambda'$, and $\Lambda''$ are formed from the primary components $\ell_i$, $\ell'_i$, and $\ell''_i$ respectively. 
We introduce the quantity
\begeqar\label{eqn:defcalQ}
\calQ^{\Lambda\Lambda'\Lambda''}=\prod_{i=1}^n\sum_{m_i,m'_i,m''_i}\calC^{\ell_i\ell'_i\ell''_i}_{m_im'_im''_i}
\calC^\Lambda_\rmM \calC^{\Lambda'}_{\rmM'} \calC^{\Lambda''}_{\rmM''},
\endeqar
where the subscripts $\rmM$, $\rmM'$, $\rmM''$ stand for collections of $\{m_i\}$, $\{m'_i\}$, and $\{m''_i\}$. Since $\calC^{\ell_i\ell'_i\ell''_i}_{m_im'_im''_i}$ has a mixture of angular momenta we write out its components explicitly.

Our goal in this work is to study the covariance matrix, which by definition involves products of spherical harmonics. For this reason, it is of use to explore products of isotropic functions and their corresponding identities.
Since the $\calY_\Lambda$ is a complete basis, it is possible to write products of two isotropic basis function with the same argument as a sum of isotropic basis function weighted by a coupling coefficient
\begeqar
\calY_\Lambda(\bfRhat)\calY_{\Lambda'}(\bfRhat)=\sum_{\Lambda''} \calE(\Lambda'')\,\calG^{\Lambda\Lambda'\Lambda''}\calY_{\Lambda''}(\bfRhat),
\endeqar
where the phase in the coefficient arises due to the conjugation property of the isotropic function $\calY_{\Lambda''}^*(\bfRhat)=\calE(\Lambda'')\calY_{\Lambda''}(\bfRhat)$ and we define $\calG^{\Lambda\Lambda'\Lambda''}$ as the generalized Gaunt integral~\citep{Cahn202010}:
%the conjugation identity gives $\calY^*_\Lambda(\bfRhat)=\calE(\Lambda'')\calY_\Lambda(\bfRhat)$. 
\begeqar\label{eqn:generalized-gaunt}
\calG^{\Lambda\Lambda'\Lambda''}&\equiv&\int d\bfRhat\, \calY_\Lambda(\bfRhat)\calY_{\Lambda'}(\bfRhat)\calY_{\Lambda''}(\bfRhat)\non
&=&(4\uppi)^{-n/2}\left[\prod_{i=1}^n\calD^{\rm P}_{\ell_i\ell'_i\ell''_i}\calC^{\ell_i\ell'_i\ell''_i}_{000}\right]\calQ^{\Lambda\Lambda'\Lambda''}.
\endeqar
From its definition we see that $\calG^{\Lambda\Lambda'\Lambda''}$ is symmetric in $\Lambda,\Lambda',\Lambda''$;
we include its explicit evaluation for $n=2,3$ and $4$ in Appendix \ref{appendix:generalized_gaunt_N234}.
%Note that if one of $\Lambda,\Lambda',\Lambda''$ is $\bfzero = \{0,0,..0\}$ the corresponding $\calY$ is simply $(4\pi)^{-N/2}$.  
%Then the orthonormality of the $\calY_\Lambda$ determines
% \begeqar
%     \calG^{\Lambda\Lambda'\Lambda''}=(4\pi)^{-N/2}.
% \endeqar

The isotropic function is expressed with arguments $\bfrhat_1, ..., \bfrhat_n$ with the canonical ordering $i=1,\ldots,n$ (index sorted from small to large in $\bfr_i$). 
%\textcolor{red}{OP: the fact that we sort from small to large $r_i$ is irrelevant, we don't need to impose it}) \jiamin{we need it for defining the "reordering". To avoid degeneracies we have used the $r_1<r_2<r_3$}\textcolor{red}{OP: We don't need it to define the reordering. We can always permute e.g. r1 and r2 to have r1 first no matter what the relevant sizes of r1 and r2 are; we're just switching variables around. Anyway, this isn't super relevant since we *do* impose r1<r2<r3 to avoid degeneracies here.}.
When we later consider the covariance, the contraction of the overdensity fields may be permuted such that the canonical ordering of the indices is no longer guaranteed. The isotropic functions with permuted arguments can be expanded in terms of the canonically ordered ones (since these latter form a complete basis) as:
\begeqar\label{eqn:perm-unperm-basis}
\calY_{\Lambda}(\bfRhat_{G})= \sum_{\Lambda'} \calB^{G^{-1}}_{\Lambda,\Lambda'} \calY_{\Lambda'}(\bfRhat'),
\endeqar
where $G$ denotes the permutation of the set $\{1,2,\ldots,n\}$. 
%\bob{should this be $(N-1)$ here?} \jiamin{yes} 
The reordering coefficient of the inverse permutation, $\calB_{\Lambda,\Lambda'}^{G^{-1}}$, can be obtained by applying the orthogonality relation %\bob{This isn't correct in general. See, for example, Sect. 1.4 of 'revision.'} \jiamin{changed.}
%\bob{I restored the $G$ in the upper index of $\calB$.} \jiamin{ok, I've changed this notation everywhere below as well.}
\begeqar\label{eqn:generalized-kronecker-delta}
\calB_{\Lambda,\Lambda'}^{G^{-1}} &\equiv& \int d\bfRhat\, \calY_{\Lambda}(\bfRhat_{G}) \calY^*_{\Lambda'}(\bfRhat')\non
&=& \sum_{\rm M}\, \calC^{\ell_{1}\ell_{2}\ell_{12}\ldots\ell_{n}}_{m_{1}m_{2}m_{12}\ldots m_{n}}\;\calC^{\ell_{G1}\ell_{G2}\ell'_{12}\ldots\ell_{Gn}}_{m_{G1}m_{G2}m'_{12}\ldots m_{Gn}} \prod_{i=1}^{n} \delta^{\rm K}_{\ell'_i\ell_{Gi^{\scaleto{-1\mathstrut}{4pt}}}},
\endeqar
where $G^{-1}$ denotes the inverse permutation of $G$. Here, products of Kronecker deltas ensure that $\Lambda$ and $\Lambda'$ have the same primary angular momenta; however, they may still differ in intermediate angular momenta. 

%\jiamin{In comparison to Bob's notation, $\Lambda$ is the intrinsic ordering  $\Lambda_G$ is the permuted ordering using the permutation rule $G$, the superscript $-1$ denotes inverse permutation. Applying the inverse permutation to the permuted order we recover the original ordering $G^{-1}[\ell_{Gi}]\equiv [\ell_{Gi}]^{-1}=\ell_i$, therefore there is not permutation appearing in the Kronecker delta in comparison to~\citet[equation (80)]{Cahn202010}, but we do need the $G$ to define the permutation rule.}\bob{The symbol $\calB$ must have $G^{-1}$ in the exponent. Otherwise it is undefined.  }

\section{$N$-point correlation functions}% in the isotropic basis}
\label{sec:npcf}
The $N$-point correlation function (NPCF) is defined as  
\begeqar\label{eqn:zeta_npoint_definition}
    \zeta(\vr_1,\vr_2,...,\vr_{N-1}) 
    \equiv \left<  \delta(\vx)\delta(\vx+\vr_1)\delta(\vx+\vr_2)\cdots\delta(\vx+\vr_{N-1})\right>,
\endeqar
where the galaxy overdensity is given by $\delta({\bfx})=n({\bfx})/\bar{n}-1$, with $n({\bfx})$ the galaxy number density with mean $\bar{n}$ and $\av{\delta}=0$. The angle bracket denotes the ensemble average of the overdensity field. 

The expectation value in Eq.~\eqref{eqn:zeta_npoint_definition} can be expanded as a sum of combinations of overdensity fields at different spatial positions.
%by applying the Wick theorem to the classical fields (such as our overdensity fields) (\jiamin{Isserlis theorem states that higher-order moments of the multivariate normal distribution can be calculated in terms of its covariance matrix, which is not our case.})\bob{I think this has nothing to do with Isserlis since we haven't imposed the assumption of a Gaussian random field}\bob{I would prefer to say that we define the connected piece by removing all the products of pairs of fields}\bob{In the case of GRF, the connected piece is zero}. \jiamin{agree. This is a reminder of myself and because I know Oliver will comment on Wick vs. Isserlis :)} 
In the $N=4$ case, the full 4PCF reads
\begeqar\label{eqn:4pcf-c-dc}
\zeta(\bfr_1, \bfr_2, \bfr_3) &=& \xi(\bfr_1)\xi(\bfr_2-\bfr_3) + \xi(\bfr_2)\xi(\bfr_1-\bfr_3) + \xi(\bfr_3)\xi(\bfr_1- \bfr_2) + \zeta^{\rm c}(\bfr_1, \bfr_2, \bfr_3)\non
&\equiv& \zeta^{\rm dc}(\bfr_1, \bfr_2, \bfr_3) + \zeta^{\rm c}(\bfr_1, \bfr_2, \bfr_3),
\endeqar
which consists of two parts.  The \textit{connected} four-point function $\zeta^{\rm c}(\bfr_1, \bfr_2, \bfr_3)$ captures the non-Gaussian part of the signal. We denote the other terms, composed of the products of two-point correlation functions, as the \textit{disconnected} part,  $\zeta^{\rm dc}(\bfr_1, \bfr_2, \bfr_3)$. For $N=4$ the disconnected terms coincides with the 2PCF that sourced by Gaussian statistics. For $N>4$ however, the disconnected piece can also receive non-Gaussian contributions, such as 2PCF$+$3PCF for the 5PCF. Our interest here is the non-Gaussianity induced by the higher order statistics. For this purpose, we employ a connected-only estimator that subtracts all the disconnected pieces at the estimator level (for details regarding the connected-only estimator, see our companion paper~\citealt{Philcox2021boss4pcf}). 

In the limit of large volumes, $V$, we can replace the ensemble average by a spatial integral by invoking ergodicity. This motivates the general NPCF estimator
\begeqar
    \hat\zeta(\bfr_1,\bfr_2,\ldots, \bfr_{N-1}) = \int \frac{d\vx}{V}\,\delta(\vx)\delta(\vx+\vr_1)\delta(\vx+\vr_2)\cdots\delta(\vx+\vr_{N-1}),
\endeqar
which is unbiased. Using orthonormality to project this onto the isotropic basis $\calY_{\Lambda}$ (using $n = N-1$) for given primary angular momenta $\L\equiv\{\ell_1,\ell_2, (\ell_{12}), ..., \ell_{N-1}\}$ as in Eq.~\eqref{eqn:f_iso_expand}, we obtain the estimator
\begeqar\label{eqn:zeta_npcf}
    \hat\zeta_\L(r_1,r_2,...,r_{N-1}) &=& \int \frac{d^3\vx}{V} \,\delta(\vx) \prod_{i=1}^{N-1} \int d\bfrhat_i \delta(\vx+\vr_i)\P_\L^*(\bfrhat_1,\bfrhat_2,..., \bfrhat_{N-1}).
%    &=&\sum_{\rm M} \calC_{\rm M}^{\Lambda} \int \frac{d\vx}{V}\,\delta(\vx)\prod_{i=1}^{N-1} \left[\int d\hr_i\,\delta(\vx+\vr_i)Y_{\ell_i m_i}^*(\hr_i)\right].
\endeqar
% In practice our estimator is binned at given directional vectors $\vr_i$, 
% \begeqar\label{eq: radial-binning-npcf}
%     \hat\zeta_\L^{\rm rbin-N} &=& {(4\uppi)^3}\int \left[\prod_i^{N-1} r_i^2dr_i\,\Theta^{n_i}(r_i)\right]\hat\zeta_\L(r_1,r_2,...,r_{N-1})\bigg/\int \left[\prod_i^{N-1} r_i^2dr_i\,\Theta^{n_i}(r_i)\right],
% \endeqar
% where $\Theta^{n_i}(r_i)$ is a binning function which is unity if $r$ is in $i$th bin and zero else. 
Explicitly, for the 4PCF, we find:% the general estimator reduces to
\begeqar\label{eqn:zeta_4pcf}
    \hat\zeta_\L(r_1,r_2,r_3) =
    \int \frac{d^3\vx}{V}\,\delta(\vx) \int d\bfrhat_1 d\bfrhat_2 d\bfrhat_3\,\delta(\vx+\vr_1)\delta(\vx+\vr_2)\delta(\vx+\vr_3)\calY^*_{\ell_1\ell_2\ell_3}(\bfrhat_1,\bfrhat_2,\bfrhat_3).
%    (-1)^{\ell_1+\ell_2+\ell_3}\sum_{m_1m_2m_3}\calC^{\ell_1\ell_2\ell_3}_{m_1m_2m_3}\int \frac{d\vx}{V}\,\delta(\vx)\prod_{i=1}^3 \left[\int d\hr_i\,\delta(\vx+\vr_i)Y_{\ell_i m_i}^*(\hr_i)\right],
\endeqar
% and now the binned estimator at radial bins $a,b,c$ is given by
% \begeqar\label{eq: radial-binning-4pcf}
%     \hat\zeta_\L^{\rm rbin-4} &=& \frac{(4\pi)^3}{v_av_bv_c}\left[\int r_1^2dr_1\,\Theta^a(r_1)\int r_2^2dr_2\,\Theta^b(r_2)\int r_3^2dr_3\,\Theta^a(r_3)\right] {\hat\zeta_\L(r_1,r_2,r_3)},
% \endeqar
% where $\Theta^{a,b,c}(r)$ is a binning function which is unity if $r$ is in bin $a$ or $b$ or $c$ and zero else. \bob{didn't define $v_{a,b,d}$}.  

Throughout this paper we make two important assumptions.
First, we work in the Gaussian limit for the covariance calculation. Even though the gravitationally-induced higher-order statistics entering the covariance in principle do not vanish, we assume they are suppressed compared to the two-point statistics.
This assumption greatly simplifies the derivation below as we will only need to consider the contractions between two overdensity fields, and thus may express results entirely in terms of the 2PCF or the power spectrum. 
%We will leave the extension of including those higher-order statistics of the covariance to the future work. 
This assumption will be addressed below by comparing the Gaussian covariance to that obtained from N-body simulations.
Second, we assume the 2PCF, and likewise the power spectrum, are isotropic.
The majority of the paper is based on this assumption, however, \S\ref{sec:numerics_and_simulations} includes a comparison between the theoretical isotropic Gaussian covariance numerical simulations including RSD, which breaks rotational invariance. %We further include an extension of the analytic covariance modeling to include RSD in Appendix \ref{appendix:cov_fc_rsd_iso}.OP: no need to repeat this!

We use the following conventions for Fourier transforms:
\begeqar
\tilde{\delta}(\mathbf{k})=\int {d}^{3} \mathbf{r}\; e^{-i \mathbf{k} \cdot \mathbf{r}} \delta(\mathbf{r}), \quad \delta(\mathbf{r})=\int_{\mathbf{k}} e^{i \mathbf{k} \cdot \mathbf{r}}\; \tilde{\delta}(\mathbf{k}),
\endeqar
where we define $ \int_{\mathbf{k}} \equiv (2\uppi)^{-3}\int d^3 \bfk$.
The 2PCF $\xi(\vr)$ and power spectrum $P(\vk)$ are related by
\begeqar\label{eqn:iso_pk_Fourier}
    \left<{\delta(\vr_i)\delta(\vr_j)} \right> = \xi(|\vr_i-\vr_j|) &=& \int_{\vk}P(\bfk)\;e^{i\vk\cdot(\vr_i-\vr_j)}.
\endeqar
Hereafter, we assume isotropy, and thus assume $P(\vk)\equiv P(k)$, with $k=|\bfk|$, and $\xi(\vr)\equiv\xi(r)$. In Appendix \ref{appendix:cov_fc_rsd_iso} we will discuss how to go beyond the assumption of an isotropic power spectrum.
%\bob{I want to understand our notation better.  Eq. (21) equates three things.  The last, $\int dx/V ...$ seems to me to indicate a finite evaluation while the <  > suggests the limiting, ultimate value, perhaps even without being limited to measurements made within the limits of our accessible universe.  I think the distinctions are important at least in Oliver's demonstration that for his definition of the connected piece there are no contributions to the covariance from auto-contracted pieces.} \jiamin{Now explained under the Eq. (21).}

\section{Derivation of the Gaussian NPCF Covariance Matrices}%Gaussian NPCF covariance in the isotropic basis}
\label{sec:cov_gaussian}
% \begin{figure}
%     \centering
%     \includegraphics[width=.7\textwidth]{figs/coord_x.pdf}
%     \caption{Coordinate convention. $\bfx$ denotes the absolute coordinate. $\bfr_i$ and $\bfr'_j$ are the relative coordinates for the unprimed and primed families, respectively. $\bfs$ is the separation vector between the two families. We take the limit that $\bfr_0=\bfr'_0=0$.}
%     \label{fig:coord_x}
% \end{figure}

% \begeqar
% \mathbb{E}_{\delta}\left[\hat{\calZ}(\bfR)\hat{\calZ}^*(\bfR')\right]
% \endeqar

The covariance matrix for the NPCF is defined as 

\begeqar\label{eqn:npcf_cov_definition}
\cov(\hat{\calZ}(\bfR), \hat{\calZ}(\bfR'))
%  &&\qquad=\int \prod_{i=0}^{N-1}\left[\frac  {d^3\bfx_id^3\bfx'_i}{V^{2N}}\right] \left<
%     \delta(\bfx_{0})\delta(\bfx_{1})...\delta(\bfx_{N-1})\delta(\bfx'_0)\delta(\bfx'_1)...\delta(\bfx'_{N-1})\right>\non
% && \qquad\qquad - \left< \delta(\bfx_{0})\delta(\bfx_{1})...\delta(\bfx_{N-1})\right> \left< \delta(\bfr'_0)\delta(\bfx'_1)...\delta(\bfx'_{N-1})\right>\non
% && \qquad\qquad - \left< \delta(\bfr_0)\delta(\bfr_{1})...\delta(\bfr_{N-1})\right> \left< \delta(\bfs+\bfr'_0)\delta(\bfs+\bfr'_1)...\delta(\bfs+\bfr'_{N-1})\right>
% &&\qquad = \int \frac {d^3\bfx d^3\bfx'}{V^2}\left<
%     \delta(\bfr_0+\bfx)\delta(\bfr_{1}+\bfx)...\delta(\bfr_{N-1}+\bfx)\delta(\bfx'+\bfr'_0)\delta(\bfx'+\bfr'_1)...\delta(\bfx'+\bfr'_{N-1})\right>\non
% && \qquad\qquad - \left< \delta(\bfx+\bfr_0)\delta(\bfx+\bfr_{1})...\delta(\bfx+\bfr_{N-1})\right> \left< \delta(\bfx'+\bfr'_0)\delta(\bfx'+\bfr'_1)...\delta(\bfx'+\bfr'_{N-1})\right>\non
&\equiv& \left< \hat{\calZ}(\bfR)\hat{\calZ}^*(\bfR')\right> - \left<\hat{\calZ}(\bfR)\right> \left<\hat{\calZ}^*(\bfR') \right>\non
&=& \int \frac{d^3\bfx}{V}\frac{d^3\bfx'}{V} \left<\prod_{i=0}^{N-1} 
    \delta(\bfx+\bfr_i)\delta(\bfx'+\bfr'_i)\right> - \left<\hat{\calZ}(\bfR)\right> \left<\hat{\calZ}^*(\bfR') \right>\non
&=& \int \frac{d^3\bfs}{V} \left<\prod_{i=0}^{N-1} 
    \delta(\bfx+\bfr_i)\delta(\bfx+\bfr'_i+\bfs)\right> - \left<\hat{\calZ}(\bfR)\right> \left<\hat{\calZ}^*(\bfR') \right>,
% &&\qquad = \int \frac {d^3\bfs}{V}\left<
%     \delta(\bfx+\bfr_0)\delta(\bfx+\bfr_{1})...\delta(\bfx+\bfr_{N-1})\delta(\bfx+\bfs+\bfr'_0)\delta(\bfx+\bfs+\bfr'_1)...\delta(\bfx+\bfs+\bfr'_{N-1})\right>\non
% && \qquad\qquad - \left< \delta(\bfx+\bfr_0)\delta(\bfx+\bfr_{1})...\delta(\bfx+\bfr_{N-1})\right> \left< \delta(\bfx+\bfs+\bfr'_0)\delta(\bfx+\bfs+\bfr'_1)...\delta(\bfx+\bfs+\bfr'_{N-1})\right>, 
\endeqar
%\textbf{Oliver: I also have some issue with this. In order to integrate over $\bfx$ and $\bfx'$ in the first line, you'd need some extra Dirac deltas to restrict the separations to be $\bfr_i$ etc. I think it really only needs an integral over $\bfs$ and $\bfs'$, where one of them can be integrated out. Also, I don't think there's any need for a limit of infinite volume here. Since we defined the NPCF in a volume $V$, the covariance also has that same volume $V$ (and also, the $\calZ$ on the LHS should have estimator hats on for that reason)} \jiamin{this equation is now cleaned up}.
where $\hat{\calZ}(\bfR)$ is the NPCF estimator with coordinates $\bfR=\{\bfr_1, \bfr_2, ..., \bfr_{N-1}\}$, with an analogous definition for $\bfR'$. Going from the second to the third line we have defined the separation vector between the primary galaxies of the two $N$-plets as $\bfs\equiv\bfx'-\bfx$, and dropped the spatial integral over $\vx$, assuming statistical homogeneity.\footnote{Strictly, we first need to apply a Poisson average to discrete tracers, giving rise to the shot noise term. Here we use abbreviated notation and replace $P(k)\to P(k)+\bar{n}^{-1}$, for number density $\bar{n}^{-1}$, when we later compare our analytic results to those from the simulations.} 
% In the large volume limit, we can replace the integral over the absolute coordinates of individual overdensity fields with the separation vector $\bfs$ between the primed and unprimed family.
We label the vertices containing $\bfr_0$ and $\bfr_0'$ as {\it primary vertices} (with $\bfr_0=\bfr_0'=\mathbf{0}$) and label the $(N-1)$ points with separations $\bfr_1, \bfr_2,\ldots,\bfr_{N-1}$ relative to the primary as {\it endpoints}.
%$\bfr_0$, and analogously for the primed fields. We term the fields with indices from one upwards {\it endpoints}. 
%Without loss of generality we can send $\bfr_0, \bfr_0'\to 0$.
In the Gaussian limit we only need to calculate contractions between pairs of overdensity fields. The NPCF covariance has $2N$ overdensity fields and thus forms $N$ pairs of contractions. 
%\bob{I dropped Wick} \jiamin{Ok.}
%zslepian: I don't think this is correct---Poisson averaging is that we have one underlying density field, but do multiple samples of it. I don't think this is what is done for the covariance, even when estimated from Patchy mocks, no?
%jiamin: I agree on the definition of Poisson average and the expectation occurs at the definition of the covariance is an ensemble average. My understanding is that when we define the correlation function for discrete sample, we first perform the Poisson average (there we see how shot noise arises), then we take the ensemble average which gives us 2PCF.

Whereas the definition of the covariance matrix given in Eq.~\eqref{eqn:npcf_cov_definition} (evaluated under the assumption of Gaussianity) includes all possible contractions of $2N$ density fields, in this section we consider only pairs that are contractions between unprimed and primed families, \textit{i.e.} between $\bfr_i$ and $\bfr_j'$. We term these contractions (and the corresponding covariance matrix contribution) ``fully-coupled'', as they fully couple the unprimed and primed families. 
Any self-contraction (\textit{i.e.} that involving contraction of two density fields within the same family, \textit{i.e.} between $\bfr_i$ and $\bfr_j$ with $i\neq j$) arises from the disconnected contributions to the NPCF. We term any covariance contribution that includes at least one self-contraction ``partially-coupled''. All such contributions vanish in the covariance of the connected-only estimator~\citep{Philcox2021boss4pcf}. This fact allows us to focus on the fully-coupled covariance terms.\footnote{With the introduction of the connected-only estimator implies that the disconnected terms can be isolated and that calculation of their associated partially-coupled covariance is not strictly needed, we provide its derivation in Appendix.~\ref{appendix:intro_cov_pc} for completeness.}

Below, we derive a general expression for the fully-coupled NPCF covariance matrix under the assumption that the density fields are Gaussian distributed. %Then we will show the fully-coupled 4PCF covariance matrix. 
In order to offer a more intuitive understanding of the coupling structure, we also present a diagrammatic approach to the calculation.

\subsection{Basic elements for the covariance}
\label{subsec:basic-cov-elements}
%In the Gaussian limit, the statistical properties of $\delta$ are fully captured by the two point statistics, which means we need to consider only the contraction of two overdensity fields. 

\begin{figure*}
    \centering
    \includegraphics[width=.93\textwidth]{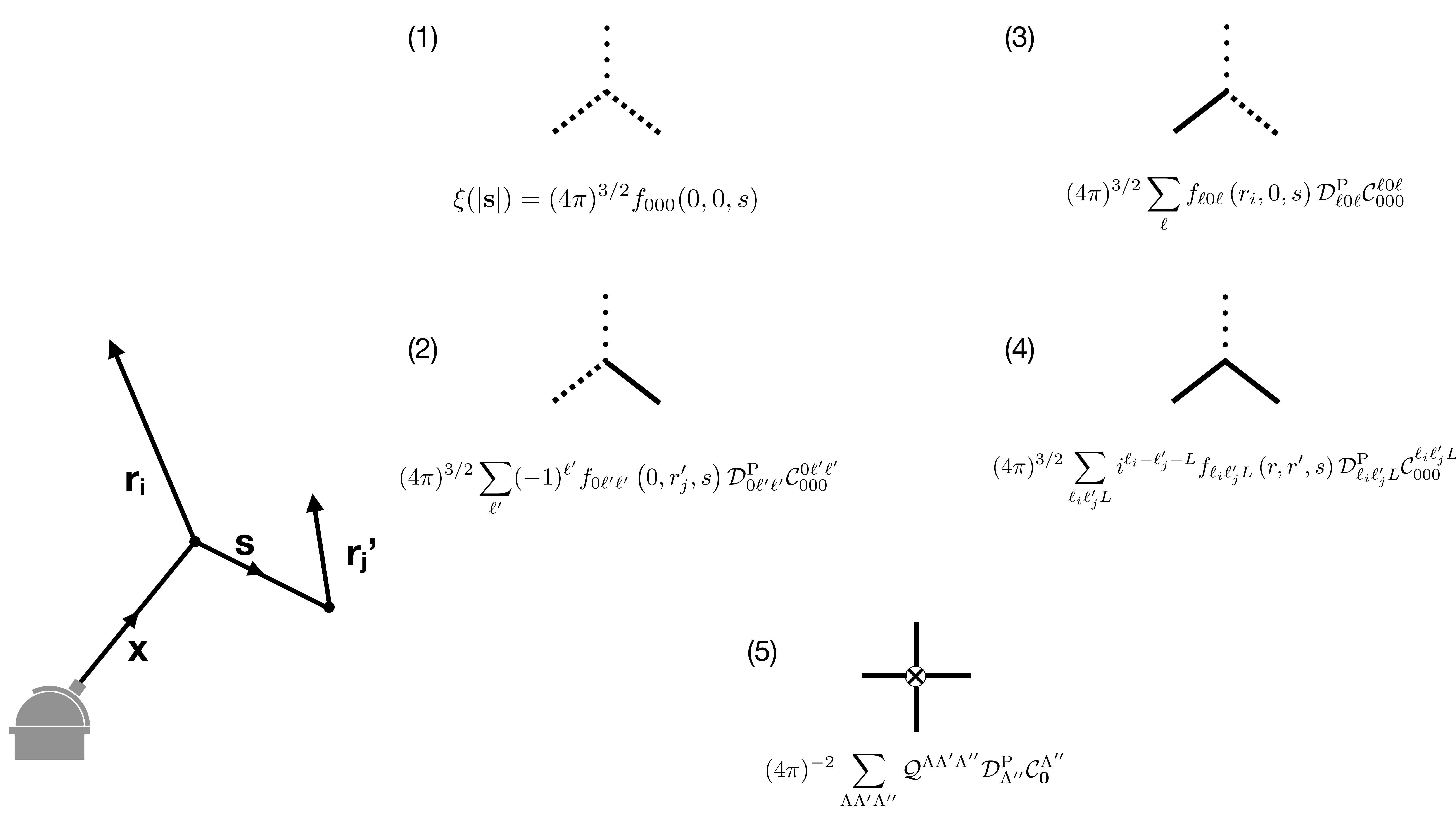}
    \caption{A diagrammatic representation of the basic elements used as building blocks for the fully-coupled (\textit{i.e.} connected) covariance. Coupling between the overdensity fields across the unprimed and primed family (corresponding to density fields from the first and second NPCFs in Eq.~\ref{eqn:npcf_cov_definition}) is represented by a tripolar structure ({\bf diagrams (1)-(4)}, cf. Eqs.~\ref{eqn:2pcf_r_rp_s}-\ref{eqn:2pcf_r_r0ps}). Each tripolar structure depends on three vectors: $\bfr_i$, $\bfr'_j$, and $\bfs$. We use dotted lines to represent the separation vector $\bfs$. Dashed lines depict primary vertices for $\bfr_0$ or $\bfr'_0$ and solid lines are for endpoints with $i$ or $j$ non-zero. {\bf Diagram (5)} is the coupling kernel arising from the rotational average over the unit vectors $\bfr$, $\bfr'$, and $\bfs$ (cf. second line in Eq.~\ref{eqn:general-contractions2}). In the $N=4$ case the coupling kernel has four legs. The lower left diagram (with the cartoon telescope) shows our coordinate convention. $\bfx$ denotes the absolute coordinate; $\bfr_i$ and $\bfr'_j$ are the relative coordinates for respectively the unprimed and primed families. $\bfs$ is the separation vector between the two families.}
    %\bob{In (5) I think you should write $(4\pi)^-{4/2}$ since this is just for N=4. Actually I think the caption is fine as is.} \jiamin{ok, changed.}}
    \label{fig:basic_coupling_elements}
\end{figure*}

We first consider the coupling between two endpoints: specifically, $\delta(\bfx+\bfr_i)$ from the unprimed family and $\delta(\bfx+\bfs+\bfr_j')$ from the primed family, with $i$ and $j$ between $0$ and $N-1$. Such a contraction is represented by the tripolar structure in Fig.~\ref{fig:basic_coupling_elements}. The primary vertices, $\bfr_0$ and $\bfr_0'$, are indexed as a convenience for keeping track of the permutations of unprimed and primed density fields; we will need such permutations later in the calculation. However, once we have computed our desired contractions in the isotropic basis, we may evaluate the result at $\bfr_0=0$ and $\bfr_0' = 0$, since we place the primary vertices at $\bfx$ and $\bfx'$ respectively. We display this approach in Fig. \ref{fig:basic_coupling_elements}.

Expanding the contraction $\langle \delta(\bfx+\bfr_i) \delta(\bfx+\bfs+\bfr'_{j}) \rangle$ in the isotropic basis, we find
\begeqar\label{eqn:2pcf_r_rp_s}
&&\langle \delta(\bfx+\bfr_i) \delta(\bfx+\bfs+\bfr'_{j}) \rangle 
\equiv \xi(|\bfr'_{j}+\bfs-\bfr_i|) \non
%&&= \int_{\bfk} P(k) e^{i \bfk \cdot (\bfr'_{j} + \bfs - \bfr_i)} \non
&&= \mathcolorbox{color26}{(4\uppi)^{3/2} \sum_{\ell_i \ell'_{j} L} i^{-\ell_i+\ell'_{j}+L} f_{\ell_i \ell'_{j} L}(r_i, r'_j, s)\; \mathcal{D}^{\rm P}_{\ell_i \ell'_j L} \;\mathcal{C}^{\ell_i \ell'_j L}_{000}} \;\mathcal{P}_{\ell_i \ell'_j L}(\bfrhat_i, \bfrhat'_j, \bfshat).
\endeqar
%where we have used the plane wave expansion %~\citep{Arfken2013mathematical} no need to give citation for elementary result
%to go from the second to the third line. 
A detailed derivation of this is in Appendix \ref{appendix:basic_elements}. The highlighted radial part corresponds to Fig.~\ref{fig:basic_coupling_elements} {\bf diagram (4)}.
To simplify what follows, we introduce the $f$-integral:
\begeqar\label{eqn:f3l-tensor}
f_{\ell_1\ell_2\ell_3}(r_1,r_2,r_3)\equiv\int\frac{k^2 dk}{2\uppi^2}P(k)j_{\ell_1}(kr_1)j_{\ell_2}(kr_2)j_{\ell_3}(kr_3),
\endeqar 
following equation (64) in~\citet{Slepian201506}. In practice, this is computed in radial bins, wherein we average each spherical Bessel function (sBF) over $r_i$ with weight $r_i^2$ (cf. Eq.~\ref{eqn:sbf_bin_avg}).\footnote{Importantly, the bin average commutes with the integral and can be done prior to the $k$ integration, which avoids performing the integral over fine radial bins.}
%. This means that the cost of computing the $f$-integrals scales as $N_k N_{\rm bins}^3$, where $N_k$ the number of $k$ sample points and $N_{\rm bins}$ the number of bins.\jiamin{we do not loop $N_k$. OP: the code does (numpy arrays are just C++ for loops) Jiamin: if the array size fits the memory there is barely difference in various $N_k$ when using numpy.sum (numpy manages to save the data in a contiguous block). However, the loop over $N_{\rm bins}^2$ do cost computational time.}} 
%We see from Eq.~\eqref{eqn:2pcf_r_rp_s} that a contraction of two overdensity fields expressed in the isotropic basis calls for sums over this $f$-integral multiplied by a Wigner 3-$j$ symbol. \

We now consider the forms of equation (\ref{eqn:2pcf_r_rp_s}) when $i$ and $j$ assume different values. There are three distinct cases.
First, we have a primary-to-primary coupling (the highlighted radial part corresponds to Fig.~\ref{fig:basic_coupling_elements}, {\bf diagram (1)}):
\begeqar
\left<\delta(\bfx+\bfr_0)\delta(\bfx+\bfs+\bfr'_0)\right>|_{r_0=r_0'=0}=\xi(|\bfr_0' + \bfs - \bfr_0|)|_{r_0=r_0'=0}=\mathcolorbox{color26}{(4\uppi)^{3/2}f_{000}(0,0,s)}\calY_{000}(0,0,\bfshat),\label{eqn:2pcf_s}
\endeqar
Second, we have a primary-to-endpoint coupling. These couplings can be obtained by taking one of $\bfr_0$ or $\bfr'_0$ and their associated angular momenta to zero.
In the first line below, the primary is unprimed and the endpoint is primed. In the second line, we give the alternate choice, easily obtained by symmetry. Below, the highlighted radial parts correspond to Fig.~\ref{fig:basic_coupling_elements}, {\bf diagrams (2) and (3)}. We have:
\begeqar\label{eqn:2pcf_r_r0ps}
&&\left<\delta(\bfx+\bfr_0)\delta(\bfx+\bfs+\bfr'_{j})\right>|_{\bfr_0=0}=\xi(|\bfr'_{j}+\bfs - \bfr_0|)|_{\bfr_0=0} 
= \mathcolorbox{color26}{(4\uppi)^{3/2}\sum_{\ell'} (-1)^{\ell'}
f_{0\ell'\ell'}(0,r'_{j},s)\calD^{\rm P}_{0\ell'\ell'}\calC^{0\ell'\ell'}_{000}}\calY_{0\ell'\ell'}(0,\bfrhat'_j, \bfshat)\\
&&\left<\delta(\bfx+\bfr_i)\delta(\bfx+\bfr'_{0}+\bfs)\right>|_{r_0'=0}
= \xi(|\bfs +\bfr'_0 - \bfr_{i}|)|_{r_0'=0}
= \mathcolorbox{color26}{(4\uppi)^{3/2}\sum_{\ell} 
f_{\ell 0\ell}(r_i,0,s) \calD^{\rm P}_{\ell 0 \ell}\calC^{\ell0\ell}_{000}}\calY_{\ell0\ell}(\bfrhat_i,0, \bfshat).
\endeqar
Finally, we have an endpoint-to-endpoint coupling , which is already given by Eq.~\eqref{eqn:2pcf_r_rp_s}. 
%To understand the coupling structure more intuitively, Fig.~\ref{fig:basic_coupling_elements} shows the basic elements we will encounter when deriving the fully-coupled covariance matrix. 
%As mentioned above, the NPCF can be specified by a primary vertex with $(N-1)$ endpoints. As we will integrate over the separation between the primary vertices, without loss of generality, we can set $\bfr_0=\bfr'_0=0$.

%OP: below should be in figure captions, not text. 
%As noted above, the contraction of two overdensity fields between the primed and unprimed family gives rise to a tripolar structure. Each leg represents one argument of the correlation function. We show $\bfs$ as a dotted line, and any endpoints as dashed lines. If the argument is $\bfr_0$ or $\bfr_0'$, we show it with a dashed line.
%As all the pair contractions we encounter do not depend on the absolute coordinate $\bfx$, we will drop the $\bfx$ argument when expressing the ensemble average of two overdensity fields $\av{\ldots}$ hereafter. 
%If the contraction happens with overdensity field being a primary vertex, we depict the leg with a dashed line.

\subsection{Fully-coupled Gaussian covariance}
\label{subsec:cov_fc}

\subsubsection{General formalism for fully-coupled Gaussian NPCF covariance}
\label{subsubsec:cov_npcf_fc}

The covariance defined in Eq.~\eqref{eqn:npcf_cov_definition} can be expanded into the isotropic basis. Using Eq.~\eqref{eqn:2pcf_r_rp_s}, each pair contraction can be written as a Fourier transform of the power spectrum, which can be expressed as products of the basic elements with tripolar structure defined in \S\ref{subsec:basic-cov-elements}:
\begeqar\label{eqn:general-contractions}
\cov(\hat{\zeta}(\bfR), \hat{\zeta}(\bfR'))
%&=&\sum_{\Lambda,\Lambda'}\calE(\Lambda')\cov_{\Lambda,\Lambda'}(\zeta(R),\zeta(R'))\calY_\Lambda(\bfRhat) \calY_{\Lambda'}(\bfRhat')\non
&=&\sum_{\Lambda,\Lambda'}\calE(\Lambda')\cov_{\Lambda,\Lambda'}(R,R')\calY_\Lambda(\bfRhat) \calY_{\Lambda'}(\bfRhat')\non
&&= \int \frac{d^3\bfs}{V} \sum_G \prod_{i=0}^{N-1} \langle \delta(\bfx+\bfr_{Gi})\delta(\bfx+\bfr'_{i}+\bfs) \rangle|_{r_{G0}=r_0'=0}\\
%  &=& \int \frac {d^3\bfs}{V} \sum_{G} \prod_{i=0}^{N-1} \int_{\bfk_{Gi}}\int_{\bfk'_i} (2\uppi)^3\,\delta_{\rm D}(\bfk_{Gi}+\bfk'_i)P(k'_i)e^{i\left(\bfk'_i\cdot(\bfr'_i+\bfs)+\bfk_{Gi}\cdot\bfr_{Gi}\right)}
 &&\qquad= \int \frac {d^3\bfs}{V} (4\uppi)^{3N/2} \sum_{G} \prod_{i=0}^{N-1} \sum_{\ell_{Gi}\ell'_i L_i} i^{-\ell_{Gi}+\ell'_i+L_i} f_{\ell_{Gi}\ell'_iL_i} (r_{Gi}, r'_i, s)\non 
 &&\quad\qquad\times\,\calD^{\rm P}_{\ell_{Gi} \ell'_i L_i} \calC^{\ell_{Gi}\ell'_iL_i}_{000} \calY_{\ell_{Gi}\ell'_iL_i}(\bfrhat_{Gi}, \bfrhat'_i, \bfshat)|_{r_{G0}=r_0'=0},\nonumber
\endeqar
where we define $\cov_{\Lambda,\Lambda'}(\zeta(R),\zeta(R'))\equiv\cov_{\Lambda,\Lambda'}(R,R')$ and use the conjugation property $\calY_{\Lambda'}^*(\bfRhat')=\calE(\Lambda')\calY_{\Lambda'}(\bfRhat')$.% In the second equality, the evaluation of the fully-coupled covariance reduces to products of the expectation of two overdensity fields, one from unprimed family and the other from the primed family, which we recognize as the basic elements in~\eqref{eqn:2pcf_r_rp_s}. 
We denote the permutation by $G$, with a total of $N!$ permutation terms.
% \begeqar\label{eqn:general-contractions_expand}
% &&\sum_{\Lambda,\Lambda'}\calE(\Lambda')\cov_{\Lambda,\Lambda'}(R,R')\calY_\Lambda(\bfRhat) \calY_{\Lambda'}(\bfRhat')\non
%  &=& \int \frac {d^3\bfs}{V} \sum_{G} \prod_{i=0}^{N-1} \int_{\bfk_i} P(k_i)e^{i\bfk_i\cdot(\bfr'_i+\bfs-\bfr_{Gi})} \non
% \endeqar
Since the basis is isotropic, we can apply Eq.~\eqref{eq:averaging} and rotationally average over $d\calR$, $d\calR'$, and $d\calS$ (with the normalization $\int d\calS = (4\pi)^{-1} \int d\hat{s}$):
\begeqar\label{eqn:general-contractions2}
&&\sum_{\Lambda,\Lambda'}\calE(\Lambda')\cov_{\Lambda,\Lambda'}(R,R')\calY_\Lambda(\bfRhat) \calY_{\Lambda'}(\bfRhat')\non
&=& \int \frac { s^2ds}{V} {4\uppi}\,(4\uppi)^{3N/2}  \sum_{G} \sum_{\calL_G\calL'\Lambda''} \prod_{i=0}^{N-1}  i^{-\ell_{Gi}+\ell'_i+L_i} f_{\ell_{Gi}\ell'_iL_i} (r_{Gi}, r'_i, s) \calD^{\rm P}_{\ell_{Gi} \ell'_i L_i} \calC^{\ell_{Gi}\ell'_iL_i}_{000}\non
&&\times \mathcolorbox{color26}{(4\uppi)^{-N/2} \calQ^{\calL_G\calL'\Lambda''} \calD^{\rm P}_{\Lambda''} \calC^{\Lambda''}_{\bfzero}}
\calY_{\calL_G}(\bfRhat_G^{(N)}) \calY_{\calL'}(\bfRhat'{^{(N)}})|_{r_{G0}=r_0'=0}\non
&=& \int \frac {s^2ds}{V} {4\uppi}\,(4\uppi)^{3N/2}  \sum_{G} \sum_{\calL_G\calL'\Lambda''} \prod_{i=0}^{N-1}  i^{-\ell_{Gi}+\ell'_i+L_i} f_{\ell_{Gi}\ell'_iL_i} (r_{Gi}, r'_i, s) \non 
&&\times \calG^{\calL_G\calL'\Lambda''} \calD^{\rm P}_{\Lambda''} \calC^{\Lambda''}_{\bfzero}
\calY_{\calL_G}(\bfRhat_G^{(N)}) \calY_{\calL'}(\bfRhat'{^{(N)}})|_{r_{G0}=r_0'=0} 
\endeqar
%\bob{I think this isn't quite right.  Maybe only the $C_0^{L....}$ needs to be changed.  See, for example the expression a couple equations later} \jiamin{corrected: $\calC^{\calL_G\calL'\Lambda''}_{\bfzero}\rightarrow\calC^{\Lambda''}_{\bfzero}$, the coefficient is from Eq.~\eqref{eq:averaging}}
where we denote $\calL_G\equiv\{\ell_{G0}, \ell_{G1}, ..., \ell_{G(N-1)}\}$, $\calL'\equiv\{\ell'_0, \ell'_1, ..., \ell'_{N-1}\}$ as the angular momenta associated with the $\bfR$ and $\bfR'$ vectors, $\Lambda''\equiv\{L_0, L_1, ..., L_{N-1}\}$ as the angular momentum associated with the separation vector $\bfs$, and ${\rm M''} = \{M_0, M_1, ..., M_N\}$. The highlighted coefficients give rise to the coupling kernel in Fig.~\ref{fig:basic_coupling_elements}, {\bf diagram (5)}.
Notice that the isotropic basis used herein has $N$ coordinates (instead of $N-1$, as in the NPCF definition of Eq.~\ref{eqn:zeta_npcf}), given that we evaluate the function at $\bfr_0=0$, $\bfr'_0=0$ with corresponding angular momentum $\ell_{G0}=0$ and $\ell'_{0}=0$. Later, we will project the covariance onto the $(N-1)$ basis; for clarity we distinguish the two with the superscript ${(N)}$. 
Since both $\calY_{\calL_G}(\bfRhat_G^{(N)})$ and $\calY_{\calL'}(\bfRhat'{^{(N)}})$ contain a factor $Y_{00}(\bfrhat_0)=(4\uppi)^{-1/2}$, we find a total prefactor $(4\uppi)^{-1}$. This cancels with our normalization convention for the rotational average. 
%Due to the normalization in the rotational average $\int d\calS = (4\pi)^{-1} \int d\hat{s}$, this $4\uppi$ cancels with the $d\calR$ and $d\calR'$ given $(N-1)$ coordinates.
%\textbf{Oliver: The $\calY_{\calL}(\hR)$ here is *different* to the $\calY_{\calL}(\hR)$ appearing on the LHS of eq. 59. In particular, it is a basis of $N$ coordinates, instead of $(N-1)$. One coordinate is 0 (due to $\ell_{G0} = 0$). To convert between, we have a factor $1/\sqrt{4\pi}$.}
The non-canonically ordered isotropic function, $\calY_{\calL_G}(\bfRhat_G^{(N)})$, can be rewritten using the reordering coefficient defined in Eq.~\eqref{eqn:generalized-kronecker-delta}:
\begeqar
\calY_{\calL_G}(\bfRhat^{(N)}_G) = \sum_{J} \calB^{G^{-1}}_{\calL_G,J} \calY_{J}(\bfRhat^{(N)}).
\endeqar
% Now, 
% \begeqar\label{eqn:general-contractions3}
% &&\sum_{\Lambda,\Lambda'}\calE(\Lambda')\cov_{\Lambda,\Lambda'}(R,R')\calY_\Lambda(\bfRhat) \calY_{\Lambda'}(\bfRhat')\non
% &=& \int \frac { s^2ds}{V} (4\uppi)^{3N/2}  \sum_{G} \sum_{\calL_G\calL'\Lambda''J} \calB^{G^{-1}}_{\calL_G,J} \prod_{i=0}^{N-1}  i^{-\ell_{Gi}+\ell'_i+L_i} f_{\ell_{Gi}\ell'_iL_i} (r_{Gi}, r'_i, s) \calD^{\rm P}_{\ell_{Gi} \ell'_i L_i} \calC^{\ell_{Gi}\ell'_iL_i}_{000}\non
% &&\times (4\uppi)^{-N/2} \calQ^{\calL_G\calL'\Lambda''} \calD^{\rm P}_{\Lambda''} \calC^{\Lambda''}_{\bfzero}
% \calY_{J}(\bfRhat) \calY_{\calL'}(\bfRhat')|_{\bfr_{G0}=0, \bfr_0' = 0} \non
% &=& \int \frac {s^2ds}{V} (4\uppi)^{3N/2}  \sum_{G} \sum_{\calL_G\calL'\Lambda''J} \calB^{G^{-1}}_{\calL_G,J}  \prod_{i=0}^{N-1}  i^{-\ell_{Gi}+\ell'_i+L_i} f_{\ell_{Gi}\ell'_iL_i} (r_{Gi}, r'_i, s) \non 
% &&\times \calG^{\calL_G\calL'\Lambda''} \calD^{\rm P}_{\Lambda''} \calC^{\Lambda''}_{\bfzero}
% \calY_{J}(\bfRhat) \calY_{\calL'}(\bfRhat')|_{\bfr_{G0}=0, \bfr_0' = 0} 
% \endeqar
Finally, we project the covariance onto the isotropic basis $\calY^*_{\Lambda}(\bfRhat)$ and $\calY^*_{\Lambda'}(\bfRhat')$, and perform an angular average over $\bfr$ and $\bfr'$. Orthogonality forces $J\to\Lambda$ and $\calL'\to\Lambda'$, giving the general form for the NPCF covariance: 
%\textbf{Oliver: The equation below is missing a factor of $4\pi$. This is because $\int d\calS = 1/(4\pi)\times \int d\hat{s}$. This cancels with the factor of $4\pi$ noted from eq. 64.}
\begeqar\label{eqn:cov_npcf_iso}
&&\cov_{\Lambda,\Lambda'}(R,R')\nonumber\\
%  &=& \int \frac {s^2 d s}{V_{\rm eff}} (4\uppi)^{3N/2} \sum_{G, \Lambda'', \calL_G} \prod_{i=0}^{N-1}  i^{\ell_{Gi}-\ell'_i-L_i}  f_{\ell_{Gi}\ell'_iL_i}(r_{Gi}, r'_i, s) \nonumber\\
%  && \,\times\, \calB^{{-1}}_{\calL_G, \Lambda}\, \calE(\Lambda')\calG^{\calL_{G}\Lambda'\Lambda''} \calD_{\Lambda''}\calC^{\Lambda''}_0\nonumber\\
&=& (4\uppi)^{3N/2}\int \frac{s^2 ds}{V}  \sum_G\sum_{\Lambda'', \calL_G}  (-1)^{\left[-\Sigma(\Lambda)-\Sigma(\Lambda')+\Sigma(\Lambda'')\right]/2} \,\calB^{G^{-1}}_{\calL_G, \Lambda}\, \calG^{\calL_{G}\Lambda'\Lambda''} \calD_{\Lambda''}\calC^{\Lambda''}_0 \prod_{i=0}^{N-1}  f_{\ell_{Gi}\ell'_iL_i}(r_{Gi}, r'_i, s)|_{r_{G0}=r_0'=0},
\endeqar
% \bob{I think you can get rid of the first equation.  } \jiamin{Ok.}
where $\Sigma(\Lambda)=\sum_i \ell_i$, $\Sigma(\Lambda')=\sum_i \ell'_i$, and $\Sigma(\Lambda'')=\sum_i L_i$. 
%\bob{perhaps we should separate the sum into two pieces, one for the Lambdas and one for the G.  it is a bit odd to combine such different things in one summation.} \jiamin{ok, done}

\subsubsection{Fully-coupled Gaussian 4PCF covariance}
\begin{figure}
    \centering
    \includegraphics[width=.9\textwidth]{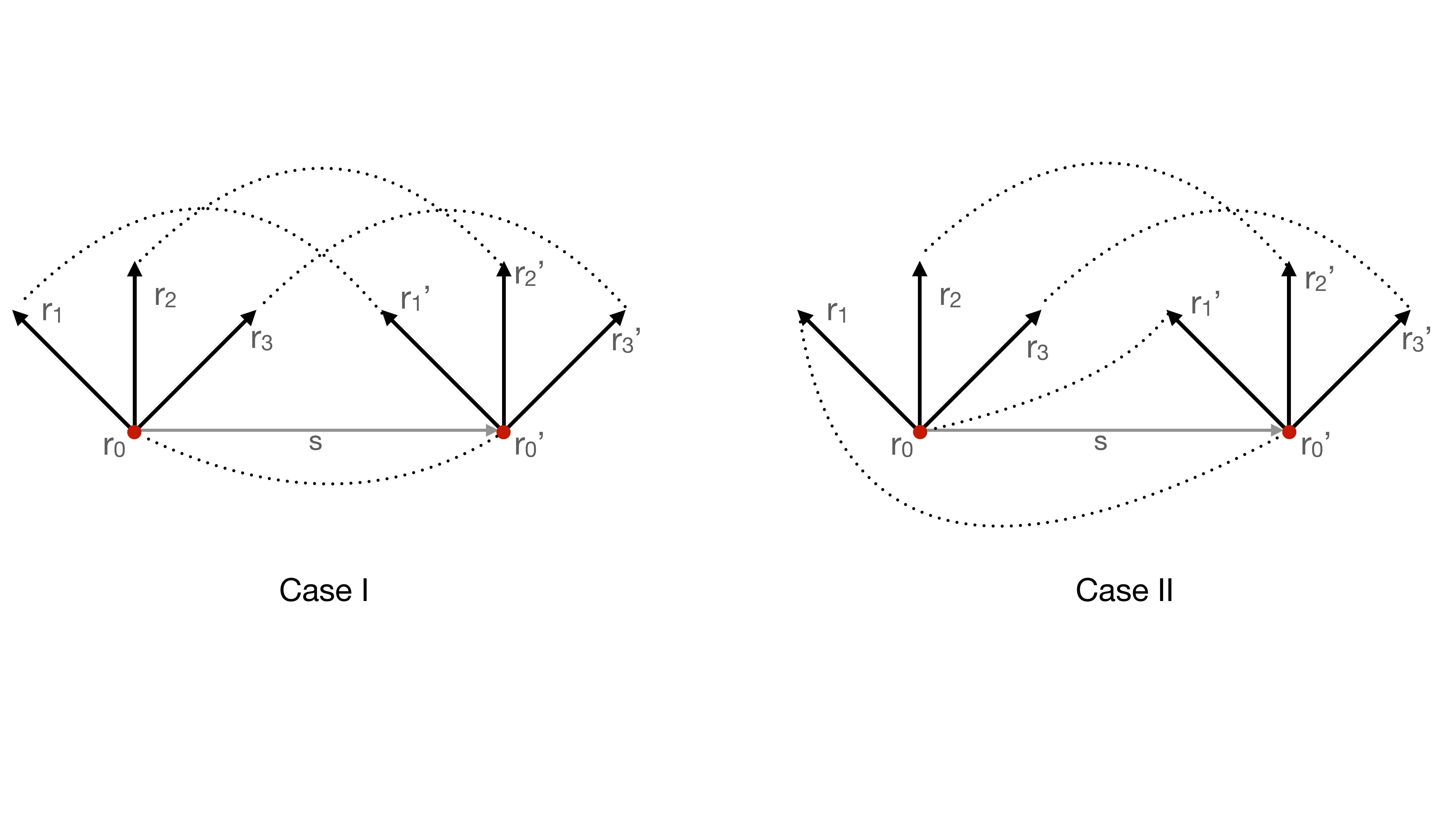}
    \caption{Schematic for the fully-coupled 4PCF covariance (\textit{i.e.} the covariance of the connected 4PCF). We split the covariance into two cases. In {\bf Case I}, the primary vertices (red dots, labelled by $r_0$ and $r'_0$) from the primed and unprimed families are mutually coupled and all the endpoints (labelled by $r_i$ and $r'_i$) are coupled. In {\bf Case II}, the primary vertices each are coupled to an endpoint from the opposite family.}
    \label{fig:tree_connected_2cases}
\end{figure}

Henceforth, we will focus on the fully-coupled covariance of the 4PCF. To derive this, we can use the general form given in Eq.~\eqref{eqn:cov_npcf_iso}; however, as an explicit verification, we construct the 4PCF covariance in a different manner.
%, making use of the building block introduced in \S\ref{subsec:basic-cov-elements}.
% The covariance for the 4PCF can be explicitly written as:
% \begeqar\label{eqn:4pcf-cov-full-explicit}
%     &&\sum_{\Lambda,\Lambda'}\calE({\Lambda'})\cov\left(\hat\zeta_\L(r_1,r_2,r_3),\hat\zeta_{\L'}(r_1',r_2',r_3')\right)\calY_{\Lambda}(\bfrhat_1,\bfrhat_2,\bfrhat_3)\calY_{\Lambda'}(\bfrhat'_1,\bfrhat'_2,\bfrhat'_3)\non 
%     &&\equiv \sum_{\Lambda,\Lambda'}\calE({\Lambda'})\cov_{\L,\L'}(R,R')\calY_{\Lambda}(\bfRhat)\calY_{\Lambda'}(\bfRhat')\non
%     % &&=\av{\hat\zeta^*_\L(r_1,r_2,r_3)\hat\zeta_\L(r_1',r_2',r_3')}\non
%     &&=\int \frac{d^3\bfs}{V} \av{\prod_{i=0}^3 \left[\int d\bfrhat_i\,d\bfrhat'_i\,\delta(\vr_i)\delta(\vr'_i+\bfs)\right]},
% \endeqar
% %\bob{I don't think this last equation is correct.} \jiamin{I corrected the coordinates $\bfr'_0\rightarrow \bfr'_0+\bfs$ and $\bfr'_i\rightarrow \bfr'_i+\bfs$}
% where we have used the conjugation condition $\P^*_\L(\bfRhat) = \calE(\Lambda)\P_\L(\bfRhat)$, therefore, we are left with only one phase term. For generality, we allow $\hat\zeta$ to be complex, which corresponds to allowing parity-odd multipoles with odd $\sum_{i=1}^3\ell_i$.
Noticing that contractions with the primary vertices lead to basis functions involving zero angular momenta, we split the fully-coupled covariance into two different cases: those involving a mutual coupling of the primary vertices $\delta(\bfr_0)$ and $\delta(\bfr'_0+\bfs)$ (upper panel of Fig.~\ref{fig:tree_connected_2cases}) and those where the primary vertices couple to the endpoints of the opposite family (lower panel of Fig.~\ref{fig:tree_connected_2cases}). 
In this decomposition, the fully-coupled covariance can be written
\begeqar \cov^{\rm (fc)}_{\L,\L'}(R,R')
    =\cov^{\rm (fc), I}_{\L,\L'}(R,R')+\cov^{\rm (fc), II}_{\L,\L'}(R,R');
\endeqar
next, we will discuss these two cases.
%for $\cov^{\rm (fc), I}_{\L,\L'}(R,R')$ and $\cov^{\rm (fc), II}_{\L,\L'}(R,R')$, respectively.

\begin{figure}
    \centering
    \includegraphics[width=.8\textwidth]{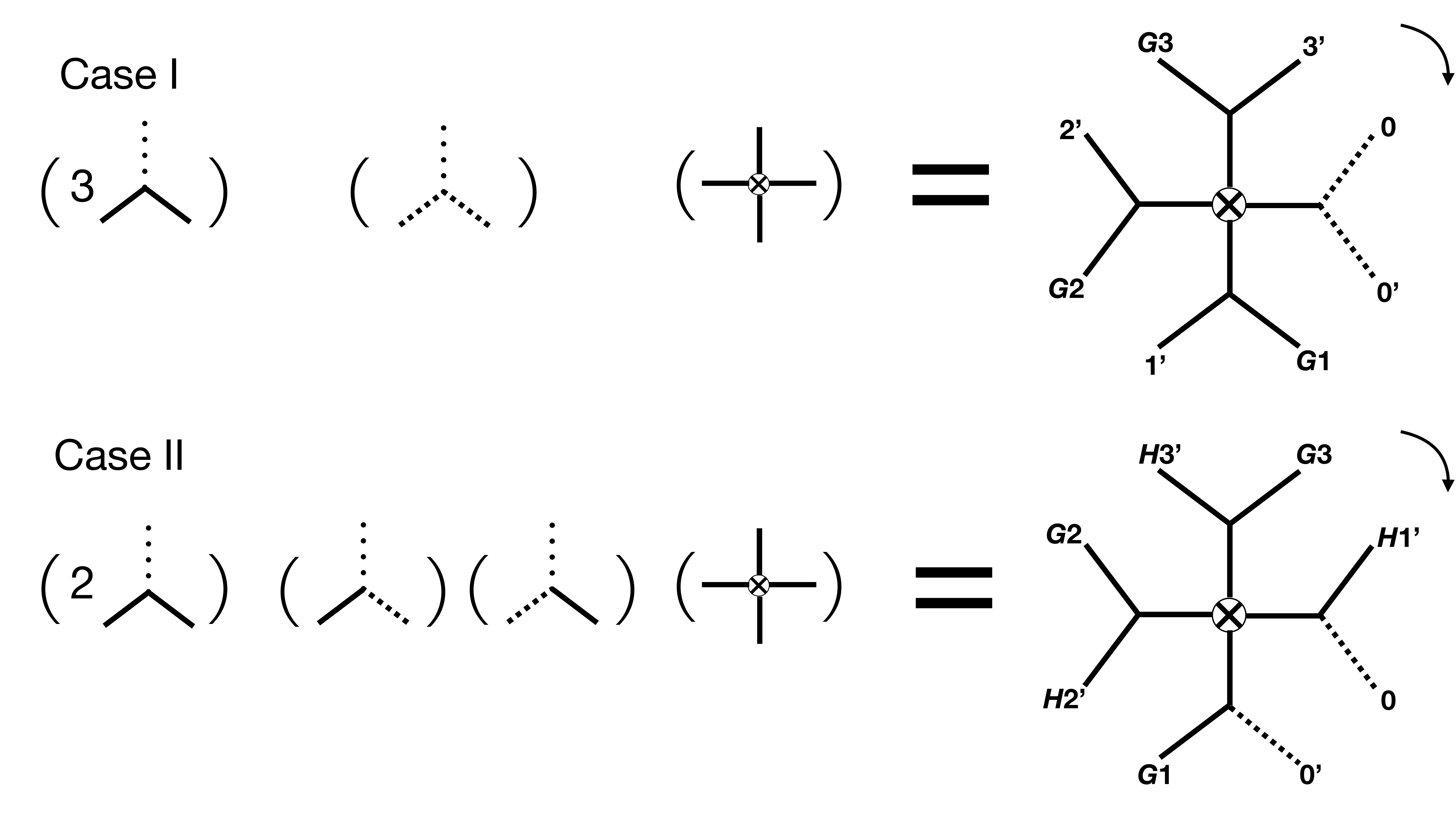}
    \caption{A diagrammatic representation of a fully-coupled covariance matrix with {\bf Case I} shown in the {\bf upper panel} and {\bf Case II} in the {\bf lower panel} (as in Fig.\,\ref{fig:tree_connected_2cases}). Each case can be broken down into two elementary structures: {\bf (a)} a tripolar structure arising from the contraction between overdensity fields from the primed and unprimed families, and {\bf (b)} a coupling kernel given by the rotational average over $\bfr$, $\bfr'$, and $\bfs$. Moreover, since the covariance involves two primary vertices (one from the primed and the other from the unprimed family), there are two dashed lines either connected to each other or connected to a solid line. %, respectively.
    All the three pieces are multiplied, summed over the angular momenta, and integrated over $s$. In this figure we use $G_i$ and $H_i$ to denote permutations. For $N=4$, the phase $(-1)^{\Sigma(\Lambda)(1-\calE_G)/2}$ or $(-1)^{\Sigma(\Lambda')(1-\calE_H)/2}$ can be directly read off from the plot as one goes around clockwise: an even permutation in the ordering of angular momenta corresponds to a positive Levi-Civita symbol and always gives a positive phase, while an odd permutation can flip the sign of the phase for parity-odd correlators. {\bf Diagrams (2) and (3)} in Fig.~\ref{fig:basic_coupling_elements} can be distinguished from each other by reading the diagram clockwise (\textit{i.e.} one cannot change one into the other by a 2D rotation in the page). The following steps are used to build the ``snowflake" diagrams on the right hand sides of the equation: {\bf (1)} take the tripolar structures and multiply them with the coupling kernel, 
    %2) project onto isotropic functions $\calY_{\Lambda}(\bfR)$ and $\calY_{\Lambda'}(\bfR')$, 
    {\bf (2)} perform an integral over the radial part $s$ of the separation vector${\bf s}$.
    %\textcolor{red}{(OP: note that the footnote doesn't work here!}\footnote{This diagram can be generalized to arbitrary $N$, however, we note that the canonical ordering introduces a non-trivial coefficient upon permutation. The extension to $N>4$ will be presented in the future work.}
    }
    \label{fig:fully_coupled_caseI_II}
\end{figure}

\paragraph*{Case I}
\label{subsubsec:cov_fc_caseI}
In this scenario the contraction of the eight density fields leads to the term
\begeqar\label{eqn:case-I-contractions}
    I_{\rm I}(\bfR,\bfR';\bfs)&\equiv&  \av{\delta(\bfx+\vr_0)\delta(\bfx+\bfs+\vr'_0)}|_{r_0=r'_0=0}\non
    &&\qquad\times\av{\delta(\bfx'+\vr_{i})\delta(\bfx'+\bfs+\vr_1')}
    \av{\delta(\bfx''+\vr_{j})\delta(\bfx''+\bfs+\vr_2')}
    \av{\delta(\bfx'''+\vr_{k})\delta(\bfx'''+\bfs+\vr_3')}\non
    &=&\sum_{G}\xi(|\bfs+\bfr'_0-\bfr_0|)\xi(|\vs+\vr'_{1}-\vr_{G1}|)\xi(|\vs+\vr'_{2}-\vr_{G2}|)\xi(|\vs+\vr'_{3}-\vr_{G3}|)|_{r_0=r'_0=0}\nonumber,
\endeqar
defining the shorthand $I_{\rm I}$ in the first line. Here, $\{i,j,k\}$ denotes a permutation of the set $\{1,2,3\}$, which does not include the primary vertices at $\bfr_0$ and $\bfr'_0$. 
There are $3! =6$ options by which to contract the remaining three density fields from the primed and unprimed families. In the second line we introduce the notation $G$ to denote a permutation, with $\{i,j,k\}=\{G1,G2,G3\}$.
The six permutations are given explicitly in Table \ref{tab: case-I-perms}.
Using the basic elements constructed in Eq.~\eqref{eqn:2pcf_r_rp_s}, we can express the product of the four 2PCFs as 
\begeqar\label{eqn:case-I-elements}
    I_{\rm I}(\bfR,\bfR';\bfs) &=& \sum_G \prod_{i=0}^3 (4\uppi)^{3/2} \sum_{\ell_{Gi} \ell'_i L} i^{-\ell_{Gi}+\ell'_i+L_i} f_{\ell_{Gi} \ell'_i L_i}(r_{Gi}, r'_i, s) \calD^{\rm P}_{\ell_{Gi} \ell'_i L_i} \calC^{\ell_{Gi} \ell'_i L}_{000} \calY_{\ell_{Gi} \ell'_i L_i}(\bfrhat_{Gi}, \bfrhat'_i, \bfshat)|_{r_{0}=r_0'=0};
\endeqar
here we denote the collection of angular momenta as $\calL_G=\{0, \ell_{G1}, \ell_{G2}, \ell_{G3}\}$, $\calL'=\{0, \ell'_{1}, \ell'_{2}, \ell'_{3}\}$ and $\Lambda''=\{0, L_{1}, L_{2}, L_{3}\}$. In principle, these should all involve intermediate angular momenta, however, the angular momentum associated with primary vertex is set to be zero, thus the intermediate momenta are uniquely defined. % and hence no longer unnecessary.

Performing a rotational average of $d\calR$, $d\calR'$, and $d\calS$ over $\bfrhat_{Gi}$, $\bfrhat'_i$, and $\bfshat$, leads to the quantity $\calQ^{\Lambda{_G}\Lambda'\Lambda''}$ and a prefactor $(4\uppi)^{-2}$ for $N=4$. When combined with the coefficients $\calD^{\rm P}_{\ell_{Gi} \ell'_i L_i}$ and $\calC^{\ell_{Gi} \ell'_i L_i}_{000}$ for $i=0,\ldots, 3$ (cf. Eq.~\eqref{eq:averaging} and Eq.~\eqref{eqn:generalized-gaunt}), we obtain the generalized Gaunt integral. The Gaunt integral for $N=4$ involves a product of two 9-$j$ symbols and intermediate angular momenta given in Eq.~\eqref{eqn:generalized-gaunt-N4}. 
However, one of the 9-$j$ symbol can be reduced due to the presence zero angular momenta, and the fully determined intermediate angular momenta: $\ell_{12}=\ell_{G1}$, $\ell'_{12}=\ell'_{1}$, and $\ell_{12}=\ell_{G1}$. The Gaunt integral in this case reads
\begeqar\label{eqn:gaunt-N4-caseI}
\calG^{\Lambda_G\Lambda'\Lambda''}
&=&(4\uppi)^{-2}\,\calD^{\rm P}_{\ell_{G1}\ell'_1L_1}
\prod_{i=0}^3\calD^{\rm P}_{\ell_{Gi}\ell'_i\L_i}\calC^{\ell_{Gi}\ell'_iL_i}_{000}\,
\nine{0}{\ell_{G1}}{\ell_{G1}}{0}{\ell'_{1}}{\ell'_{1}}0{L_1}{L_1}
\nine{\ell_{G1}}{\ell_{G2}}{\ell_{G3}}{\ell'_1}{\ell'_2}{\ell'_3}{{L}_{1}}{{L}_2}{{L}_3}\non
&=&(4\uppi)^{-2}
\prod_{i=0}^3\calD^{\rm P}_{\ell_{Gi} \ell'_i \L_i} \calC^{\ell_{Gi} \ell'_i L_i}_{000}\,
\nine{\ell_{G1}}{\ell_{G2}}{\ell_{G3}}{\ell'_1}{\ell'_2}{\ell'_3}{{L}_{1}}{{L}_2}{{L}_3},
\endeqar
where $\calD^{\rm P}_{\ell_{G1} \ell'_1 L_1}$ in the first line is cancelled by the first 9-$j$ symbol, leaving only one 9-$j$ symbol in the second line. Here we introduce a Levi-Civita symbol, defined by $\mathcal{E}_{G} = 1$ if $\{G1,G2,G3\}$ is an even permutation of $\{1,2,3\}$ and $-1$ otherwise. The values of $\mathcal{E}_{G}$ for each permutation $G$ are given in Table \ref{tab: case-I-perms}. Practically, this leads to a prefactor of $(-1)^{\ell_1+\ell_2+\ell_3}$ if the permutation is odd, and unity otherwise. For the even-parity $\Lambda$ this phase does not play a role, but it is of importance for odd parity $\Lambda$.

% next we perform rotation average $d\calR$, $d\calR'$, and $d\calS$ over $\bfrhat_{Gi}$, $\bfrhat'_i$, and $\bfshat$ by using the result from Eq.~\eqref{eq:averaging}. For $N=4$ this leads to a factor $(4\uppi)^{-2}$ and we arrive at
% \begeqar\label{eqn:case-I-elements}
%     \av{...}_{\rm I} &=& \sum_G (4\uppi)^{6} \prod_{i=0}^3 \sum_{\ell_{Gi} \ell'_i L_i} i^{\ell_{Gi}-\ell'_i-L_i} f_{\ell_{Gi} \ell'_i L_i}(r_{Gi}, r'_i, s) \calD^{\rm P}_{\ell_{Gi} \ell'_i L_i} \calC^{\ell_{Gi} \ell'_i L}_\bfzero \non &&\,\times\,(4\uppi)^{-2}\sum_{\Lambda{_G}\Lambda'\Lambda''}\calQ^{\Lambda{_G}\Lambda'\Lambda''}\calD^{\rm P}_{\Lambda''}\calC^{\Lambda''}_\bfzero\calY_{\Lambda_G}(\bfRhat^{(4)}_G)\calY_{\Lambda'}(\bfRhat'^{(4)})\non
%     &=& \sum_G (4\uppi)^{6} \prod_{i=0}^3 \sum_{\ell_{Gi} \ell'_i L_i} i^{\ell_{Gi}-\ell'_i-L_i} f_{\ell_{Gi} \ell'_i L_i}(r_{Gi}, r'_i, s)  \sum_{\Lambda{_G}\Lambda'\Lambda''}\calD^{\rm P}_{\Lambda''}\calC^{\Lambda''}_\bfzero\calG^{\Lambda{_G}\Lambda'\Lambda''}\calY_{\Lambda_G}(\bfRhat^{(4)}_G)\calY_{\Lambda'}(\bfRhat'^{(4)}),
% \endeqar
% we use the superscript $^{(4)}$ to emphasize that the coordinate includes $\bfr_0=\bfr'_0$ and each of them give rise to a factor of $(4\uppi)^{-1/2}$. Meanwhile, replacing the angular average $d\bfshat$ by rotational average $d\calS$ leads to a normalization factor $4\uppi$ and they are both cancelled.
%\bob{can't we avoid this by referring to Eq. (34)?} \jiamin{I shorten the step discussing rotational average for both Case I and II}

%Now we can make use of the general result in Eq.~\eqref{eqn:cov_npcf_iso}. 

Using Eq.~\eqref{eqn:perm-unperm-basis}, we can restore the canonical ordering in $\bfRhat_G=\{\bfrhat_{G1}, \bfrhat_{G2}, \bfrhat_{G3}\}$. For the 4PCF covariance, the reordering coefficient $\calB^{G^{-1}}_{\calL_G, J}$ for $(N-1)=3$ involves only a phase and the product three of Kronecker deltas:
\begeqar\label{eqn:calB_3n}
\calB^{G^{-1}}_{\calL_G,J} = (-1)^{\Sigma(\Lambda)(1-\calE_G)/2}\prod_{i=1}^3\delta^{\rm K}_{j_i\ell_{Gi^{\scaleto{-1\mathstrut}{4pt}}}}.
\endeqar
Performing angular averages over $\bfRhat$ and $\bfRhat'$ allows us to set $J\to\Lambda$ and pick out the coefficients of the isotropic basis $\calY_{\Lambda}(\bfRhat)$ and $\calY_{\Lambda'}(\bfRhat')$. Altogether, we arrive at the final form for Case I:
\begeqar\label{eqn:final-cov-caseI}
    \mathrm{Cov}_{\L,\L'}^{\rm{(fc)}, I}(R,R') &=& (4\uppi)^4 \sum_G (-1)^{\Sigma(\Lambda)(1-\calE_G)/2}\sum_{L_1L_2L_3} \calD^{\rm P}_{L_1L_2L_3}\,\mathcal{C}^{L_1L_2L_3}_{000} \begin{Bmatrix} \ell_{G1} & \ell_{G2} & \ell_{G3}\\ \ell_1' & \ell_2' & \ell_3'\\ L_1 & L_2 & L_3\end{Bmatrix}\non
    &&\,\times\, \prod_{i=1}^3 \left[(-1)^{(-\ell_{Gi}-\ell_i'+L_i)/2}\,\calD^{\rm P}_{\ell_i\ell'_iL_i}\mathcal{C}^{\ell_{Gi}\ell'_i L_i}_{000} 
    \int \frac{s^2ds}{V}\xi(s)f_{\ell_{Gi}\ell_i'L_i}(r_{Gi},r_i',s) \right].
\endeqar
% \bob{probably should put in some brackets to make clear which factors get multiplied in the product and which only appear once.} \jiamin{ok, rearranged and added brackets}

For illustration, we consider the limit where the correlation function $\xi(\bfs)$ becomes a Dirac delta function, and the power spectrum consequently becomes unity. This limit enables a direct evaluation of both Eq.~\eqref{eqn:case-I-contractions} and its representation Eq.~\eqref{eqn:final-cov-caseI}, providing a useful cross-check of our calculation.

From Eq.~\eqref{eqn:case-I-contractions}, we see that $\xi(\bfs) \to \delta_{\rm D}^{[3]}(\bfs)$ implies that $\bfs \to \mathbf{0}$. Consequently, we have that $\bfr_1' \to \bfr_{G1}$,  $\bfr_2' \to \bfr_{G2}$, $\bfr_3' \to \bfr_{G3}$. 
We now consider the representation in terms of $f$-integrals. For the first, taking $P(k) \to 1$ gives
%Case I in the limit that $\xi(\bfs)$ becomes a Dirac delta function, \textit{i.e.} $\bfs \to 0$, and its corresponding power spectrum simply being one. \bob{I don't like this.  First, Eq. (41) doesn't make sense.  The integral over k cannot depend in any way on k.  Moreover, it is dimensionally wrong.  Eq. (42) is missing a Kronecker delta of ell and ell'.  I don't think this exercise teaches us anything.} In this limit, the $f$-integrals simplify as
\begeqar
f_{000}(0,0,s) = \int \frac{k^2dk}{2\uppi^2} j_0(ks) = \frac{1}{4\uppi s^2} \delta_{\rm D}^{[1]}(s).
\endeqar
This is simply a representation of the 3D Dirac delta function with spherical symmetry, expected since $f_{000}(0,0,s) = \xi(s)$.
%Unsurprisingly, this is simply telling us that if $P(k) \to 1$, $\xi(s)$ is indeed a delta function.

The other $f$-integrals can be similarly evaluated in the limit $s \to 0$ (and again, $P(k)= 1$). We have
\begeqar\label{eqn:f2l-dirac}
\lim_{s\to 0} f_{\ell_{Gi}\ell'_i L}(r_{Gi},r'_i,s) = \lim_{s\to 0} \int \frac{k^2dk}{2\uppi^2} j_{\ell_{Gi}}(kr_{Gi})j_{\ell'_i}(kr'_i) j_L(ks) = \int \frac{k^2dk}{2\uppi^2} j_{\ell_{Gi}}(kr_{Gi})j_{\ell'_i}(kr'_i)\; =
\frac{1}{4\uppi r_{Gi}r'_i}\delta_{\rm D}^{[1]}(r_{Gi}-r'_i)\delta_{\ell_{Gi} \ell'_i}^{\rm K}.
\endeqar
For the first equality, we have noted that, as $s \to 0$, only $j_0$ is non-zero, meaning $L \to 0$ and hence $\ell \to \ell'$ due to the 3-$j$ symbol in equation (\ref{eqn:final-cov-caseI}). We recognize this integral as a Dirac delta function, as before.

%in the second equation the spherical Bessel function is non zero only when $L=0$ and therefore the integral reduces to product of two spherical Bessel functions. 

\begin{figure}
    \centering
    \includegraphics[width=.8\textwidth]{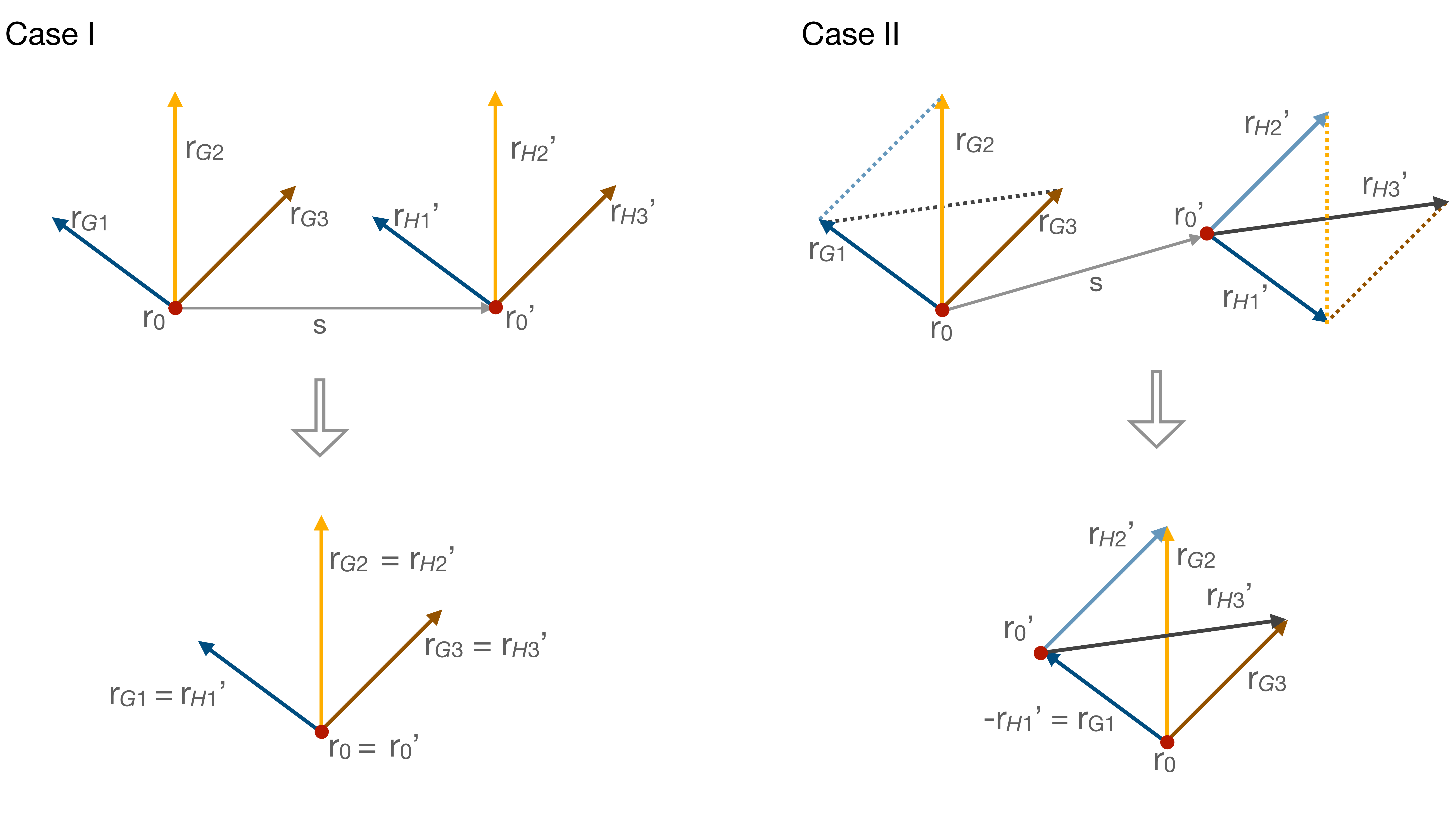}
    \caption{Covariance calculation for the 4PCF in the limit of zero separation (\textit{i.e.} where where $\xi$ becomes a Dirac delta function). 
    %\textcolor{red}{(OP: maybe say limit of zero separation, for people who haven't read the text)}. 
    {\bf Left column:} In Case I, this limit implies that the two tetrahedra overlap at their origin with $\bfs \to 0$, $\bfr_1' \to \bfr_{G1}$,  $\bfr_2' \to \bfr_{G2}$, and $\bfr_3' \to \bfr_{G3}$. {\bf Right column:} the same limit in Case II implies that the two tetrahedra also overlap but with one of the primary vertices sitting on an endpoint from the other family. Consequently, we find $\bfs \to \bfr_{G1}$, $\bfs \to -\bfr'_{H1}$, $\bfs \to \bfr'_{H2}-\bfr_{G2}$, and $\bfs \to \bfr'_{H3}-\bfr_{G3}$.}
    \label{fig:delta_limit_CaseI_II}
\end{figure}

As shown in. Fig.~\ref{fig:delta_limit_CaseI_II}, this result implies that, in the limit of uniform power spectra, the covariance for two tetrahedra is non-vanishing only when (1) they have zero separation length and one of their vertices is coincident and (2) their sides are the same lengths; \textit{i.e.} when one tetrahedron can be perfectly rotated in 3D to overlap with the other.

\begin{table}
    \centering
    \begin{tabular}{ccc|r}
        G1 & G2 & G3 & $\mathcal{E}_{G}$\\\hline
        1 & 2 & 3 & 1\\
        1 & 3 & 2 & -1\\
        2 & 3 & 1 & 1\\
        2 & 1 & 3 & -1\\
        3 & 1 & 2 & 1\\
        3 & 2 & 1 & -1\\
    \end{tabular}
    \caption{Explicit forms of the six permutations appearing in the Case I covariance terms. These arise from the various options for contracting density fields in Eq.~\eqref{eqn:case-I-contractions}. Each term involves a contraction between $\vr_{Gi}$ and $\vr'_i$. We additionally give the Levi-Civita permutation factor $\mathcal{E}_{G}$ for each.}
    \label{tab: case-I-perms}
\end{table}

\paragraph*{Case II}
\label{subsubsec:cov_fc_caseII}
Here, we consider sets of contractions that involve couplings between primary vertices and endpoints across the two families. Each is of the form
\begeqar\label{eqn:case-II-contractions}
    I_{\rm II}(\bfR,\bfR';\bfs)
    &\equiv&\av{\delta(\bfx+\vr_{i})\delta(\bfx+\bfs+\bfr'_0)}\av{\delta(\bfx'+\bfr_0)\delta(\bfx'+\bfs+\vr_{i'}')}|_{r_{0}=r_0'=0}\non
    &&\times\,\av{\delta(\bfx''+\vr_j)\delta(\bfx''+\bfs+\vr_{j'}')}\av{\delta(\bfx'''+\vr_k)\delta(\bfx'''+\bfs+\vr_{k'}')}\non
    &=&\sum_{G,H}\xi(\vs-\vr_{G1})\xi(\vs+\vr'_{H1})\xi(\vs-\vr_{G2}+\vr'_{H2})\xi(\vs-\vr_{G3}+\vr'_{H3}),
\endeqar
where $\{i,j,k\}$ and $\{i',j',k'\}$ are permutations of the set $\{1,2,3\}$. We write the two sets of the permutations as $\{i,j,k\} = \{G1,G2,G3\}$, $\{i',j',k'\}=\{H1,H2,H3\}$, where one set follows a cyclic permutation, due to the explicit contraction with the primary vertex. Given the symmetry among the pair ordering, \textit{i.e.} $\{j,j'\}\leftrightarrow\{k,k'\}$, we can always fix the permutation of one set of endpoints and let the other set explore all permutations. Here we choose $G$ to follow cyclic permutation (giving rise to a factor of three), with $H$ being a standard permutation including six terms. In total, there are 18 permutations in this scenario.
%since we have three choices of primed vertex with which to contract $\delta(\bfr_0)$, three choices of unprimed vertex with which to contract $\delta(\bfr'_0)$ and two ways to contract the remaining fields. 
For clarity, we write them explicitly in Table \ref{tab: case-II-perms}. As before, the primary vertices at $\bfr_0$ and $\bfr'_0$ are not permuted.

\begin{table}
    \centering
    \begin{tabular}{ccc|ccc|cr}
        G1 & G2 & G3 & H1 & H2 & H3 & $\mathcal{E}_{G}$ & $\mathcal{E}_{H}$\\\hline
        1 & 2 & 3 & 1 & 2 & 3 & 1 & 1 \\
        1 & 2 & 3 & 1 & 3 & 2 & 1 & -1 \\
        1 & 2 & 3 & 2 & 1 & 3 & 1 & 1 \\
        1 & 2 & 3 & 2 & 3 & 1 & 1 & -1 \\
        1 & 2 & 3 & 3 & 1 & 2 & 1 & 1 \\
        1 & 2 & 3 & 3 & 2 & 1 & 1 & -1 \\\hline
        2 & 3 & 1 & 1 & 2 & 3 & 1 & 1 \\
        2 & 3 & 1 & 1 & 3 & 2 & 1 & -1 \\
        2 & 3 & 1 & 2 & 1 & 3 & 1 & 1 \\
        2 & 3 & 1 & 2 & 3 & 1 & 1 & -1 \\
        2 & 3 & 1 & 3 & 1 & 2 & 1 & 1 \\
        2 & 3 & 1 & 3 & 2 & 1 & 1 & -1 \\\hline
        3 & 1 & 2 & 1 & 2 & 3 & 1 & 1 \\
        3 & 1 & 2 & 1 & 3 & 2 & 1 & -1 \\
        3 & 1 & 2 & 2 & 1 & 3 & 1 & 1 \\
        3 & 1 & 2 & 2 & 3 & 1 & 1 & -1 \\
        3 & 1 & 2 & 3 & 1 & 2 & 1 & 1 \\
        3 & 1 & 2 & 3 & 2 & 1 & 1 & -1 \\
    \end{tabular}
    \caption{Explicit forms of the 18 permutations appearing in the Case II covariance terms. These arise from the various options for contracting density fields in Eq.~\eqref{eqn:case-II-contractions}, in particular the contraction of $\bfr+\vr_{G1}$ with $\bfr'_0$, $\bfr$ with $\bfr_0'+\vr_{H1}'$, $\bfr_0+\vr_{G2}$ with $\bfr_0'+\vr'_{H2}$ and $\bfr_0+\vr_{G3}$ with $\bfr_0'+\vr'_{H3}$ (noting the symmetry of the final two terms). We additionally give the permutation factors $\mathcal{E}_{G}$ and $\mathcal{E}_{H}$ for each.}
    \label{tab: case-II-perms}
\end{table}

Including the basic covariance elements, we can write:
\begeqar\label{eqn:case-II-contractions-tmp1}
    I_{\rm II}(\bfR,\bfR';\bfs)
    &=&\sum_{G,H} \,
(4\uppi)^{3/2}\,\sum_{\ell_{G1}L_1} 
f_{\ell_{G1} 0\ell_{G1}}(r_{G1},0,s) \calD^{\rm P}_{\ell_{G1}0\ell_{G1}}\calC^{\ell_{G1}0\ell_{G1}}_{000}\calY_{\ell_{G1}0\ell_{G1}}(\bfrhat_{G1},0,\bfshat) \non
&&\, \times\,(4\uppi)^{3/2}\sum_{\ell'_{H1}L_1} (-1)^{\ell'_{H1}}
f_{0\ell'_{H1}\ell'_{H1}}(0,r'_{H1},s)\calD^{\rm P}_{0\ell'_{H1}\ell'_{H1}}\calC^{0\ell'_{H1}\ell'_{H1}}_{000} \calY_{0\ell'_{H1}\ell'_{H1}}(0,\bfrhat'_{H1}, \bfshat) \non
&&\,\times\, \prod_{i=2}^3(4\uppi)^{3/2}\sum_{\ell_{Gi} \ell'_{Hi} L_i} i^{-\ell_{Gi}+\ell'_{Hi}+L_i} f_{\ell_{Gi} \ell'_{Hi} L_i}(r_{Gi}, r'_{Hi}, s) \calD^D_{\ell_{Gi} \ell'_{Hi} L_i} \calC^{\ell_{Gi} \ell'_{Hi} L_i}_{000} \calY_{\ell_{Gi} \ell'_{Hi} L_i}(\bfrhat_{Gi}, \bfrhat'_{Hi}, \bfshat),
\endeqar
where the collection of angular momenta is $\calL_{G}=\{\ell_{G1}, 0, \ell_{G2}, \ell_{G3}\}$, $\calL'_{H}=\{0, \ell'_{H1}, \ell'_{H2}, \ell'_{H3}\}$, and $\Lambda''=\{\ell_{G1},\ell'_{H1},L_2,L_3\}$.

To restore the canonical ordering for $\bfRhat_G=\{\bfrhat_{G1}, \bfrhat_{G2}, \bfrhat_{G3}\}$ and  $\bfRhat'_H=\{\bfrhat'_{H1}, \bfrhat'_{H2}, \bfrhat'_{H3}\}$, we use again the reordering coefficients, in the form
\begeqar\label{eqn:calB_caseII}
\calB^{G^{-1}}_{\calL_G,J} = (-1)^{\Sigma(\Lambda)(1-\calE_G)/2}\prod_{i=1}^3\delta^{\rm K}_{j_i\ell_{Gi^{\scaleto{-1\mathstrut}{4pt}}}},\qquad \calB^{H^{-1}}_{\calL'_H, J'} = (-1)^{\Sigma(\Lambda')(1-\calE_H)/2}\prod_{i=1}^3\delta^{\rm K}_{j'_i\ell'_{Hi^{\scaleto{-1\mathstrut}{4pt}}}}.
\endeqar
Since we restrict $G$ to cyclic permutations, $\calB^{G^{-1}}_{\calL_G,J}$ is merely a Kronecker delta with a trivial phase. Additionally, the phase factor $\Sigma(\Lambda')$ does not play a role for even parity $\Lambda$, but is of importance for odd-parity $\Lambda$. 
%\bob{Restored index $G$ for $\calB$} \jiamin{We should also do the RHS $H$ for $\calB$}

As before, we proceed by performing a rotational average over $d\calR$, $d\calR'$, and $d\calS$, which leads to a generalized Gaunt integral, involving two 9-$j$ symbols, and a sum over intermediate angular momenta. 
% \begeqar
%     \av{...}_{\rm II}
%     &=&\sum_{G,H} \,
% (4\uppi)^{6}\,\sum_{\calL_{G}\calL'_{H}\Lambda''} (-1)^{\ell'_{H1}}i^{\ell_{G2}+\ell_{G3}-\ell'_{H2}-\ell'_{H3}-L_2-L_3}\\
% && \times\,f_{\ell_{G1} 0\ell_{G1}}(r_{G1},0,s)f_{0\ell'_{H1}\ell'_{H1}}(0,r'_{H1},s) f_{\ell_{G2} \ell'_{H2} L_2}(r_{G2}, r'_{H2}, s) f_{\ell_{G3} \ell'_{H3} L_3}(r_{G3}, r'_{H3}, s)\non
% &&\, \times\, \calD^{\rm P}_{\ell_{G1}0\ell_{G1}}\calD^{\rm P}_{0\ell'_{H1}\ell'_{H1}}  \calD^D_{\ell_{G2} \ell'_{H2} L_2}\calD^D_{\ell_{G3} \ell'_{H3} L_3} \calC^{\ell_{G1} 0\ell_{G1}}_\bfzero \calC^{0\ell'_{H1}\ell'_{H1}}_\bfzero \calC^{\ell_{G2} \ell'_{H2} L_2}_\bfzero\calC^{\ell_{G3} \ell'_{H3} L_3}_\bfzero\non &&\times\,(4\uppi)^{-2}\calQ^{\calL_{G}\calL'_{H}\Lambda''}\calD^{\rm P}_{\Lambda''}\calC^{\Lambda''}_\bfzero\calY_{\calL_{G}}(\bfRhat^{(4)}_{G}) \calY_{\calL_{H}}({{\bfRhat_{H}}^{'(4)}})\non
%     &=&\sum_{G,H} \,
% (4\uppi)^{6}\,\sum_{\calL_{G}\calL'_{H}\Lambda''} (-1)^{\ell'_{H1}}i^{\ell_{G2}+\ell_{G3}-\ell'_{H2}-\ell'_{H3}-L_2-L_3}\\
% && \times\,f_{\ell_{G1} 0\ell_{G1}}(r_{G1},0,s)f_{0\ell'_{H1}\ell'_{H1}}(0,r'_{H1},s) f_{\ell_{G2} \ell'_{H2} L_2}(r_{G2}, r'_{H2}, s) f_{\ell_{G3} \ell'_{H3} L_3}(r_{G3}, r'_{H3}, s)\non
% &&\times\,\calG^{\calL_{G}\calL'_{H}\Lambda''}\calD^{\rm P}_{\Lambda''}\calC^{\Lambda''}_\bfzero\calY_{\calL_{G}}(\bfRhat^{(4)}_{G}) \calY_{\calL_{H}}({{\bfRhat_{H}}^{'(4)}}),
% \endeqar
%Nominally we need to work with the product of these two 9-$j$ symbols and sum over the intermediate angular momentum $\ell_{12}$. 
As before, the presence of zero angular momenta simplifies the intermediate coefficients, such that $\ell_{12}=\ell_{G1}$, $\ell'_{12}=\ell'_{H1}$, and consequently, $\ell''_{12}\equiv L_{1}$. We do not need to consider permutation of the angular momenta $L$ because their allowed range is fixed once the unprimed $\ell_i$ and the primed angular momenta $\ell_i'$ are explicitly given (due to the triangular inequality). With these considerations, the generalized Gaunt integral for $N=4$ can be simplified as:
\begeqar\label{eqn:case-II-contractions-tmp2}
\calG^{\calL_G\calL'_H\Lambda''}
&=&(4\uppi)^{-2} {(\calD^{\rm P}_{\ell_{G1}\ell'_{H1}0})}^2 \prod_{i=1}^3\left[\calD^{\rm P}_{\ell_{Gi}\ell'_{Hi}L_i}\right]\,\calC^{\ell_{G1}0\ell_{G1}}_{000} \calC^{0\ell'_{H1}\ell'_{H1}}_{000}\calC^{\ell_{G2}\ell'_{2}L_2}_{000}\calC^{\ell_{G3}\ell'_{3}L_3}_{000} \non
&&\,\times\,\nine{\ell_{G1}}0{\ell_{G1}}0{\ell'_{H1}}{\ell'_{H1}}{\ell_{G1}}{\ell'_{H1}}{L_{1}}
\nine{\ell_{G1}}{\ell_{G2}}{\ell_{G3}}{\ell'_{H1}}{\ell'_{H2}}{\ell'_{H3}}{{L}_{1}}{{L}_2}{{L}_3}\non
&=& (4\uppi)^{-2} (-1)^{\ell_{G1}+\ell'_{H1}}\calD^{\rm P}_{\ell_{G2} \ell'_{H2} L_2}\calD^{\rm P}_{\ell_{G3} \ell'_{H3} L_3}\,\calC^{\ell_{G2}\ell'_{2}L_2}_{000}\calC^{\ell_{G3}\ell'_{3}L_3}_{000} \, \nine{\ell_{G1}}{\ell_{G2}}{\ell_{G3}}{\ell'_{H1}}{\ell'_{H2}}{\ell'_{H3}}{{L}_{1}}{{L}_2}{{L}_3},
\endeqar
where the first 9-$j$ symbol yields a factor of ${\left(\calD^{\rm P}_{\ell_{G1}\ell'_{H1}}\right)}^{-2}$. The two 3-$j$ symbols involving zero angular momentum get cancelled with ${\calD^{\rm P}_{\ell_{G1}\ell'_{H1}}}$, giving rise to an overall phase factor.

From the definition of the coefficients we find
\begeqar\label{eqn:calDcalC_caseII}
\calD^{\rm P}_{\Lambda''}\calC^{\Lambda''}_\bfzero&=&(-1)^{L_{1}}\sqrt{(2{\ell}_{G1}+1)(2{\ell'}_{H1}+1)(2{\ell_1''}+1)(2{\ell_2''}+1)(2{\ell_3''}+1)}\,\times\,\calC^{\ell_{G1}\ell'_{H1}L_1}_{000}\calC^{L_{1}L_{2}L_3}_{000}.
\endeqar
We proceed by combining Eqs.~(\ref{eqn:calB_caseII}-\ref{eqn:calDcalC_caseII}), inserting these expressions into the definition of the covariance, and projecting out the coefficients proportional to $\calY_{\Lambda}(\bfRhat)$ and $\calY_{\Lambda'}(\bfRhat')$. Noting that $\ell_{G1}+\ell'_{H1}+L_1$ must be an even integer (else $\calC^{\ell_{G1}\ell'_{H1}L_1}_{000}$ is zero), this factor can be dropped from the overall phase.
Altogether we arrive at the final form for Case II:
\begeqar\label{eq: final-cov-II}
    \mathrm{Cov}_{\L,\L'}^{\rm{(fc)},II}(R,R') &=& (4\uppi)^4\sum_{G,H}(-1)^{\Sigma(\Lambda')(1-\calE_H)/2}%(-1)^{\Sigma(\Lambda)(1-\calE_G)/2}\non (Jiamin: this phase is always one if G is a cyc. permutation.)
    \sum_{L_1L_2L_3}\calD^{\rm P}_{L_1L_2L_3}\calC^{L_1L_2L_3}_{000}\, \begin{Bmatrix} \ell_{G1} & \ell_{G2} & \ell_{G3}\\ \ell_{H1}' & \ell_{H2}' & \ell_{H3}'\\ L_1 & L_2 & L_3\end{Bmatrix}\non
    &&\,\times\,\prod_{i=1}^3\left[
    (-1)^{(-\ell_{Gi}-\ell'_{Hi}+L_i)/2}\,
    \calD^{\rm P}_{\ell_{Gi}\ell'_{Hi}L_i} \calC^{\ell_{Gi}\ell'_{Hi}L_i}_{000} \right]\non
    &&\,\times\, \int\frac{s^2ds}{V}f_{\ell_{G1}0\ell_{G1}}(r_{G1},0,s)f_{0\ell_{H1}'\ell_{H1}'}(0,r_{H1}',s)f_{\ell_{G2}\ell_{H2}'L_2}(r_{G2},r_{H2}',s)f_{\ell_{G3}\ell_{H3}'L_3}(r_{G3},r_{H3}',s).
\endeqar
%\bob{need some brackets to indicate what is included in the product} \jiamin{Ok}
%\bob{Again, I would drop this paragraph.}

As before, if we take the limit that the 2PCF is a Dirac delta function, $\xi(\bfs-\bfr_{G1})\to \delta^{\rm [3]}_{\rm D}(\bfs-\bfr_{G1})$ implies the limit $\bfs \to \bfr_{G1}$. Recalling $P(k)= 1$, the $f$-integral associated with the second correlation function becomes
\begeqar
\lim_{\bfs \to \bfr_{G1}} f_{0\ell'_{H1}\ell'_{H1}}(0, r'_{H1}, s) = \lim_{\bfs \to \bfr_{G1}} \int \frac{k^2dk}{2\uppi^2} j_{\ell} (kr_{G1}) j_{\ell} (kr'_{H1}) = \frac{1}{4\uppi r_{G1}r'_{H1}}\delta_{\rm D}^{[1]} (r_{G1}-r'_{H1}).
\endeqar
In addition, %in the limit that correlation function approaches delta functions 
we have $\bfs\to \bfr_{Gi}-\bfr'_{Hi}$ for $i=2, 3$. In this case, the resulting integral of three spherical Bessel functions can be simplified using Eq.\,(3.21) of~\citet{Mehrem1991}, which we do not duplicate here. However, the former work shows the result to be zero unless the three vectors $\bfs$, $\bfr_{Gi}$, and $\bfr'_{Hi}$ form a closed triangle, coinciding with our delta function assumption.\footnote{For a similar discussion for the 3PCF covariance of the limit that $\xi$ becomes a Dirac delta function, see \S6.3 in~\citealt{Slepian201506}.} This result is unsurprising because the Dirac delta function can be written as an integral of a product of spherical Bessel functions. It is interesting to consider the physical picture (cf. Fig.~\ref{fig:delta_limit_CaseI_II}). When the correlation functions approach delta functions in case II, the two tetrahedra also overlap but with their primary vertices sitting on the endpoint of that side, in particular, that side of the tetrahedra must have the same length as the separation vector of each family.

%\bob{again, I would drop any particular evaluation for an assumed behavior of the power spectrum.}

Notably, Case I and Case II have similar mathematical structure, with essentially no differences induced by distinguishing between the primary vertices and the endpoints. Combining both cases allows us to recover the general form (cf.~Eq.~\ref{eqn:cov_npcf_iso}) including all $4!=24$ permutation terms.\footnote{We additionally note that all the above derivations could be performed in the spherical harmonics basis and would have the same results. We will not repeat this derivation here.}

\section{Numerical implementation and comparison with simulations}
\label{sec:numerics_and_simulations}
\subsection{Implementation of the connected covariance}
The ingredients for the analytic covariance calculation from Eq.~\eqref{eqn:cov_npcf_iso} comprise the $f$-integrals, a set of coefficients including the product of $\calD_{\Lambda}^{\rm P}$, Wigner 3-$j$, Wigner 9-$j$ symbols, and the phase.
In practice, we compute all these elements using \textsc{Python}. For efficiency, with the evaluation of the Wigner 3-$j$ and 9-$j$ symbols performed using the \textsc{Sympy} package. We pre-compute the $f$-integrals for each radial bin, as well as the coupling coefficients, before assembling the covariance. These are stored in dictionary format and loaded during the calculation. To compute the $f$-integral, which involves fine binning in $k$ and $s$, we use an analytic form for the bin-averaged spherical Bessel functions (cf. Eq.~\ref{eqn:sbf_bin_avg}), which is exact and speeds up the implementation. We use $5,000$ points in $k\in \left[10^{-4}, 5\right]\, {\rm Mpc}^{-1} h$ and $4,100$ points in $s\in \left[10^{-5}, 10^3\right]h^{-1}\, {\rm Mpc}$ with both linearly spaced. We choose these ranges and grid sizes such that on the one hand, the arrays fit in the same memory block managed by \textsc{NumPy}, and on the other hand, they cover the integration range of interest with sufficiently small grid size. Given that our aim is to measure the 4PCF up to $\ell_{\rm max}=4$, we compute the $f$-integrals up to $\ell=8$ (considering $L=\ell+\ell'$). 
To verify the numerical evaluation and implementation of the bin-averaged $f$-integral, we compared the resulting forms to an analytic solution for the integral of a product of three spherical Bessel functions \citep{Fabrikant2013}, modified to accommodate for the bin-averaging. This is discussed in Appendix \ref{appendix:fabrikant_rbin}.

\subsection{Comparison with \logn simulations}
\label{subsec:lognormal}
We now compare theoretical covariance to those extracted from simulations. First, we use a set of $1,000$ \logn mocks at redshift $z=2$ with a number density of $\sim 1.5 \times\, 10^{-4}\, [h^{-1}{\rm Mpc}]^{-3}$ and volume $V = 3.9\, [{\rm Gpc}/h]^3$.\footnote{While it may seem more prudent to construct simulations that match our assumption of Gaussianity, this is non-trivial, since we require a discrete density field. In principle, one could use a set of discrete particles which are assigned the Gaussian random field value as weights. However, this approach does not reproduces covariance correctly, since it puts multiple galaxies at the same position and effectively enhances the shot noise.}
The \logn mocks are generated using \nbk \citep{Hand2018}, where the overdensity fields are evolved according to the Zel’dovich approximation (zeroth-order Lagrangian perturbation theory)~\citep{Schneider1995}. We prepare mocks in both real- and redshift space in order to investigate  the impact of RSD on the covariance. The input linear power spectrum is generated with the cosmological parameters $\left\{\Omega_{\rm m},\, \Omega_{\rm b}h^2,\, h, \,n_{\rm s},\, \sigma_8 \right\} = \left\{0.31,\, 0.022, \,0.676,\, 0.97, \, 0.8 \right\}$ with a linear bias $b_1=1.8$. The 4PCFs are measured using the \textsc{encore} code~\footnote{\url{https://github.com/oliverphilcox/encore}} at 10 radial bins centered at $r_{\rm bin}=\{27, 41, ..., 153\}\, \mpch$ with a bin width of $14\,\mpch$.
In this setup, these \logn mocks have a low level of non-Gaussianity due to the high redshift and have a relatively high shot noise.

%\bob{you need to explain a litte - at least say that it gives RSD}   

\begin{figure}
     \centering
     \begin{subfigure}[b]{0.32\textwidth}
         \centering
         \includegraphics[width=\textwidth]{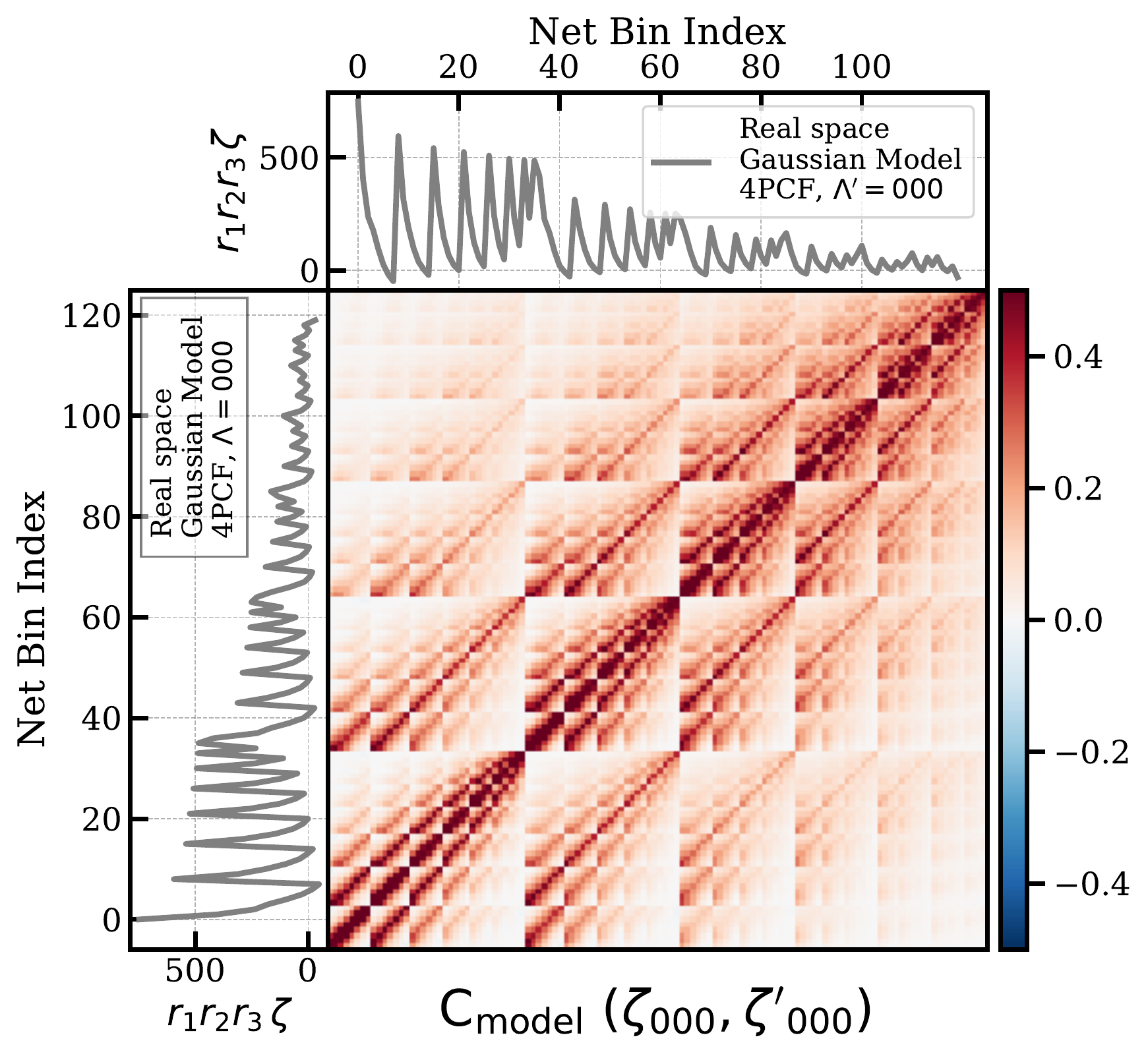}
         \caption{Analytic correlation matrix}
         \label{fig:logn_real_cov_model_000_000}
     \end{subfigure}
     \hfill
     \begin{subfigure}[b]{0.32\textwidth}
         \centering
         \includegraphics[width=\textwidth]{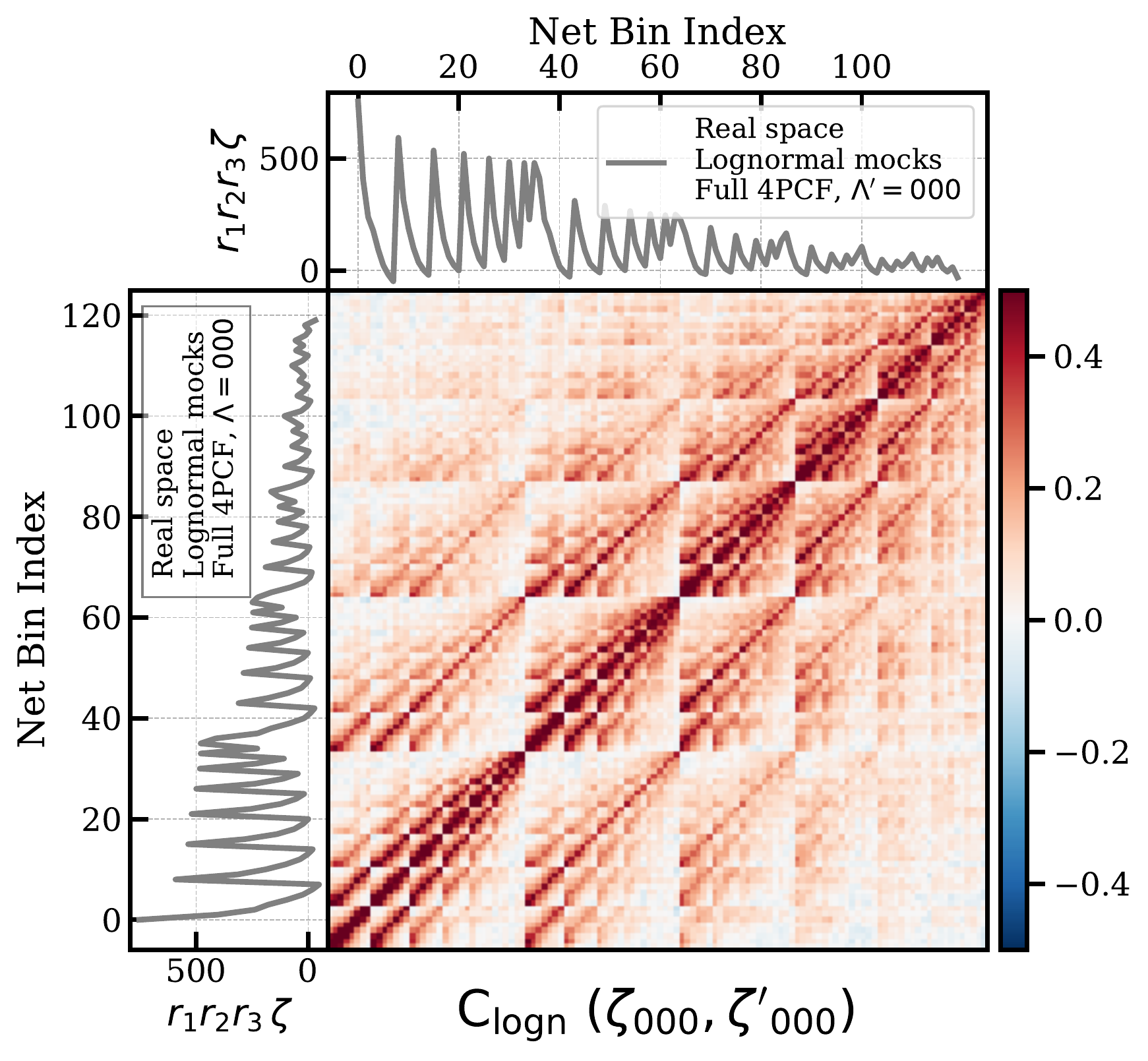}
         \caption{Lognormal correlation matrix}
         \label{fig:logn_real_cov_mock_000_000}
     \end{subfigure}
     \hfill
     \begin{subfigure}[b]{0.28\textwidth}
         \centering
         \includegraphics[width=\textwidth]{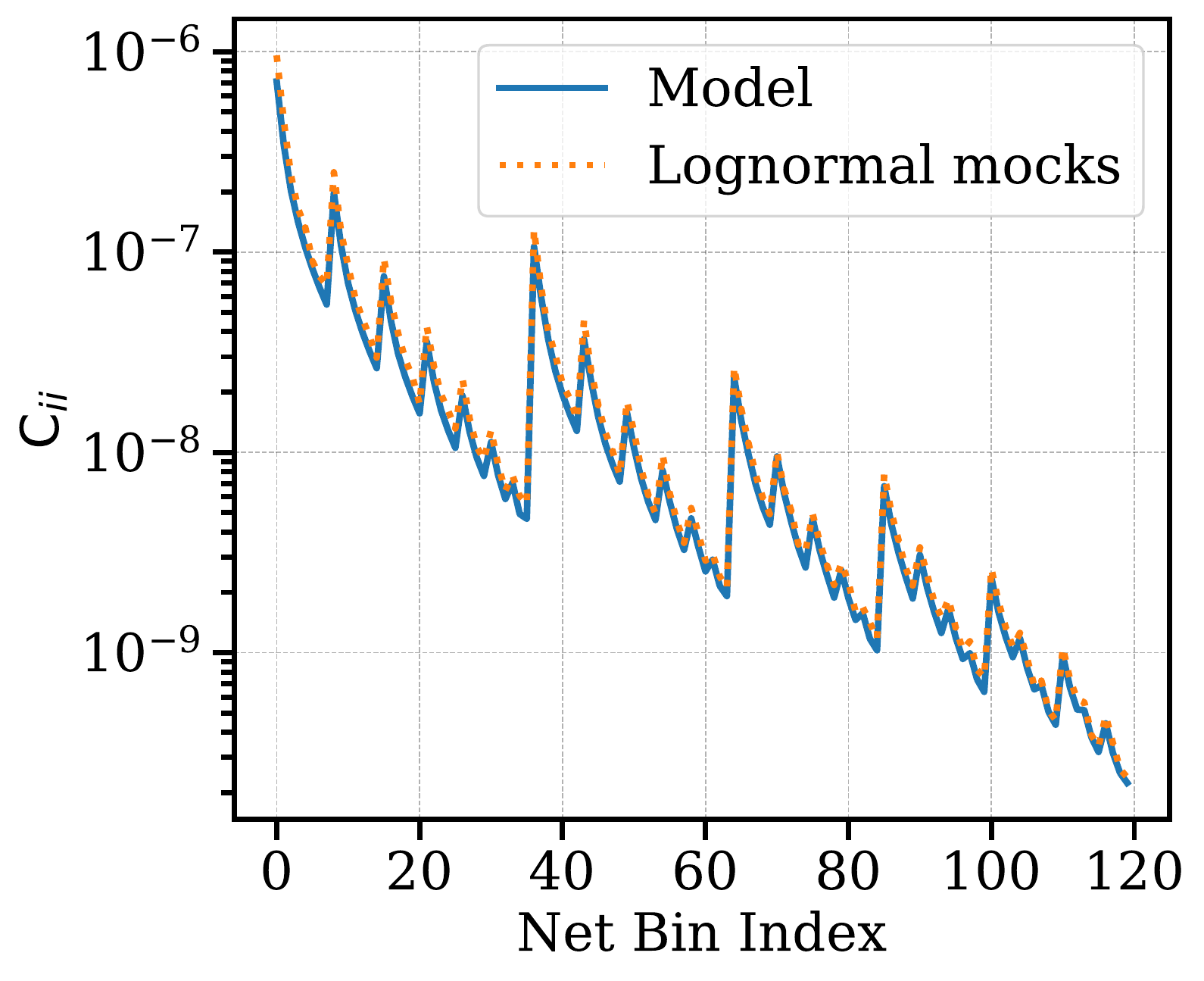}
         \caption{Comparison of diagonals}
         \label{fig:logn_real_diag_000_000}
     \end{subfigure}
    \caption{Comparison of the analytic and sample covariance matrices for a set of lognormal simulations. The first and second panels show the comparison of the correlation matrix (defined by $\mathsf{M}_{ij}=\mathsf{C}_{ij}/\sqrt{\mathsf{C}_{ii} \mathsf{C}_{jj}}$) for angular momenta $\{\Lambda, \Lambda'\}= \{000, 000\}$ in real space. Fig.\,\ref{fig:logn_real_cov_model_000_000} gives the model prediction for the fully-coupled 4PCF correlation matrix, and the panels above and to the left show the (disconnected) Gaussian 4PCF model in real space. 
    %\bob{I changed this} \jiamin{ok, reads good} 
    The horizontal and vertical axes indicate 120 radial bins, ordered so that $r_1<r_2<r_3$.  This gives rise to the block structure in the matrix and the saw-tooth shape of the correlation functions.
    Fig.\,\ref{fig:logn_real_cov_mock_000_000} shows the correlation matrix estimated from $1,000$ \logn mocks, with the extended panels showing the measured full 4PCF from the \logn mocks in real space. Fig.\,\ref{fig:logn_real_diag_000_000} shows a comparison of the diagonal elements of the two covariance matrices; we note the values (vertical axis) are logarithmically scaled.}
\label{fig:compare1_logn_real_000_000}
\end{figure}
The sample covariance estimated from mock simulations is defined as
\begeqar
\mathsf{C}_{\rm mock} = \frac{1}{N_{\rm mock}-1}
\sum_{i=1}^{N_{\rm mock}} \left(\zeta^{(i)}-\bar{\zeta}\right)\left(\zeta^{(i)}-\bar{\zeta}\right)^{\rm T},
\endeqar
where the data vector $\zeta^{(i)}$ (with dimension $N_{\rm bins}$) is the 4PCF measured from the $i$-th mock simulation, and $\bar{\zeta}$ is the mean over all $N_{\rm mock}$ realizations. Since the mean is estimated from the mocks themselves, the definition includes the prefactor $(N_{\rm mock}-1)^{-1}$.

When computing the $f$-integrals in real space, we use the same input power spectrum that was used to generate the \logn mocks. In redshift space the power spectrum is additionally multiplied by the isotropic Kaiser factor $(b^2+2fb/3+f^2/5)/b^2$, with $f$ being the logarithmic derivative with respect to scale factor of the linear growth rate~\citep{Kaiser1987}. In both cases, we damp the power spectra by $\exp\left(-(k/k_0)^2\right)$ to avoid numerical issues, setting $k_0=1\, [{\rm Mpc}^{-1}\, h]$.
We find that the shot noise term is sensitive to the precise form of exponential damping function. For the \logn mocks, which feature a large shot noise, we observe better agreement between theory and simulations when the shot noise damping is not included.

Fig.~\ref{fig:compare1_logn_real_000_000} shows a comparison between the theoretical and sample covariance from the \logn mocks for angular momenta $\{\Lambda, \Lambda'\}= \{000, 000\}$ in real space. 
%\bob{I added this sentence:} The analytic calculation uses as its power spectrum the power spectrum that was the basis for the the lognormal mocks. \jiamin{Is this not the same as the next sentence?}  
The 2D plot in the first panel shows the model prediction for the fully-coupled 4PCF correlation matrix $\mathsf{M}$, where the correlation matrix is the covariance matrix $\mathsf{C}$ normalized by its diagonal terms, \textit{i.e.} $\mathsf{M}_{ij}=\mathsf{C}_{ij}/\sqrt{\mathsf{C}_{ii} \mathsf{C}_{jj}}$. We arrange the radial bins in the following manner: we start by fixing bins in $r_1$ and $r_2$ and loop over $r_3$, then move to the next radial bin in $r_2$ at the same fixed $r_1$ and again loop over $r_3$, before move to the next bin in $r_1$. This is repeated until all possible radial binning combinations are explored; this specific way of arranging the bins is denoted as the net bin index. During this process we force the radial bin arrangement to be $r_1<r_2<r_3$. In total, we have $C_{10}^3=10!/(7!\,3!)=120$ radial bins.
The radial bin arrangement also leads to the block structure in the covariance matrix.

The second panel of Fig.~\ref{fig:compare1_logn_real_000_000} shows the measurement from $1000$ \logn mocks in real space with the inset showing the measurements of the full 4PCF from Gaussian mocks. Comparing the first and the second panel, we can see that the analytic covariance is able to capture the off-diagonal features. The covariance for $\{\Lambda, \Lambda'\}= \{000, 000\}$ is mostly positive as a result of the auto-covariance for the angular momenta themselves. The third panel shows a comparison of the diagonal elements of these two matrices in log-scale. 
The extended panels at the top and right of the first panel in Fig.~\ref{fig:logn_real_cov_model_000_000} show the Gaussian 4PCF model in real space, where the (disconnected) Gaussian 4PCF consists of a product of two 2PCF (see appendix A in~\citet{Philcox2021boss4pcf} for a derivation ). Since the 2PCF is approximately given by a declining power law, combination with our radial bin arrangement leads to the saw-tooth shape of the 4PCF. The extended panels at the top and right of the second panel of Fig.~\ref{fig:logn_real_cov_mock_000_000} are the measured \textit{full} 4PCF (includes both connected and disconnected term) in real space. They both assist the visualization of the block structure of the correlation matrices.

%Fig. 5
\begin{figure}
     \centering
     \begin{subfigure}[b]{0.35\textwidth}
         \centering
         \includegraphics[width=1\textwidth]{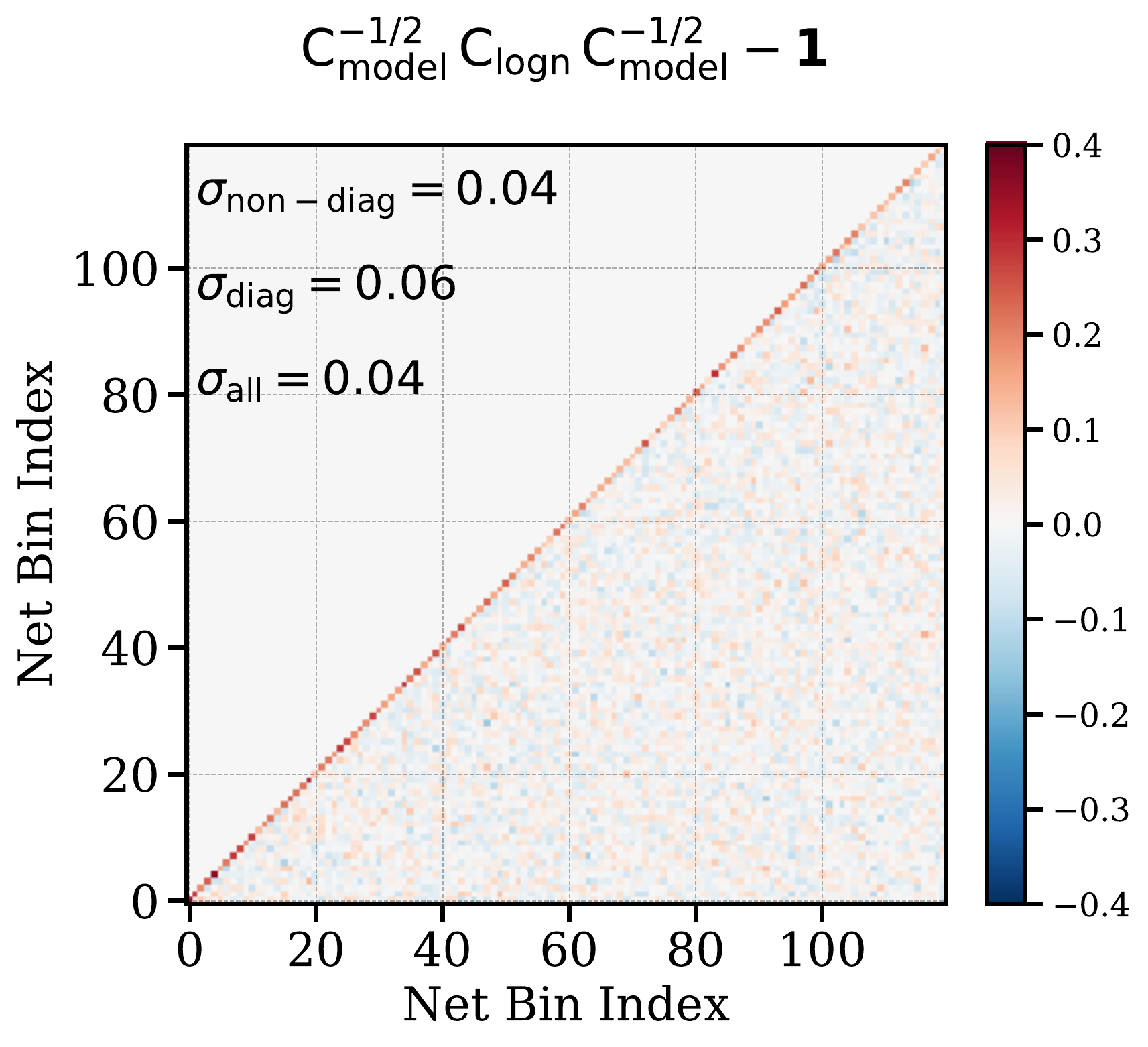}
         \caption{Half-inverse test $\{\Lambda, \Lambda'\}= \{000, 000\}$}
         \label{fig:half_inv_logn_real_000_000}
     \end{subfigure}
    %  \hfill
     \begin{subfigure}[b]{0.35\textwidth}
         \centering
         \includegraphics[width=1\textwidth]{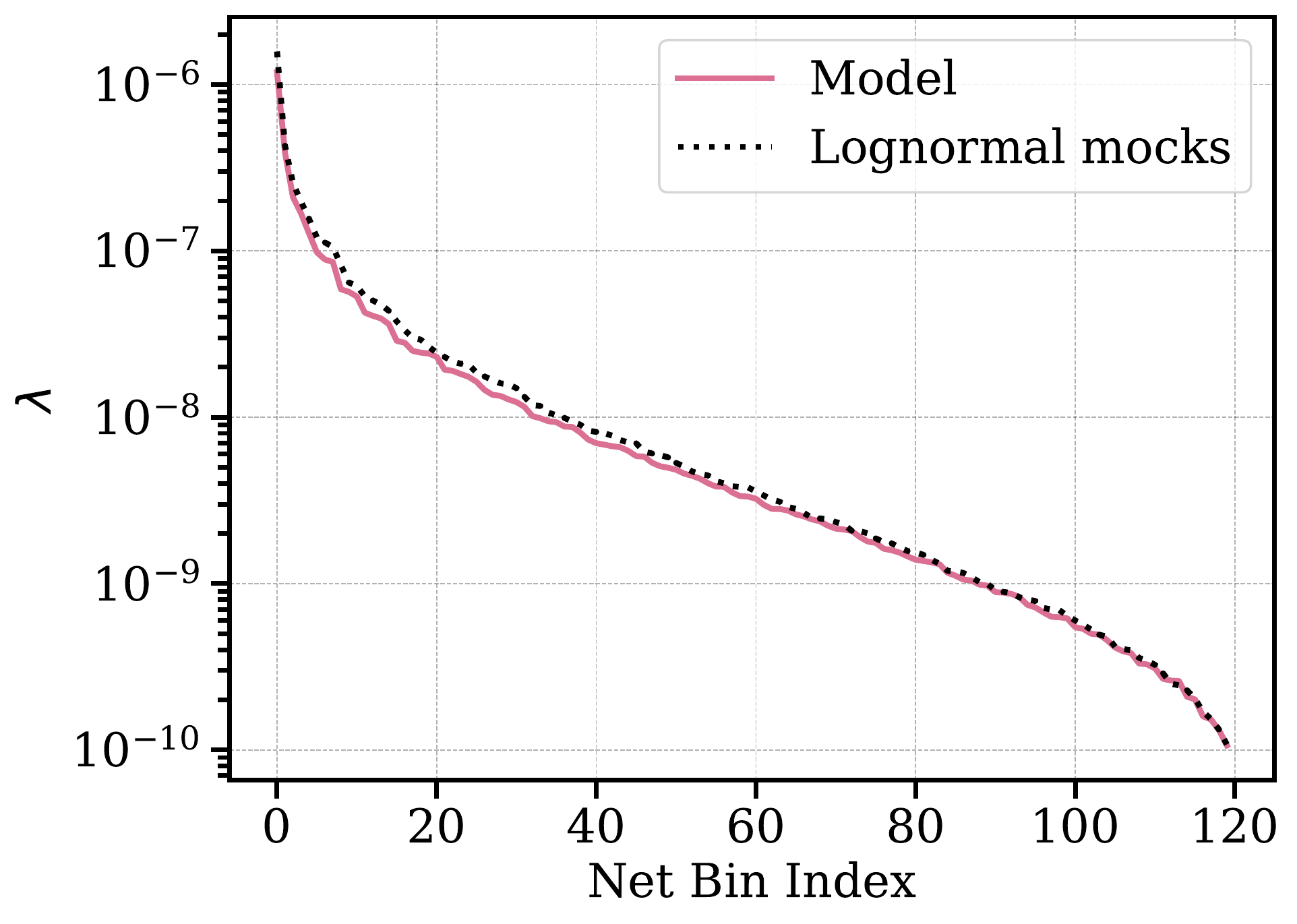}
         \caption{Eigenvalue $\{\Lambda, \Lambda'\}= \{000, 000\}$}
         \label{fig:eigval_logn_real_000_000}
     \end{subfigure}
    % %  \hfill
    %  \begin{subfigure}[b]{0.24\textwidth}
    %      \centering
    %      \includegraphics[width=1\textwidth]{figs/half_inv_4pcf_connected_022x033_model_logn_real.pdf}
    %      \caption{Half-Inverse test}
    %      \label{fig:half_inv_logn_real_000_000}
    %  \end{subfigure}
    % %  \hfill
    %  \begin{subfigure}[b]{0.25\textwidth}
    %      \centering
    %      \includegraphics[width=1\textwidth]{figs/evals_4pcf_connected_022x033_model_logn_real.pdf}
    %      \caption{Eigenvalue}
    %      \label{fig:eigval_logn_real_022_033}
    %  \end{subfigure}
    \caption{{\bf Left panel:} half-inverse test comparing the lognormal simulations and the analytic covariance, both of which are shown in Fig.\,\ref{fig:compare1_logn_real_000_000}. If the covariance matrices agree, both the mean and the off-diagonal elements should be noisy fluctuations around zero. For clarity, we plot only the lower triangle, and give the standard deviation for the off-diagonal elements, $\sigma_{\rm non-diag}$, for the diagonal elements, $\sigma_{\rm diag}$, and for all elements combined, $\sigma_{\rm all}$. {\bf Right panel:} comparison between the eigenvalues of the analytic covariance (solid curve) and \logn mock covariance (dotted curve).}
    \label{fig:compare2_logn_real_000_000}
\end{figure}
In order to quantify the similarity between the Gaussian model prediction and the mock measurements, we perform a test, which we label as `the half-inverse test'. This considers the matrix
\begeqar
\mathsf{S}\equiv\mathsf{C}_{\rm model}^{-1/2}\mathsf{C}_{\rm mock}\mathsf{C}_{\rm model}^{-1/2} - \mathds{1},
\endeqar
where $\mathds{1}$ is the identity matrix. If the two covariances were identical $\mathsf{S}$ which would vanish~\citep{Deadman2013}.  Fig.~\ref{fig:compare2_logn_real_000_000} shows the half-inverse test in the left panel, with the eigenvalues of the 4PCF covariance 
%\bob{You really mean covariance, not the correlation matrix, rght? }\jiamin{yes} 
inferred from the model (solid blue curve) and the mocks (dotted black curve) shown in the right panel.
%We label the standard deviation of the off-diagonal elements $\sigma_{\rm non-diag}$, the diagonal elements $\sigma_{\rm diag}$, and the combined $\sigma_{\rm all}$. OP: should be in caption
If the analytic and sample covariance matrices agree, the half-inverse matrix should follow a Wishart distribution~\citep{wishart28,Anderson2009} and we expect the standard deviation of half-inverse matrix elements to scale as $1/\sqrt{N_{\rm mock}}$$\sim$$0.03$, where $N_{\rm mock}=1,000$ is the number of mocks.
%\bob{is this the standard deviation of the elements of covariance or the correlation matrix?  I would expect it to be the latter.} \jiamin{the std is for the covariance matrix and the half inverse test uses covariance matrix.}\bob{I believe the std will vary as $1/\sqrt{n}$, but the $0.03$ suggests an actual size, which I don't believe.} \jiamin{yes, std does vary with $1/\sqrt{N_{\rm mock}}$. 0.03 is not a fixed number, it's just because we used 1000 mocks in this test.} 
The standard deviation of the diagonal elements should be two times larger than that of the off-diagonal ones, since the expression for the variance of a Wishart distribution contains a Kronecker delta for matrix elements $i=j$. 

For the \logn mocks, the mean of the half-inverse matrix elements is $\left< \mathsf{S}\right>=2.3\times 10^{-3}$, much smaller than their standard deviation.
%\bob{What is the mean of a matrix?} \jiamin{the mean of the covariance matrix is at the order of $\sim 10^{-9}$. But here we are really interested in the mean of the half inverse $\mathsf{S}$ matrix, because it quantifies if the model and sample covariance are identical or not.}
However, we observe a residual in the diagonal terms; indeed, the mean of these is $0.180$. If we decompose the theoretical covariance into its diagonal eigenvalue matrix $\mathsf{D}$ and a unitary matrix $\mathsf{V}$ of eigenvectors, we can write $\mathsf{C}_{\rm model}^{-1/2}=\mathsf{V}\mathsf{D}^{-1/2}\mathsf{V}^{-1}$. If the eigenbasis of the analytic covariance is close enough to the mock-estimated one, the half-inverse test reduces to the ratio between the eigenvalues of the two covariances. Here, we see that the eigenvalues of the model covariance are slightly lower than those of the mock covariance. A possible explanation for this residual is that the \logn mocks have intrinsically high shot noise, which can generate non-Gaussian (but Poissonian) terms in the covariance that require modeling beyond the Gaussian approximation. 
Another possibility arises from the choice of input power spectrum. Here, we used the power spectrum which generated the \logn mock, instead of that measured from the \logn mocks. Due to the lognormal transformation of the density fields, and post Zel'dovich evolution, the two spectra could differ slightly.%, the two power spectra could differ due to the lognormal transformation of the density field as well as the density evolution given by the Zel'dovich approximation.

%Fig.6
\begin{figure}
     \centering
     \begin{subfigure}[b]{0.26\textwidth}
         \centering
         \includegraphics[width=\textwidth]{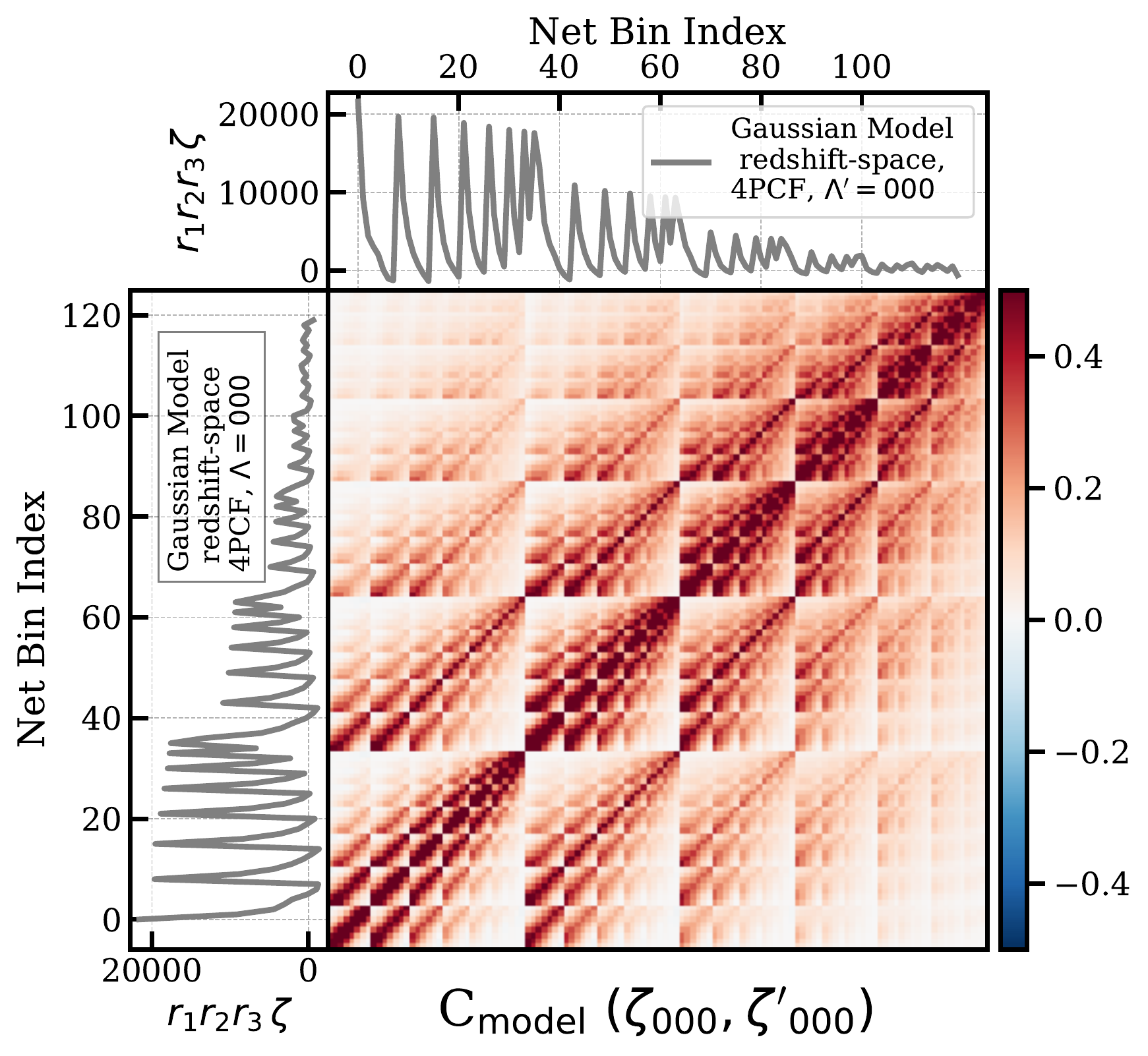}
         \caption{Analytic correlation matrix}
         \label{fig:logn_zs_cov_model_000_000}
     \end{subfigure}
    %  \hfill
     \begin{subfigure}[b]{0.26\textwidth}
         \centering
         \includegraphics[width=\textwidth]{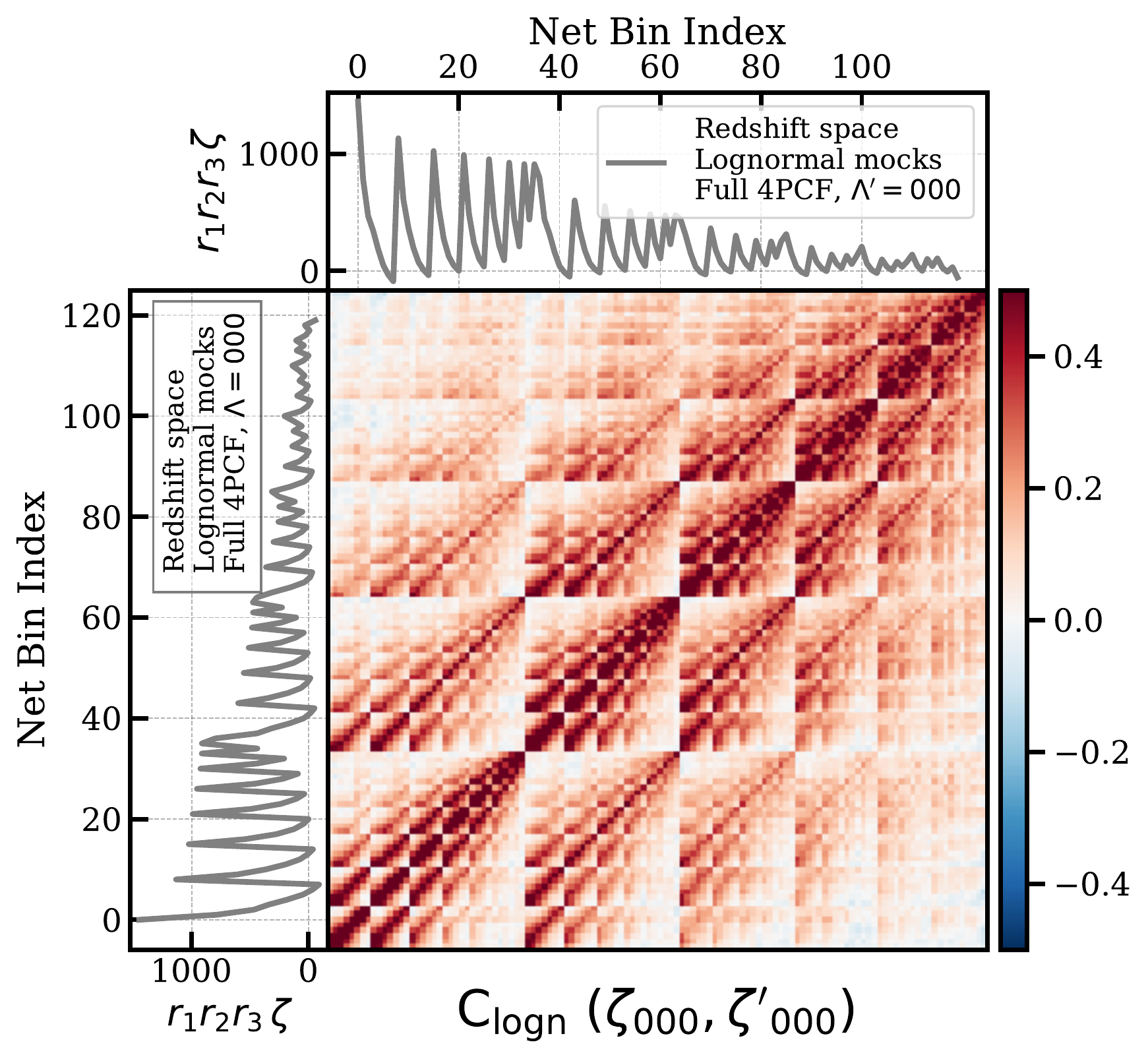}
         \caption{Lognormal correlation matrix}
         \label{fig:logn_zs_cov_mock_000_000}
     \end{subfigure}
    %  \hfill
     \begin{subfigure}[b]{0.23\textwidth}
         \centering
         \includegraphics[width=\textwidth]{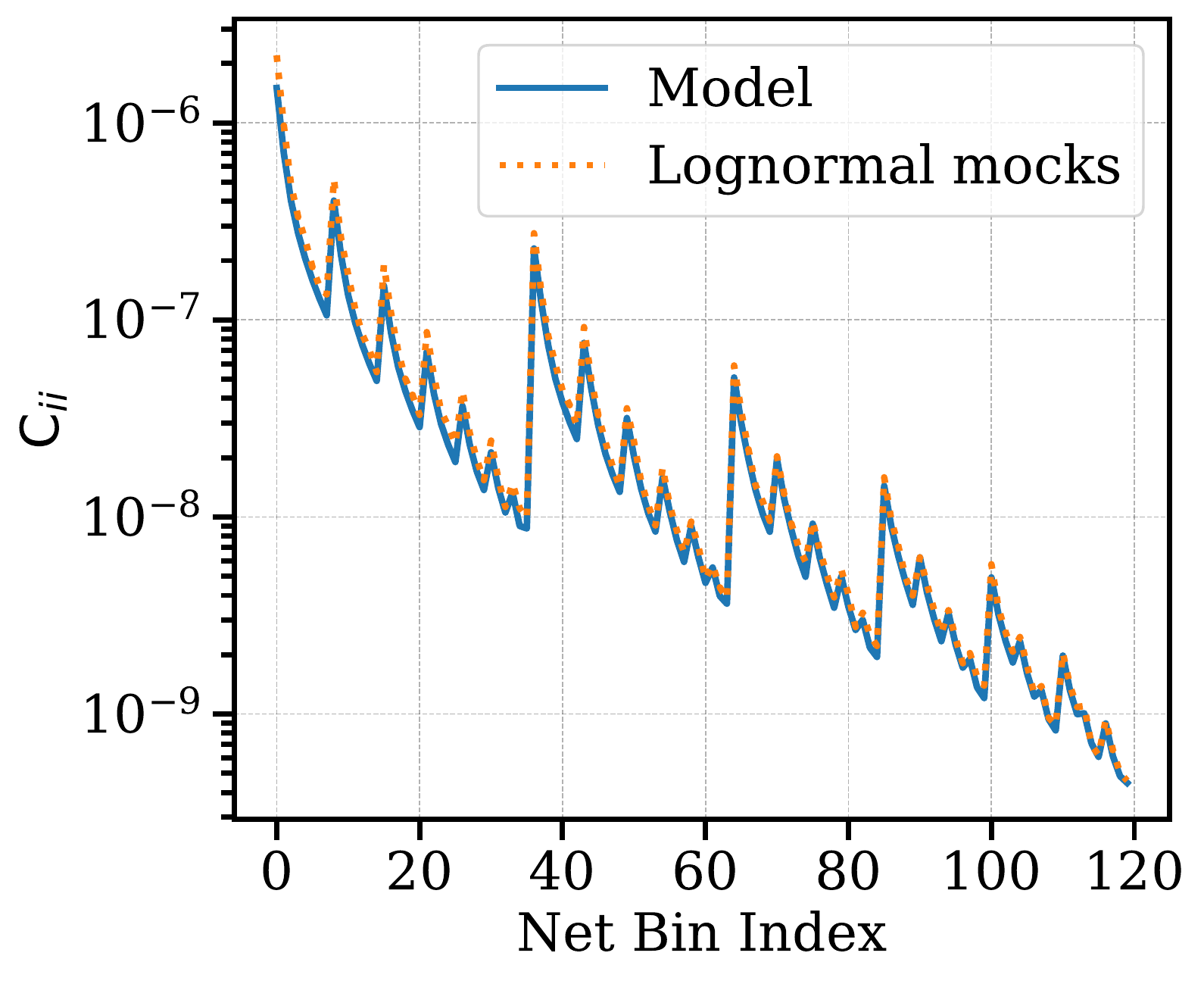}
         \caption{Comparison of diagonals}
         \label{fig:logn_zs_diag_000_000}
     \end{subfigure}
     \begin{subfigure}[b]{0.23\textwidth}
         \centering
         \includegraphics[width=1\textwidth]{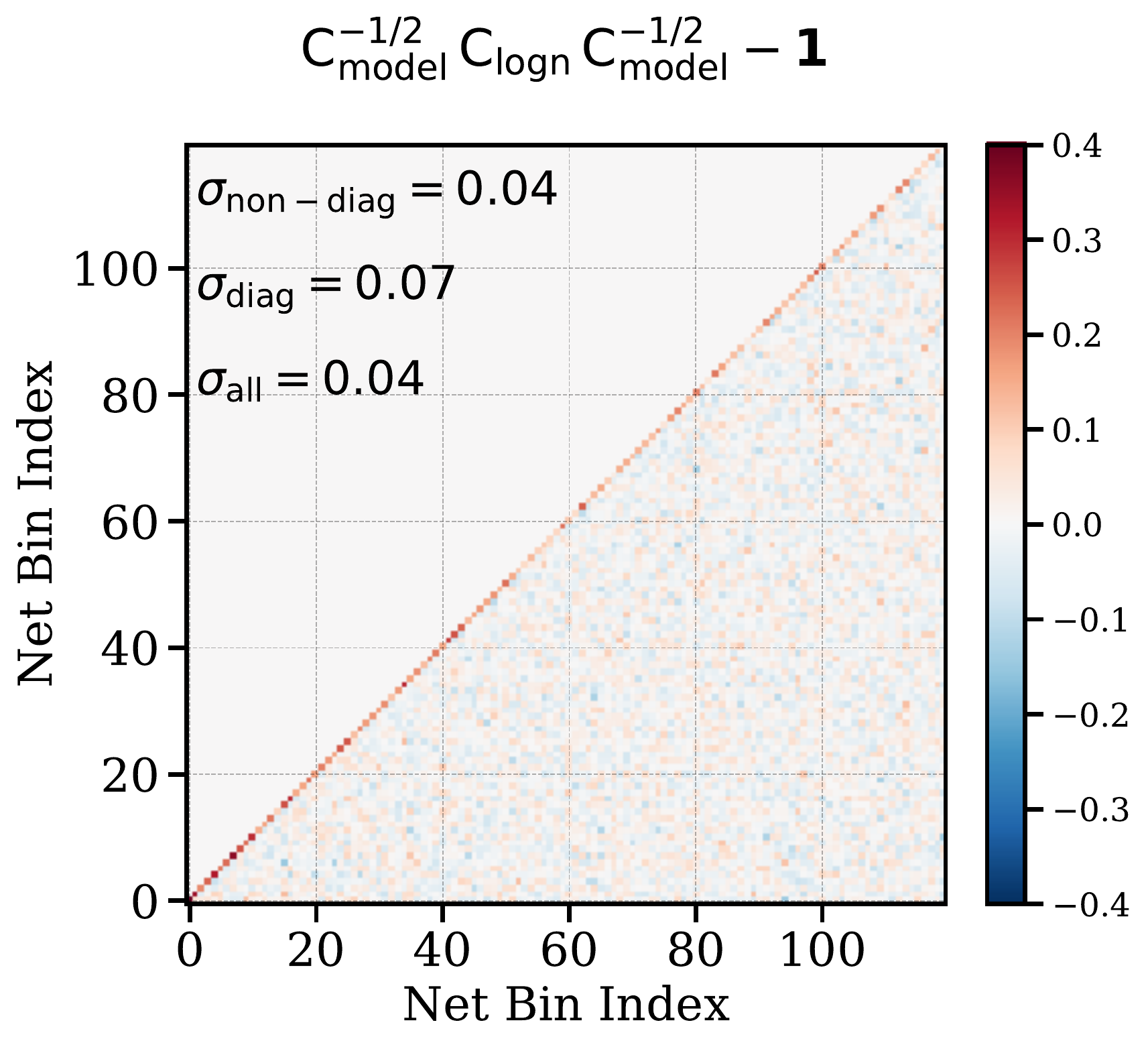}
         \caption{Half-inverse test}
         \label{fig:half_inv_logn_zs_000_000}
     \end{subfigure}
    \caption{As Fig.\,\ref{fig:compare1_logn_real_000_000}, but comparing the analytic and lognormal covariances in redshift space. Fig.\,\ref{fig:compare1_logn_zs_000_000} gives the results of the half-inverse test in the same format as Fig.\,\ref{fig:half_inv_logn_real_000_000}. The model also works well in redshift space in that it shows comparable coupling structure for the correlation matrices and the diagonal elements of the covariances. However, we do observe a residual in the diagonal elements of the matrix for the half-inverse test.}
%    The first and second panels show a comparison of correlation matrix $\mathsf{M}$ for angular momenta $\{\Lambda, \Lambda'\}= \{000, 000\}$ in redshift space. The first panel is the model prediction for the fully-coupled 4PCF correlation matrix, the extended panels show the Gaussian 4PCF model in redshift space. The second panel shows the correlation matrix estimated from $1000$ \logn mocks, the extended panels are the measurements of the full 4PCF from the \logn mocks in redshift space. The third panel shows a comparison of the diagonal elements of the two covariance matrices in log-scale. The fourth panel is the half inverse test for the model and sample covariance.}
\label{fig:compare1_logn_zs_000_000}
\end{figure}

Fig.~\ref{fig:compare1_logn_zs_000_000} is similar to Fig.~\ref{fig:compare1_logn_real_000_000}, but shows a comparison between the two sets of covariances in redshift space. Compared to the real space correlation matrix, we see that RSD slightly enhances the off-diagonal structure for $\{\Lambda, \Lambda'\}= \{000, 000\}$. The agreement in the diagonal elements and the half-inverse test are of the similar level compared to the real space test, with similar diagonal residual found in the half-inverse test as well. Although our numerical implementation of the 4PCF covariance ignores higher order angular momentum contribution arising from RSD, \footnote{See Appendix \ref{appendix:cov_fc_rsd_iso} for a more rigorous treatment of this effect} this comparison shows that the RSD effect can be largely accounted for by simply modeling the covariance using an input power spectrum equal to the RSD monopole. Finally, we note that the RSD doubles the amplitude of the Gaussian 4PCF model and the full 4PCF measured from the mocks in the extended panels of Fig.~\ref{fig:logn_zs_cov_model_000_000} and Fig.~\ref{fig:logn_zs_cov_mock_000_000}. These quantities are dominated by the two-point statistics and the increase in the amplitude is approximately given by the Kaiser factor to the fourth power. %As before, both the extended panels are intended to guide the eye for the block structure of the correlation matrix. 

\subsection{Comparison with \quijote simulations}
\label{subsec:quijote}
To further understand the non-Gaussianity arising from gravitational evolution and to test the validity of our Gaussian assumption, we compare the theoretical covariance formalism to the sample covariance measured from the \quijote halo catalogues.\footnote{\url{https://quijote-simulations.readthedocs.io/en/latest/halos.html}} Each of the \quijote simulations has a box size of $V = 1.0\, [h^{-1}\,{\rm Gpc}]^3$, a fiducial cosmology $\left\{\Omega_{\rm m},\, \Omega_{\rm b},\, h, \,n_{\rm s},\, \sigma_8 \right\} = \left\{0.3175,\, 0.049, \,0.6711,\, 0.9624, \, 0.834 \right\}$, zero neutrino mass, and is at redshift $z=0.5$~\citep{Quijote_sims}. 
%\bob{Should't we say what value of z each set of mocks corresponds to?} \jiamin{good point, added.}

We test our algorithm on $100$ \quijote halo catalogues created from $1,024^3$ cold dark matter (CDM) particles. Halos are identified using a particle number cut $N_{\rm particle}>150$ per halo, which corresponds to $M_{\rm cut}=1.2\times 10^{13}\, [h^{-1}\,M_{\odot}]$. This gives two times lower shot noise compared to the \logn mocks. As before, the catalogues are prepared both in real and redshift space, and we use the same radial binning.
%The 4PCFs are measured at the same 10 radial bins as the \logn mocks. We first verified the algorithm on the low resolution version and obtained good agreement. 
The $f$-integral is constructed from the power spectrum monopole measured from the \quijote halo catalogues for both real and redshift space.
For this set of simulations we applied exponential damping to both the power spectrum and shot noise.

\begin{figure}
     \centering
     \begin{subfigure}[b]{0.32\textwidth}
         \centering
         \includegraphics[width=\textwidth]{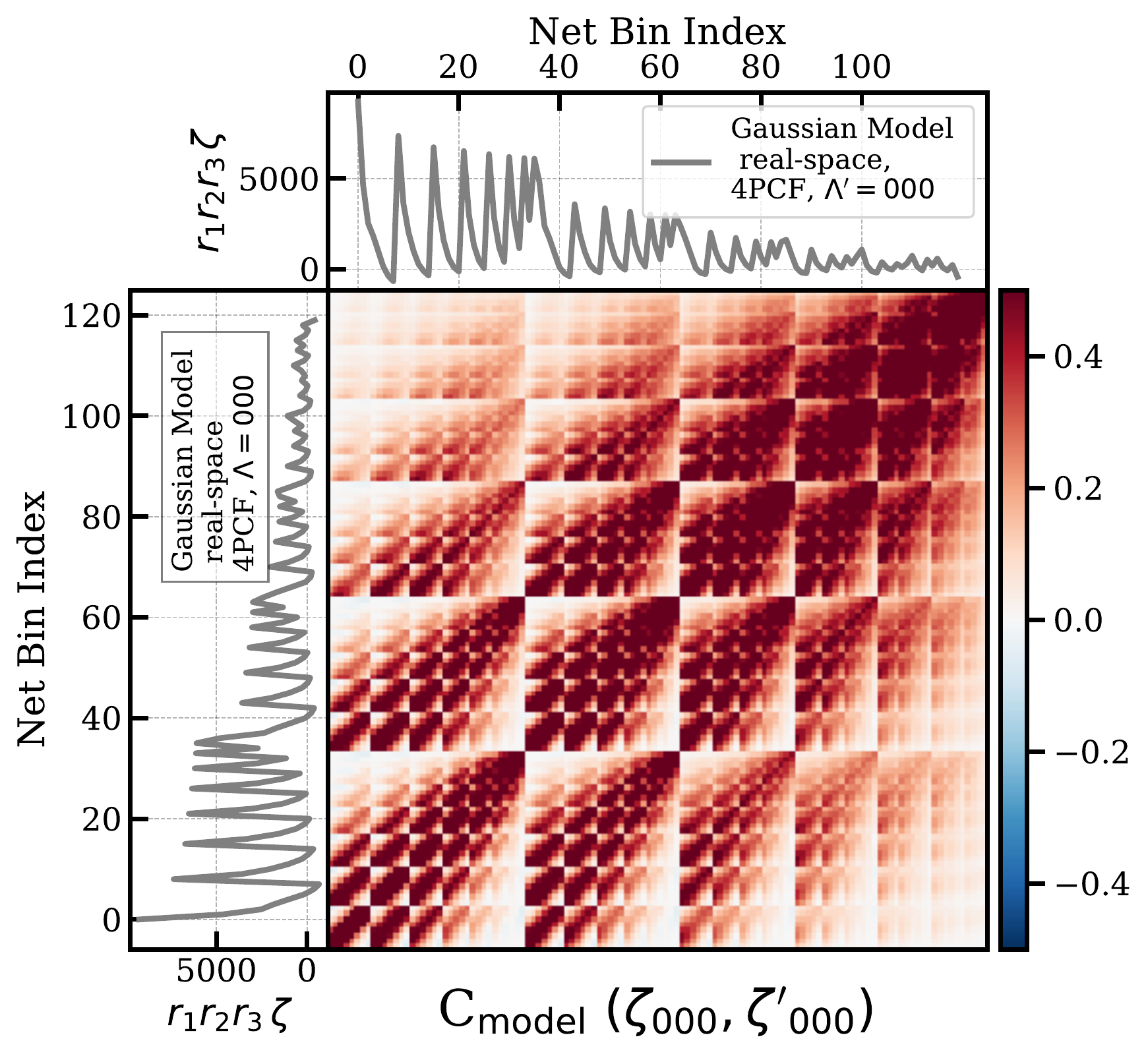}
         \caption{Analytic correlation matrix}
         \label{fig:qjt_halo_hr_real_cov_model_000_000}
     \end{subfigure}
     \hfill
     \begin{subfigure}[b]{0.32\textwidth}
         \centering
         \includegraphics[width=\textwidth]{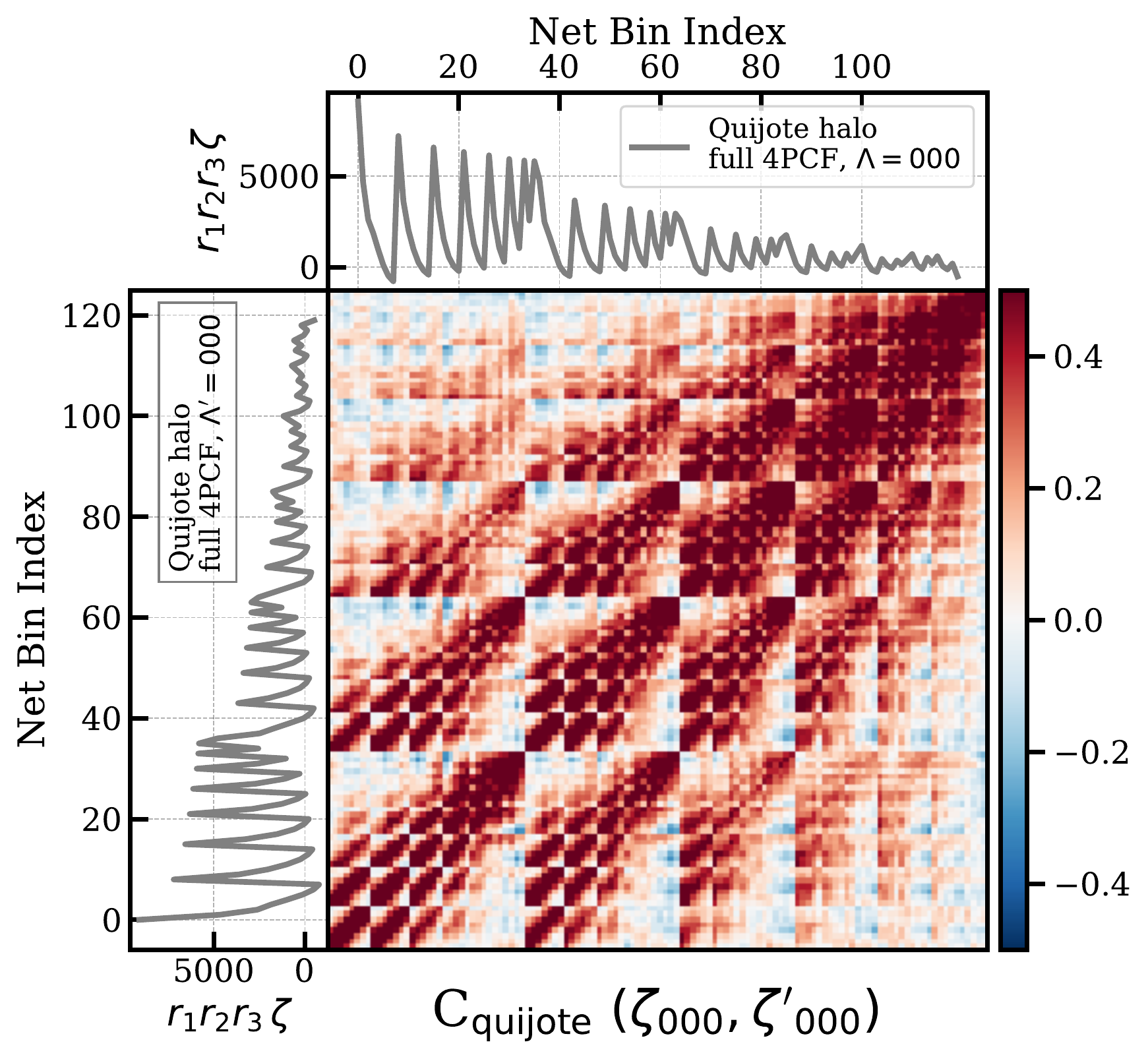}
         \caption{\quijote correlation matrix}
         \label{fig:qjt_halo_hr_real_cov_mock_000_000}
     \end{subfigure}
     \hfill
     \begin{subfigure}[b]{0.28\textwidth}
         \centering
         \includegraphics[width=\textwidth]{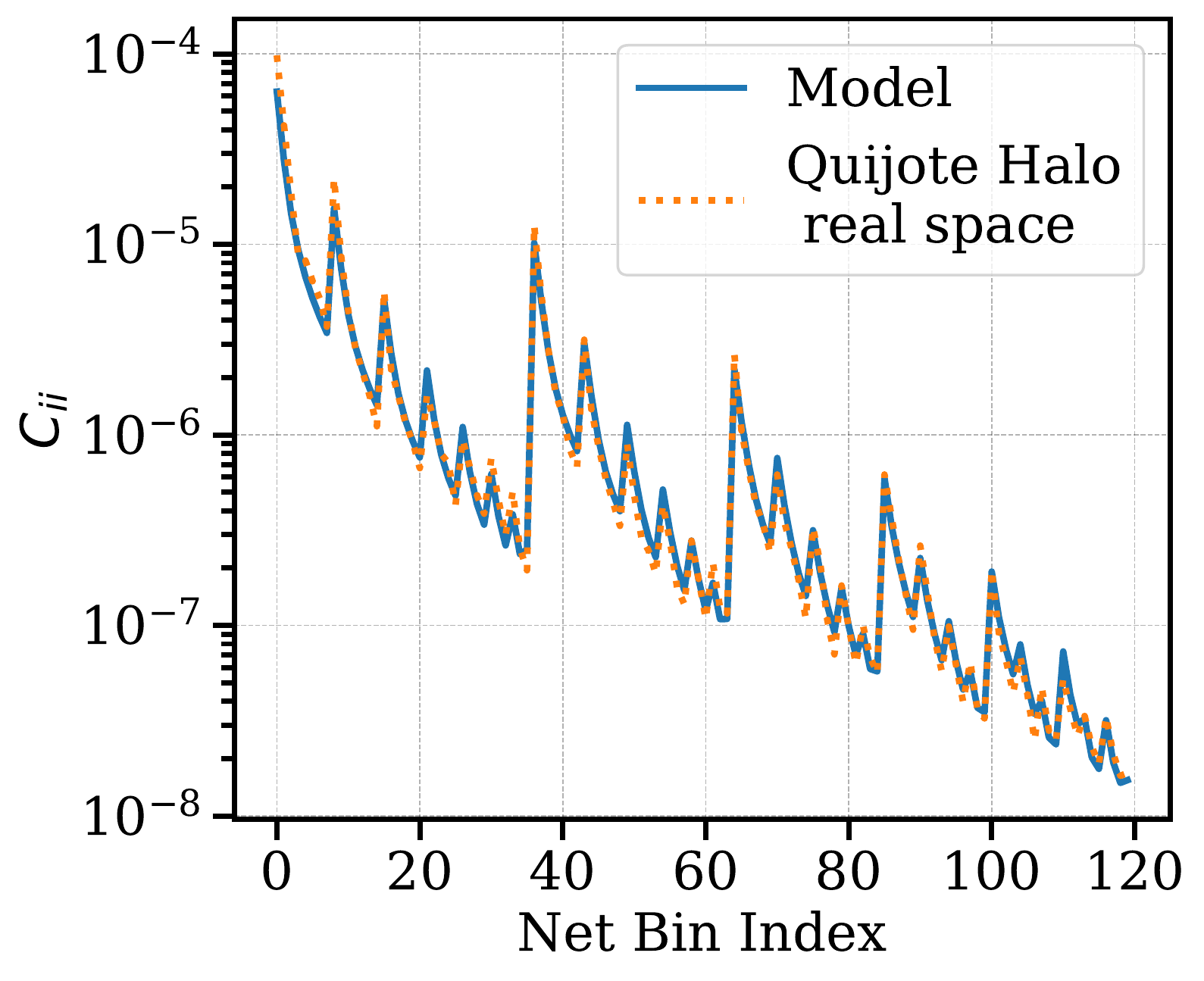}
         \caption{Comparison of diagonals}
         \label{fig:qjt_halo_hr_real_diag_000_000}
     \end{subfigure}
    \caption{As Fig.~\ref{fig:compare1_logn_real_000_000}, but for \quijote halo catalogue in real space, using 1,000 simulations. For the simulations with non-negligible non-Gaussianity, the model can adequately predict various features of the correlation matrix, with a good match for the diagonal elements of the covariance as well.
    %. Left panel: model prediction for the correlation matrix. Middle: correlation matrix estimated from the $100$ \quijote halo catalogue. Right panel: diagonal elements of the model covariance (solid blue) and sample covariance (dotted orange).
    }
    \label{fig:compare1_qjt_halo_hr_real_000_000}
\end{figure}

\begin{figure}
     \centering
     \begin{subfigure}[b]{0.32\textwidth}
         \centering
         \includegraphics[width=\textwidth]{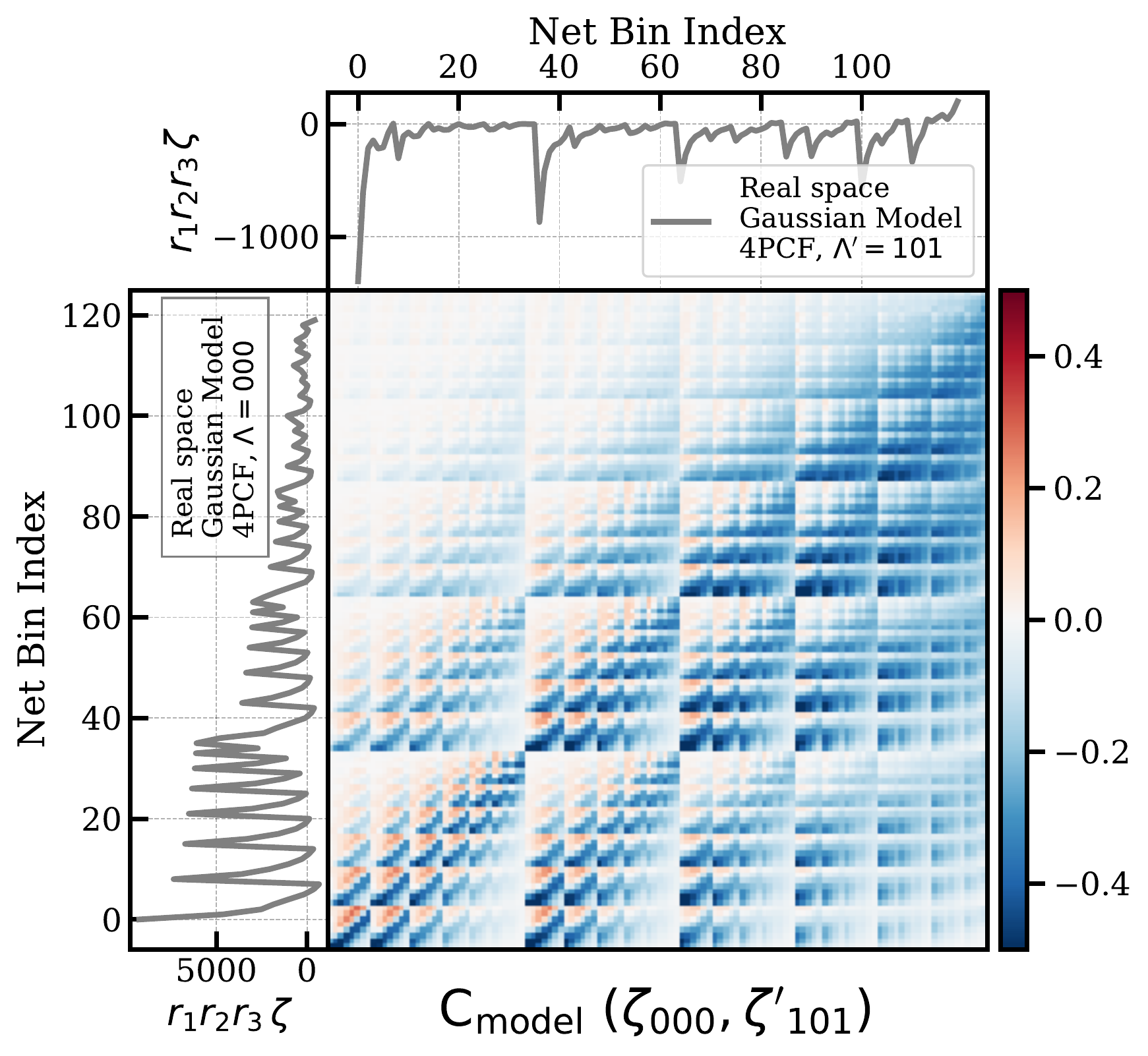}
         \caption{Analytic correlation matrix}
         \label{fig:qjt_halo_hr_real_cov_model_000_101}
     \end{subfigure}
     \hfill
     \begin{subfigure}[b]{0.32\textwidth}
         \centering
         \includegraphics[width=\textwidth]{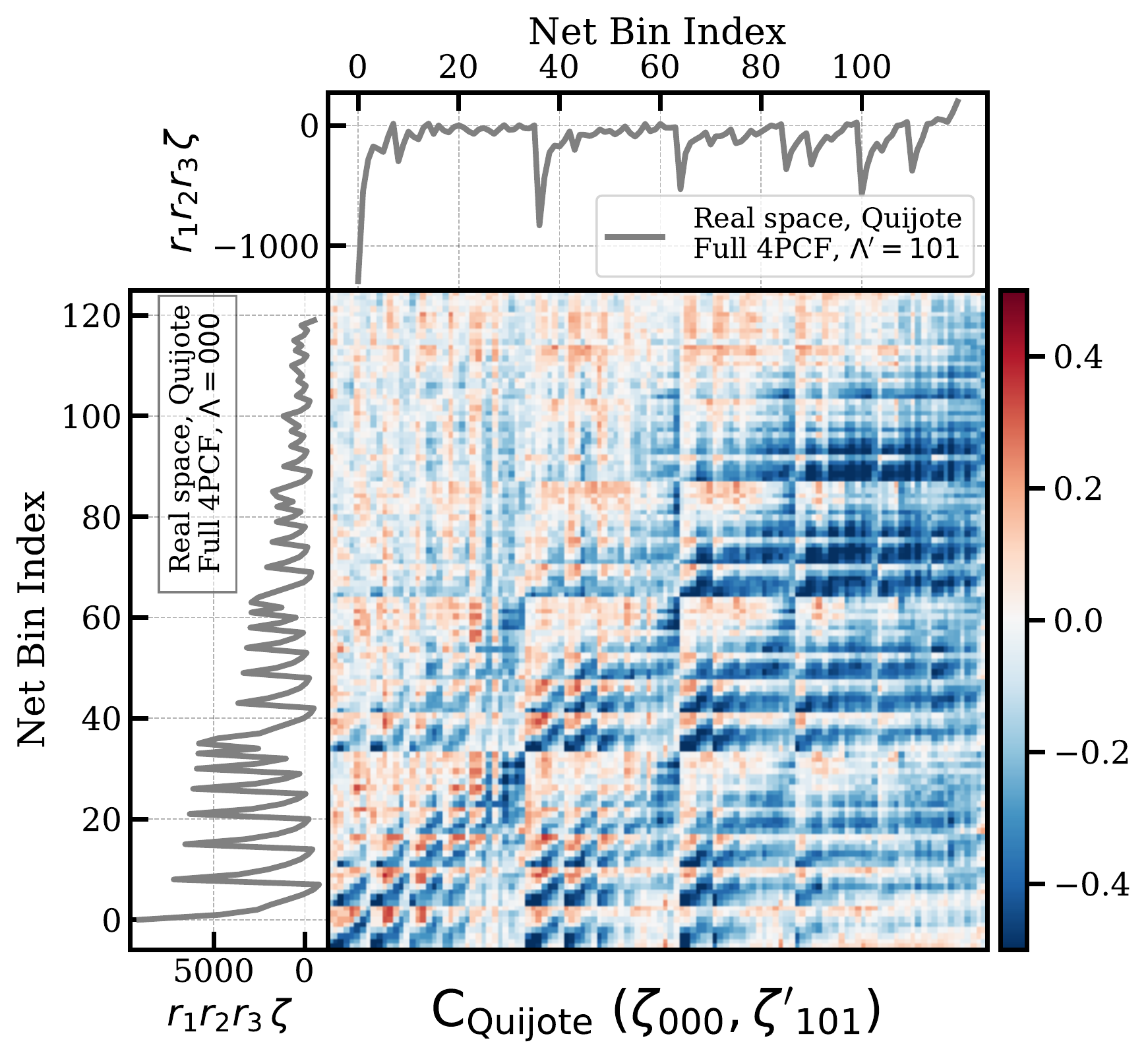}
         \caption{\quijote correlation matrix}
         \label{fig:qjt_halo_hr_real_cov_mock_000_101}
     \end{subfigure}
     \hfill
     \begin{subfigure}[b]{0.28\textwidth}
         \centering
         \includegraphics[width=\textwidth]{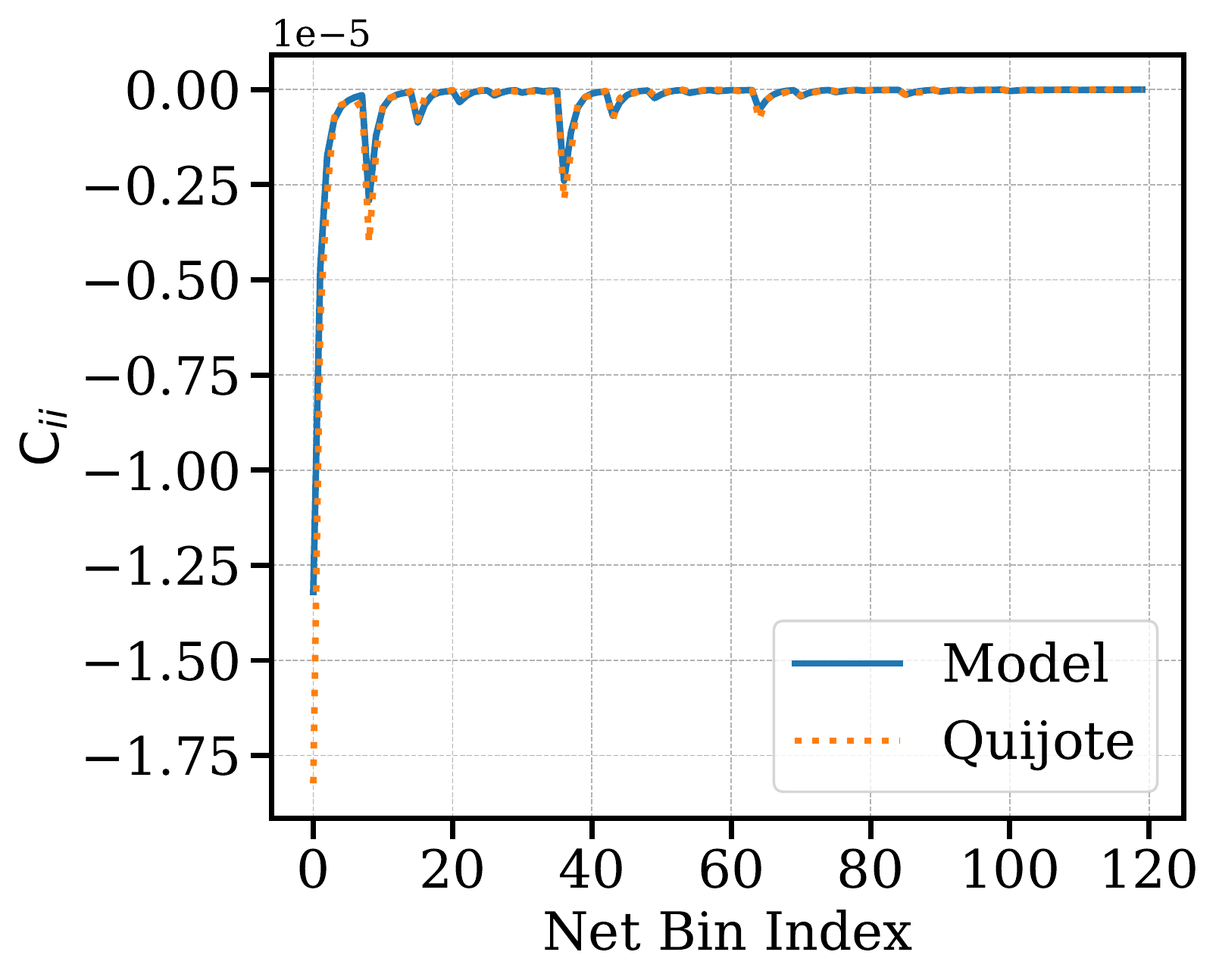}
         \caption{Comparison of diagonals}
         \label{fig:qjt_halo_hr_real_diag_000_101}
     \end{subfigure}
    \caption{As Fig.~\ref{fig:compare1_logn_real_000_000}, but for \quijote halo catalogue in real space. Here, we plot the cross-covariance with angular momenta $\{\Lambda, \Lambda'\}= \{000, 101\}$.
%    Left panel: theoretical prediction for the correlation matrix. Middle: correlation matrix estimated from the $100$ \quijote halo catalogue. Right panel: diagonal elements of the model covariance (solid blue) and sample covariance (dotted orange).
    }
    \label{fig:compare1_qjt_halo_hr_real_000_101}
\end{figure}
Fig.~\ref{fig:compare1_qjt_halo_hr_real_000_000} shows a comparison for $\{\Lambda, \Lambda'\}= \{000, 000\}$ in real space. Again, we see a positive matrix, but this time with an enhanced off-diagonal feature, due to the lower shot noise (approximately less by a factor of two than that of the \logn mocks).
Fig.~\ref{fig:compare1_qjt_halo_hr_real_000_101} gives a comparison for the cross order $\{\Lambda, \Lambda'\}= \{000, 101\}$ in real space. Again, the analytic correlation matrix is able to capture the features in the off-diagonal elements seen in the mocks.
The overall negative structure in the correlation matrix is due to the anti-correlation between the 4PCF $\zeta_{000}$ and $\zeta_{101}$. Since we correlate two different angular distributions we expect the structure of the covariance to be asymmetric. The right panel shows the diagonal elements of the cross covariance for the theoretical model and the \quijote simulation; here, the model covariance slightly underpredicts the covariance diagonal at the small scales seen at the peaks of the saw-tooth shape, but overall the ratio between the sample and mock covariance oscillates around unity with a mean $\av{\mathsf{C}^{\rm mock}_{ii}/\mathsf{C}^{\rm model}_{ii}}\sim 0.96$.

\begin{figure}
     \centering
     \begin{subfigure}[b]{0.45\textwidth}
         \centering
         \includegraphics[width=.7\textwidth]{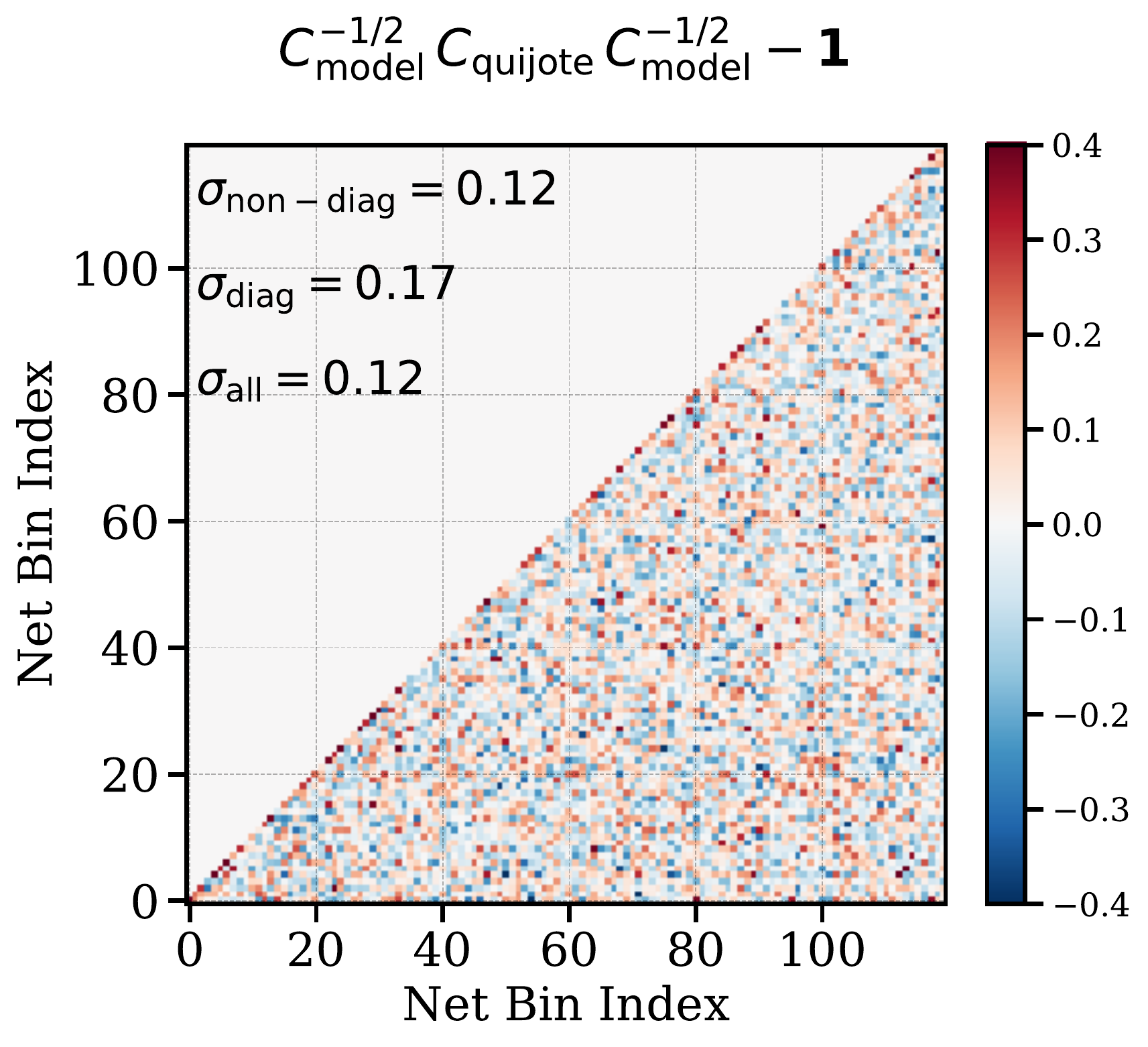}
         \caption{$\{\Lambda, \Lambda'\}= \{000, 000\}$}
         \label{fig:half_inv_qjt_halo_hr_zs_000_000}
     \end{subfigure}
     \begin{subfigure}[b]{0.45\textwidth}
         \centering
         \includegraphics[width=.7\textwidth]{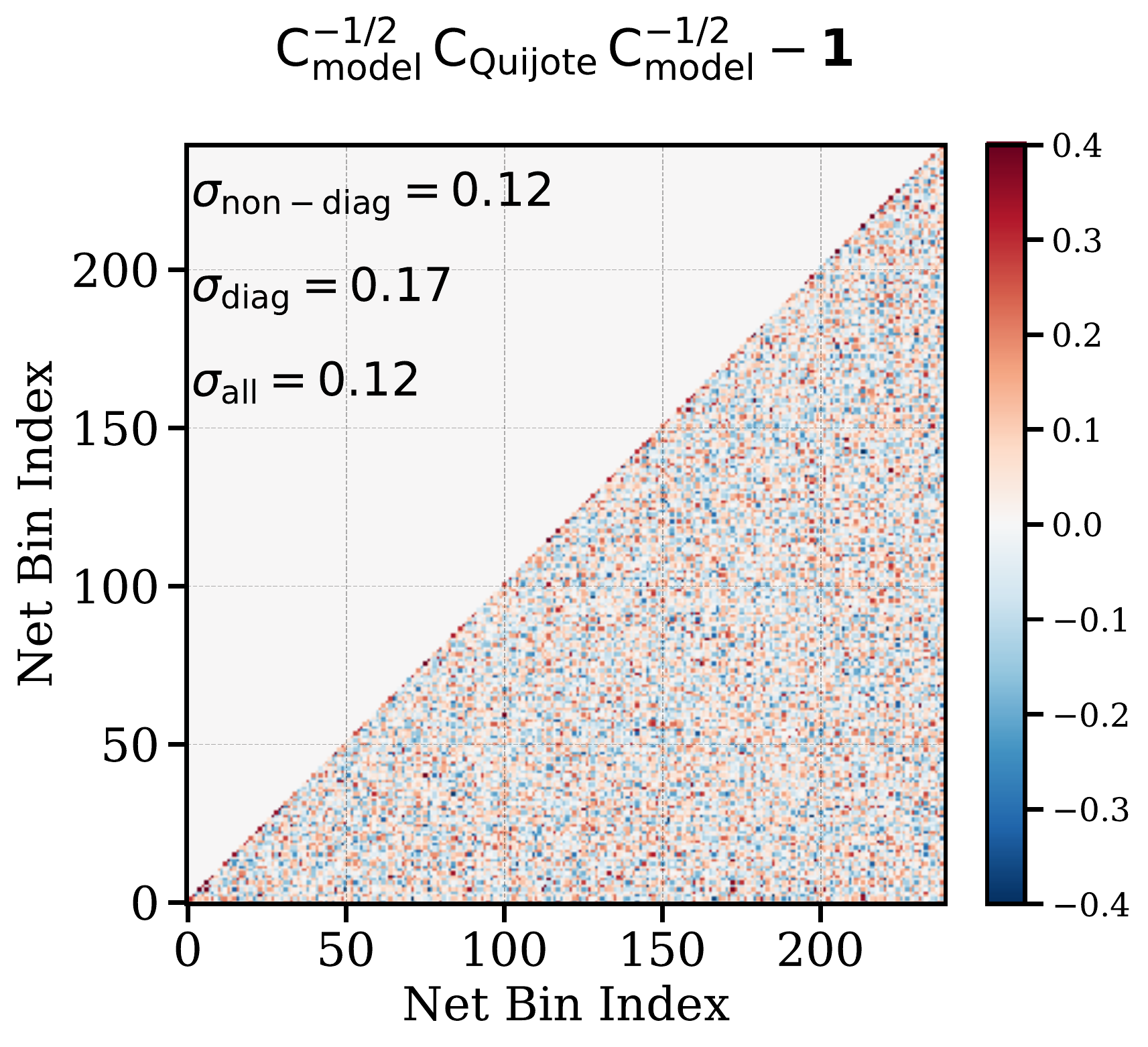}
         \caption{$\{\Lambda, \Lambda'\}= \small{\{000, 000\}+\{101, 000\}+\{000, 101\}+\{101, 101\}}$}
         \label{fig:half_inv_qjt_halo_hr_zs_000_101}
     \end{subfigure}
    \caption{Half-inverse test for the analytic covariance and sample covariance of the \quijote halo catalogue in real space, in the format of Fig.\,\ref{fig:half_inv_logn_real_000_000}. {\bf Left panel:} angular momenta $\{\Lambda, \Lambda'\}= \{000, 000\}$. {\bf Right panel:} cross-covariance with angular momenta $\{\Lambda, \Lambda'\}= \{000, 101\}$. For comparison we show the full matrix with $\{\Lambda, \Lambda'\}= \{000, 000\}+\{000, 101\}+\{101, 000\}+\{101, 101\}$. The standard deviations for respectively the off-diagonal elements, $\sigma_{\rm non-diag}$, the diagonal elements, $\sigma_{\rm diag}$, and all the elements, $\sigma_{\rm all}$, are given in the insets.}
    \label{fig:compare2_qjt_halo_hr_real}
\end{figure}
To quantify the similarity between the model predictions and simulations, we again utilize the half-inverse test. The left panel in Fig.~\ref{fig:compare2_qjt_halo_hr_real} shows the results for $\{\Lambda, \Lambda'\}= \{000, 000\}$, while the right panel shows $\{\Lambda, \Lambda'\}= \{000, 101\}$, both of which are in real space. In order to invert the cross covariance, we build a full matrix which includes the auto-covariance $\{\Lambda, \Lambda'\}= \{000, 000\}$ and $\{\Lambda, \Lambda'\}= \{101, 101\}$, which doubles the size of the matrix. In this case, we do not observe any residuals in the diagonal of the matrix.
Given $100$ halo catalogues, the standard deviation is expected to be of order $1/\sqrt{100} = 0.1$, matching that found from the data. %Also, we confirm from these tests that the standard deviation depends only on the number of mocks not on the length of the data vector (degrees of freedom).

\begin{figure}
     \centering
     \begin{subfigure}[b]{0.26\textwidth}
         \centering
         \includegraphics[width=\textwidth]{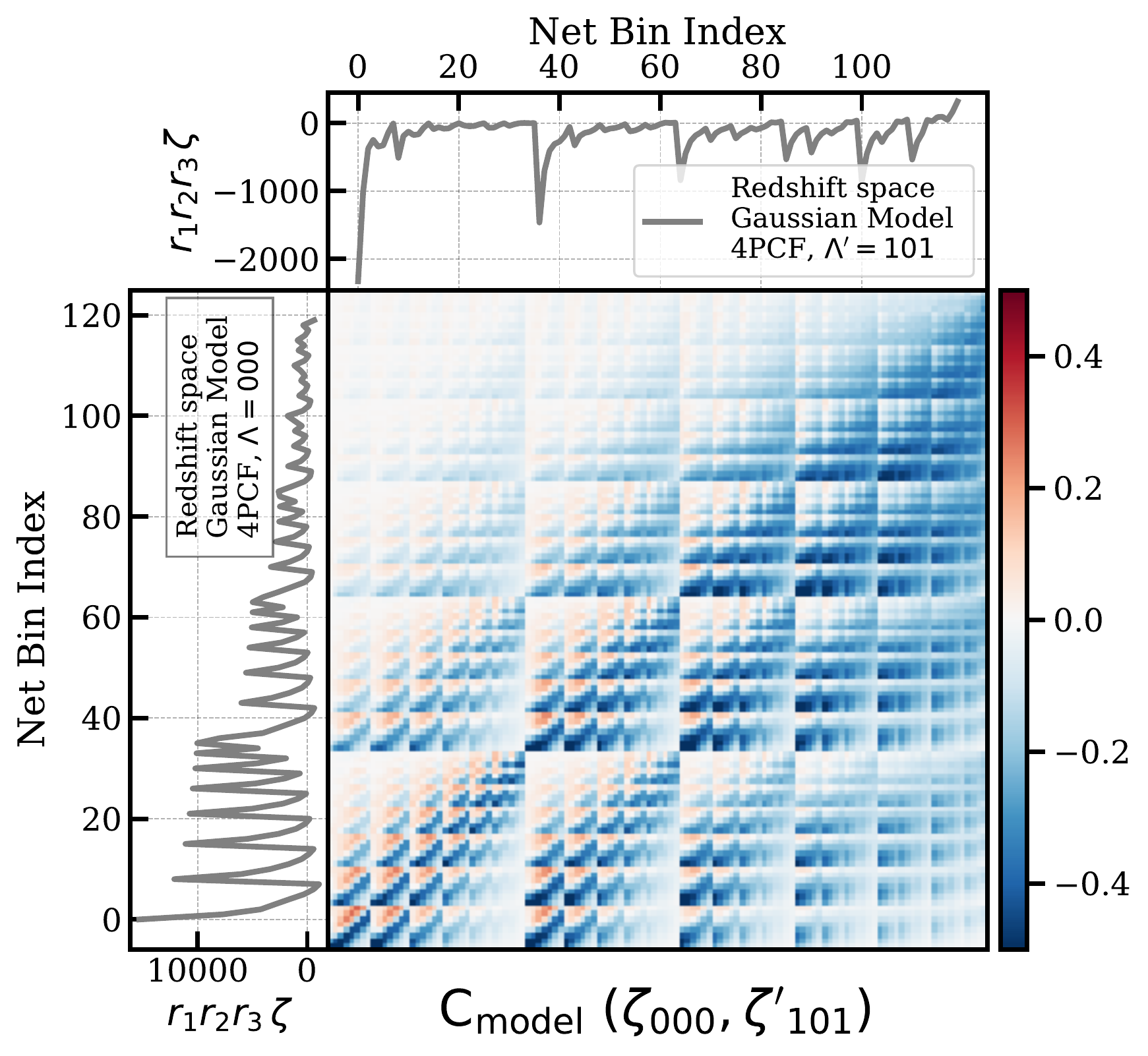}
         \caption{Analytic correlation matrix}
         \label{fig:qjt_halo_hr_zs_cov_model_000_101}
     \end{subfigure}
    %  \hfill
     \begin{subfigure}[b]{0.26\textwidth}
         \centering
         \includegraphics[width=\textwidth]{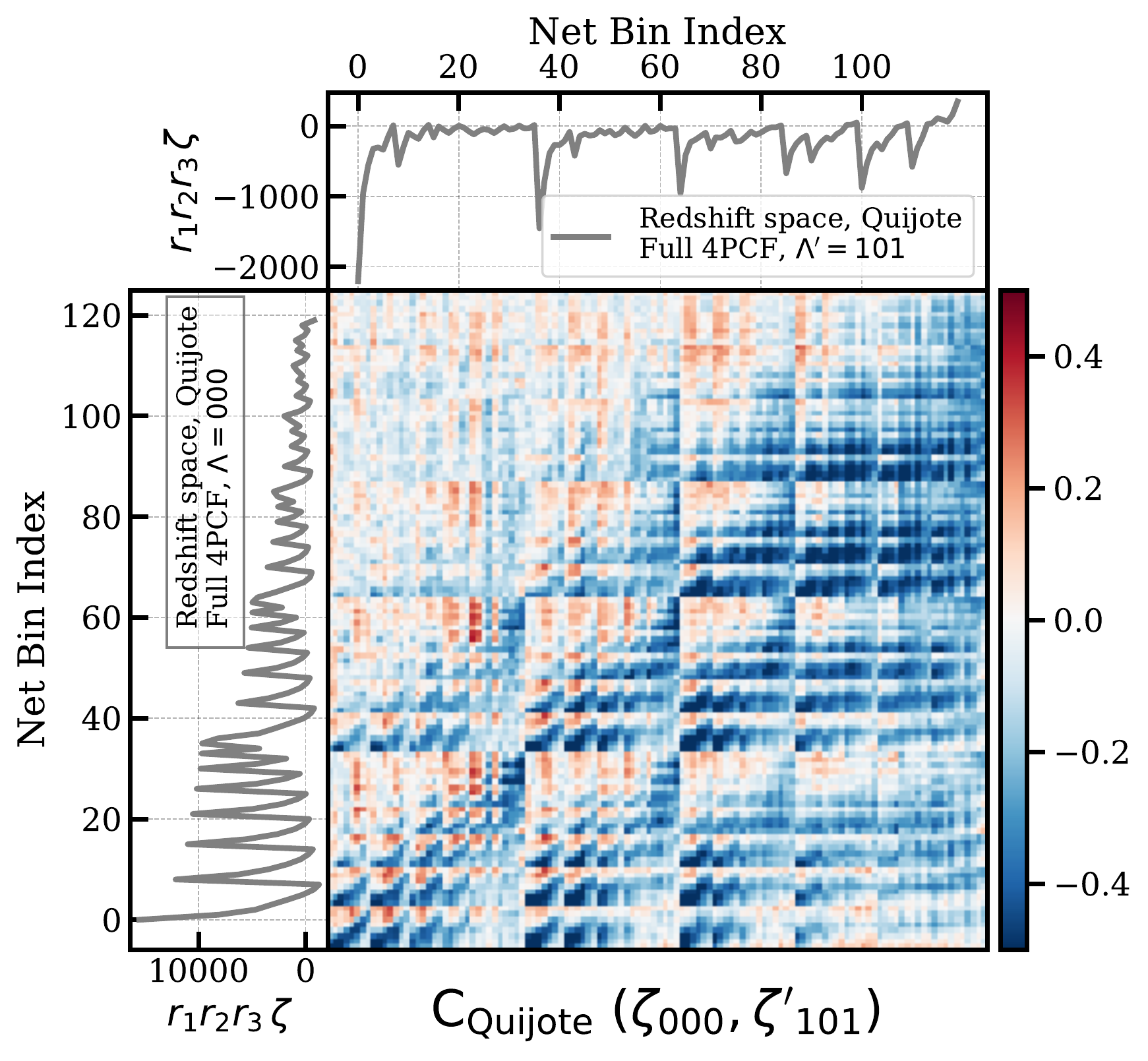}
         \caption{\quijote correlation matrix}
         \label{fig:qjt_halo_hr_zs_cov_mock_000_101}
     \end{subfigure}
    %  \hfill
     \begin{subfigure}[b]{0.23\textwidth}
         \centering
         \includegraphics[width=\textwidth]{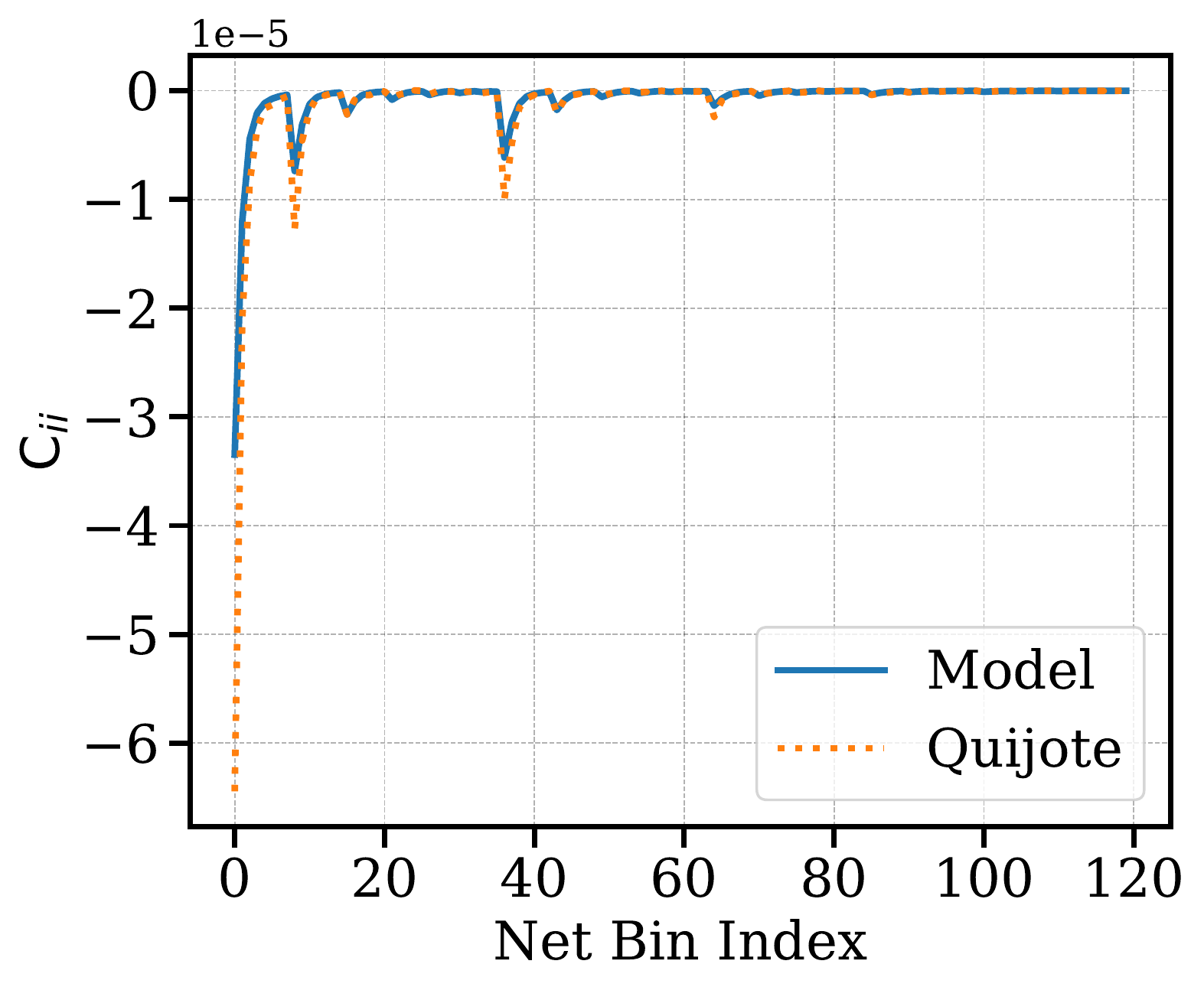}
         \caption{Comparison of diagonals}
         \label{fig:qjt_halo_hr_zs_diag_000_101}
     \end{subfigure}
     \begin{subfigure}[b]{0.23\textwidth}
         \centering
         \includegraphics[width=1\textwidth]{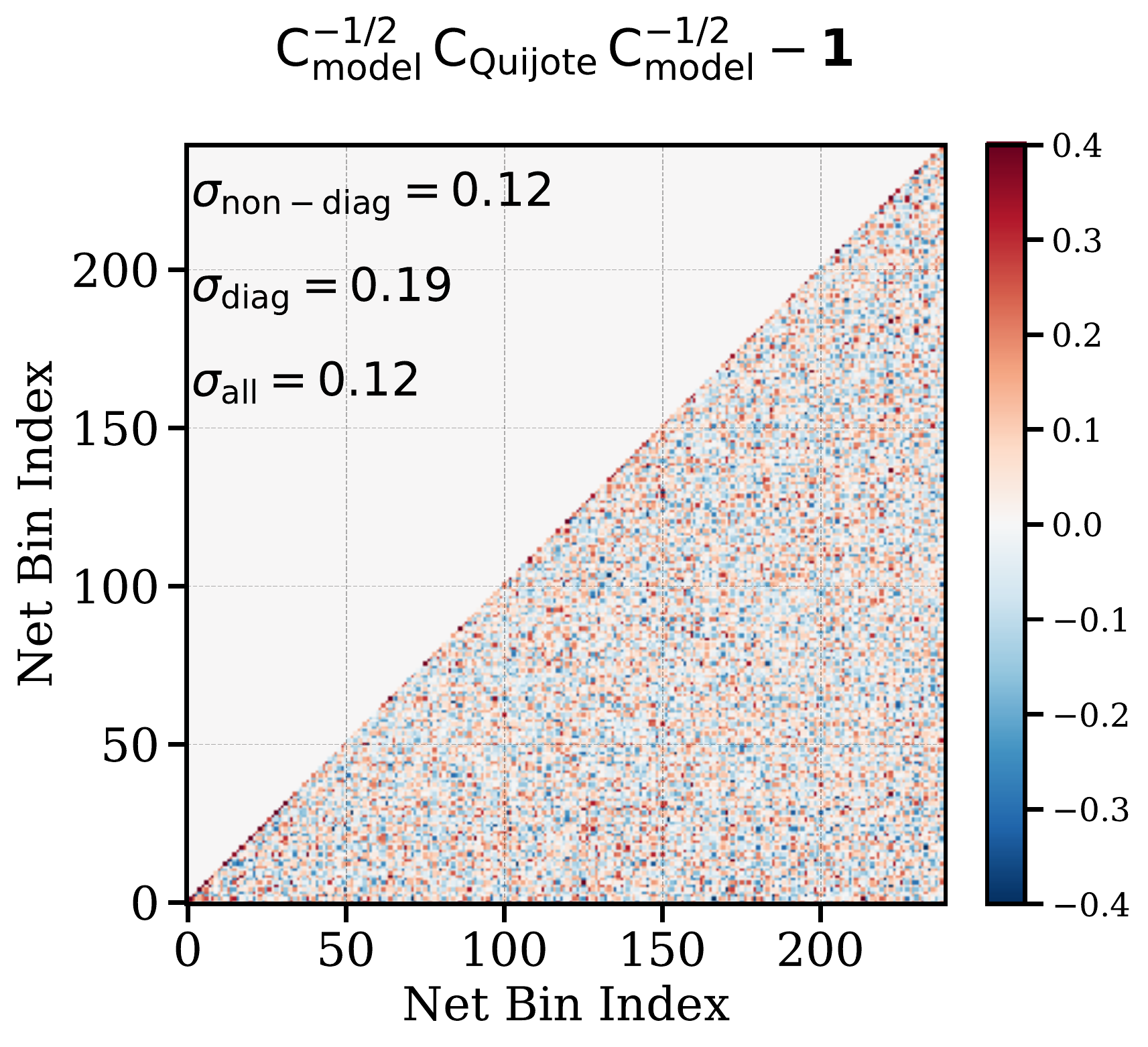}
         \caption{Half-inverse test}
         \label{fig:half_inv_qjt_halo_hr_zs_000_101}
     \end{subfigure}
    \caption{As Fig.~\ref{fig:compare1_logn_zs_000_000}, but for \quijote halo catalogue with angular momenta $\{\Lambda, \Lambda'\}= \{000, 101\}$ including RSD. The analytic covariance well describes the structure of the sample covariance in this scenario.}% in redshift space.}
\label{fig:compare1_qjt_halo_hr_zs_000_101}
\end{figure}

In addition we also perform a comparison for $\{\Lambda, \Lambda'\}= \{000, 101\}$ in redshift space, shown in Fig.~\ref{fig:compare1_qjt_halo_hr_zs_000_101}. Compared to the real space, RSD enhances the diagonals by a factor of $\sim 2.3$ for this cross covariance term, but its overall shape is almost unaffected. 
%\bob{but isn't this a serious matter?  Why is the factor this big? Isn't this the biggest problem for our analytical approach?  the real data have RSD.} \jiamin{Not really, we see a good agreement between the model and mock covariance for this cross term in real and redshift space. The covariance diagonal is enhanced by $\sim 2.3$, but this test is telling us that even without including RSD modeling (using our existing numerical implementation), it is pretty good to account for the RSD effect simply by inputting the power spectrum monopole.}
From the right panel, we see that the diagonal elements of the theoretical covariance slightly under-predict those estimated from \quijote simulation at small scales, but the mean of the ratio is close to unity, with $\av{\mathsf{C}^{\rm mock}_{ii}/\mathsf{C}^{\rm model}_{ii}}\sim 1.04$. This is also demonstrated in panel (d), showing no residual from the half-inverse test. In principle, we could extend our model to include RSD effects as described in Appendix \ref{appendix:cov_fc_rsd_iso}; we leave this effort to future for work.% We leave the extension of the numeric code by including the higher order Legendre multipoles to the future work.

\subsection{Comparison with the \mdpatchy mocks}
\label{subsec:patchy}
Finally, to test the impact of the non-uniform survey geometry, we compare our Gaussian covariance model to a set of \mdpatchy mocks~\citep{Kitaura201603,RodriguezTorres201604} produced for the Sloan Digital Sky Survey (SDSS) Baryon Oscillation Spectroscopic Survey (BOSS) Data Release (DR) 12~\citep{Dawson2013,Alam201507}. In this test, we focus on the set of \patchy mocks that match the galaxy clustering
of the BOSS Constant Stellar Mass (CMASS) Luminous Red Galaxy (LRG) sample at an effective redshift $z_{\rm eff}=0.57$ in the North Galactic Cap (NGC). The mock catalogues were constructed using the {\it Planck} cosmology $\left\{\Omega_{\rm m},\, \Omega_{\rm b},\, h, \,n_{\rm s},\, \sigma_8 \right\} = \left\{0.307115,\, 0.048206, \,0.6777,\, 0.9611, \, 0.8288 \right\}$. 

For simulations in a cubic box, the volume, $V$, entering the theoretical covariance is simply given by the box size, and the number density is the ratio between the number of particles (galaxies or halos) and the volume. For a sample with survey geometry and a radial selection function, we generalize the volume and number density estimator of~\citet{Wadekar202012, Putter201204}:
\begeqar
V_{\rm eff} = \frac{\big[\int d^3 r\, n^4(\bfr) {w}^4(\bfr)\big]^2}{\int d^3 r\, n^8(\bfr){w}^8(\bfr)},\qquad \bar{n}_{\rm eff} = \frac{\int d^3 r\, n^8(\bfr) {w}^8(\bfr)}{\int d^3 r\, n^7(\bfr){w}^8(\bfr)},
\endeqar
where $n(\bfr)$ is the number density of the sample as a function of redshift and ${w}(\bfr)$ is the galaxy weight (including both systematic and FKP weights~\citep{Feldman1994}). 
To calculate this, we apply the default weights provided in the \patchy mocks. These are given by\footnote{\url{http://www.skiesanduniverses.org/page/page-3/page-15/page-9/}}
\begeqar
{w}_{\rm tot} = {w}_{\rm fkp}\cdot {w}_{\rm veto}\cdot{w}_{\rm fiber\, collision},
\endeqar
where the FKP weight is ${w}_{\rm fkp}=\left(1+10^4 (h^{-1}\,\mathrm{Mpc})^{3}\cdot n(\bfr)\right)^{-1}$, ${w}_{\rm veto}$ is a binary indicating whether the object is excluded by veto mask or not, and ${w}_{\rm fiber\, collision}$ is a fiber collision weight.
For \patchy NGC, we obtain $\bar{n}=3.2\times 10^{-4}\, (h^{-1}\,\mathrm{Mpc})^{-3}$ and $V^{-1}_{\rm eff}=1.9\, (h^{-1}\,\mathrm{Gpc})^3$.
We caution however that this is only an approximation and does not fully account for the survey geometry, even for the 2PCF covariance~\citep{Wadekar202012}. 

The input power spectrum is measured from the \patchy mocks then fitted using the Effective Field Theory of Large Scale Structure~\citep[EFT;][]{Carrasco201209,Baumann201207} including one-loop bias, RSD, counterterms and infrared resummation~\citep{Senatore201409,Senatore201511,Ivanov202005}, implemented using the \textsc{class-pt} code~\citep{Chudaykin202009}.
The 4PCF is measured from $999$ \patchy mocks with random catalogues $50\times$ larger than data, and the same radial binning scheme as before. As above, we apply an exponential damping to the power spectrum and shot noise. 

\begin{figure}
    \centering
    \includegraphics[width=.4\textwidth]{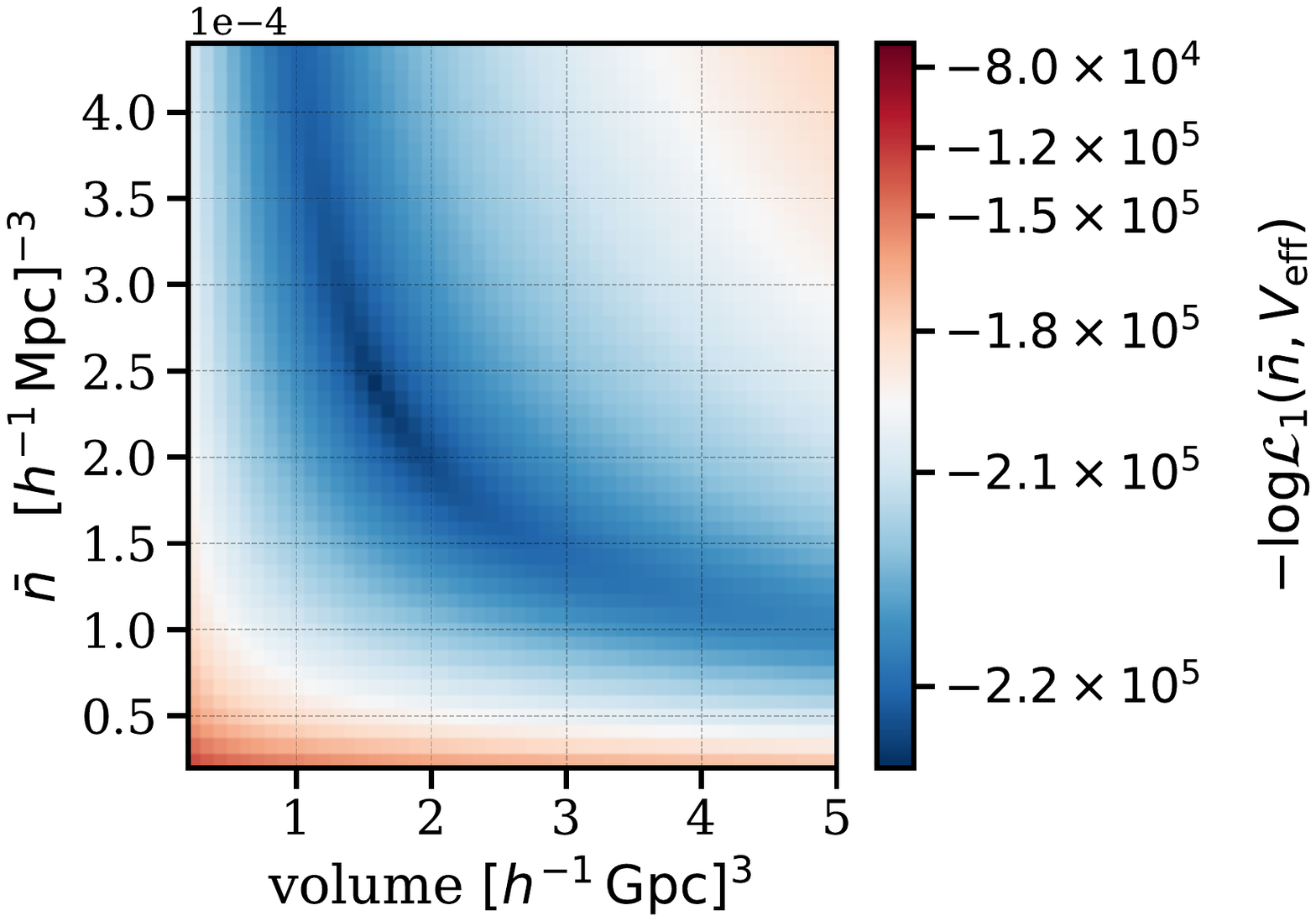}
    \caption{Log-likelihood for the parameters $\bar n$ and $V_{\rm eff}$ obtained from fitting the analytic covariance to the sample covariance of 1,000 \patchy mocks (including redshift-space effects and non-uniform survey geometry). The likelihood is constructed using the KL-divergence, as in Eq.\,\ref{eqn:mlogL1}.}
    \label{fig:mlogL1_2D}
\end{figure}

In order to mitigate the the window function effect, we further consider an optimization of the effective number density and survey volume. Our motivation is that increasing the number density is an approximate way to incorporate non-Gaussianity and {effective volume is a leading order correction of the survey geometry}. To compute this, we create a 2D grid of parameters, scanning over both the number density and the effective volume. We maximize a log-likelihood based on the Kullback-Leibler (KL) divergence using the expected Wishart distribution for mock covariances~\citep{Kullback1951} following~\citet{OConnell201607,Philcox201911}. This has the advantage that it only requires the analytic covariance to be inverted. The log-likelihood involves both the Gaussian covariance and the sample covariance measured from \patchy mocks:
\begeqar\label{eqn:mlogL1}
-\log \mathcal{L}_{1}(\bar{n}, V_{\rm eff})=\frac{N_{\text {mock }}}{2}\left[\operatorname{Tr}\left(\mathsf{C}^{-1}_{\mathrm{model}}(\bar{n}, V_{\rm eff}) \mathsf{C}_{\text {mocks }}\right)-\log \operatorname{det} \mathsf{C}^{-1}_{\mathrm{model}}(\bar{n}, V_{\rm eff})\right]+\cdots
\endeqar
%with the model precision matrix $\Psi_{\rm model} = \mathsf{C}_{\rm model}^{-1}$. 
As a test, we optimized the likelihood for the \patchy NGC region using $\{\Lambda,\Lambda'\}=\{000,000\}$. The 2D-grid was constructed using $\bar{n}\in [0.2, 4.4]\times 10^{-4}\, (h^{-1}\,\mathrm{Mpc})^{-3}$ with an interval of $2\times 10^{-5} \,(h^{-1}\,\mathrm{Mpc})^{-3}$ and $V^{-1}_{\rm eff}\in [0.2, 5]\, (h^{-1}\,\mathrm{Gpc})^3$ in 40 volume bins. Fig.~\ref{fig:mlogL1_2D} shows a 2D interpolation of the log-likelihood. The degeneracy direction shows an inverse scaling relation between the number density and volume; this is as expected since lowering the number density increases the shot noise, which increases the overall amplitude of the covariance, but can be suppressed by a higher volume. For the \patchy NGC region,\footnote{We also performed the same fitting procedure for the SGC, obtaining $\bar{n}=2.5\times 10^{-4}\, (h^{-1}\,\mathrm{Mpc})^{-3}$ and $V^{-1}_{\rm eff}=0.49\, (h^{-1}\,\mathrm{Gpc})^3$.} the optimized number density and volume are respectively given by $\bar{n}=2.4\times 10^{-4}\, (h^{-1}\,\mathrm{Mpc})^{-3}$ and $V^{-1}_{\rm eff}=1.57\, (h^{-1}\,\mathrm{Gpc})^3$.
%\footnote{We also performed the same fitting procedure for SGC and obtain $\bar{n}=2.1\times 10^{-4}\, (h^{-1}\,\mathrm{Mpc})^{-3}$ and $V^{-1}_{\rm eff}=0.49\, (h^{-1}\,\mathrm{Gpc})^3$.}

\begin{figure}
     \centering
     \begin{subfigure}[b]{0.32\textwidth}
         \centering
         \includegraphics[width=\textwidth]{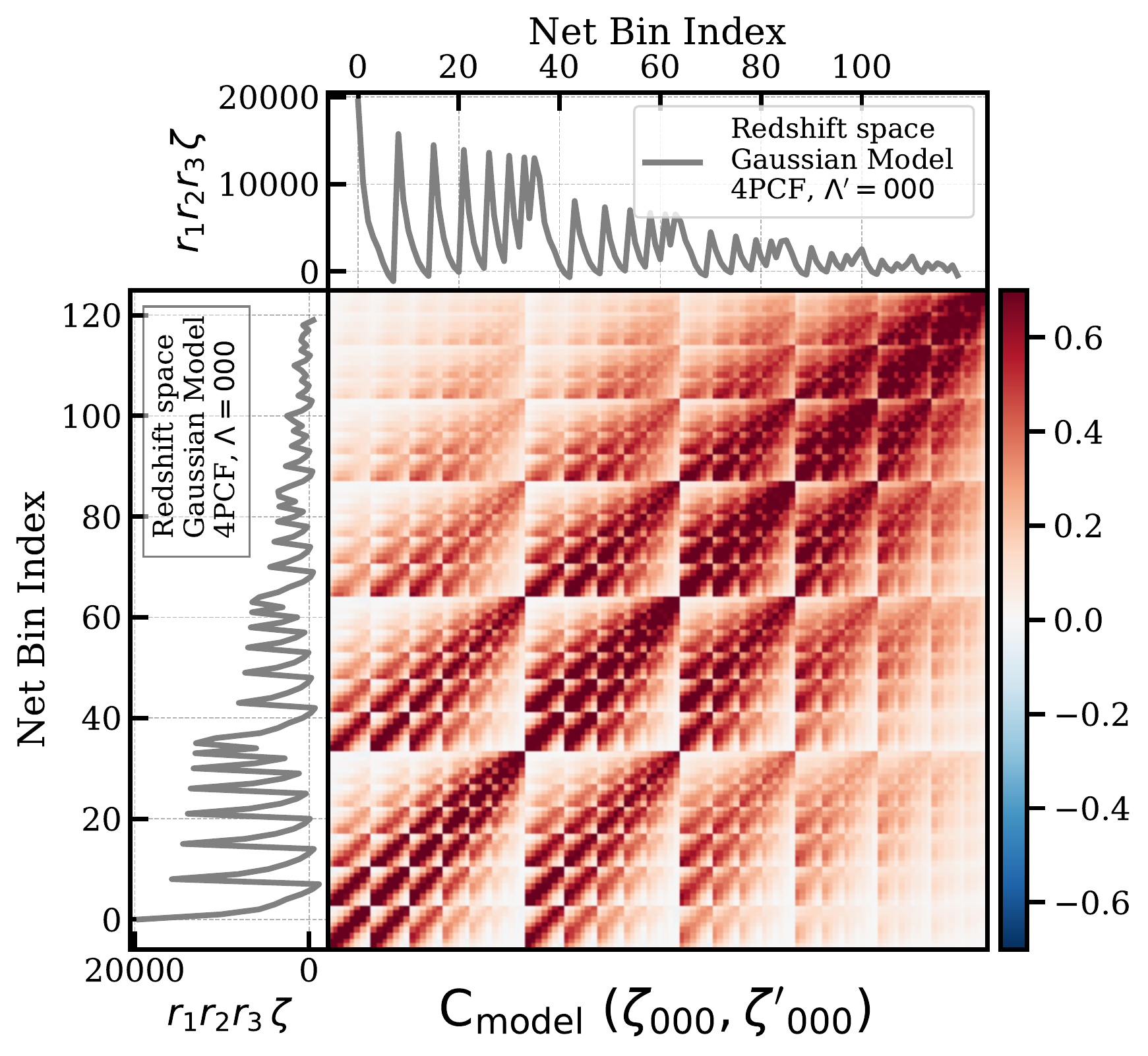}
         \caption{Analytic correlation matrix}
         \label{fig:patchy_ngc_cov_model_000_000}
     \end{subfigure}
     \hfill
     \begin{subfigure}[b]{0.32\textwidth}
         \centering
         \includegraphics[width=\textwidth]{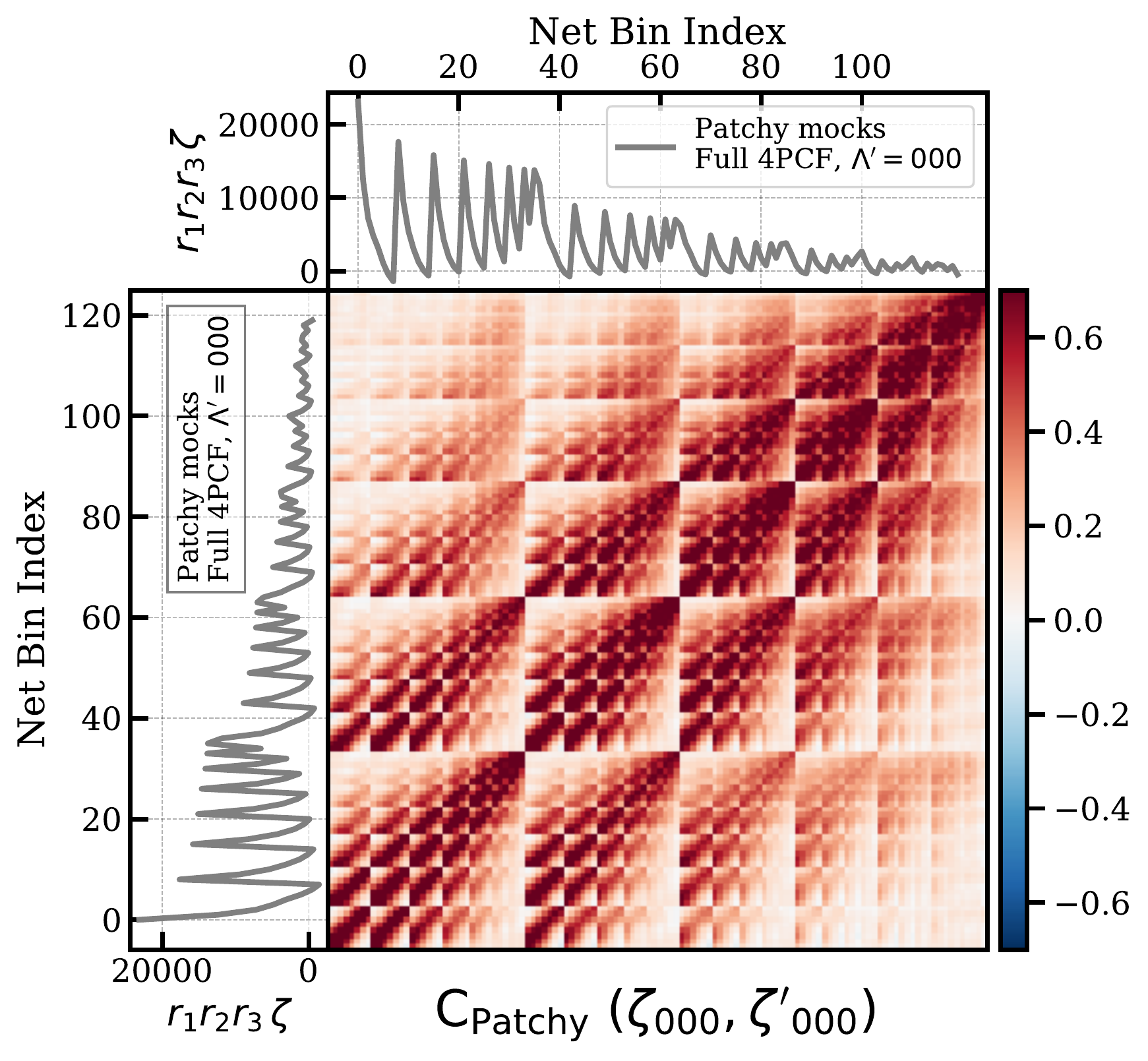}
         \caption{\patchy correlation matrix}
         \label{fig:patchy_ngc_cov_mock_000_000}
     \end{subfigure}
     \hfill
     \begin{subfigure}[b]{0.28\textwidth}
         \centering
         \includegraphics[width=\textwidth]{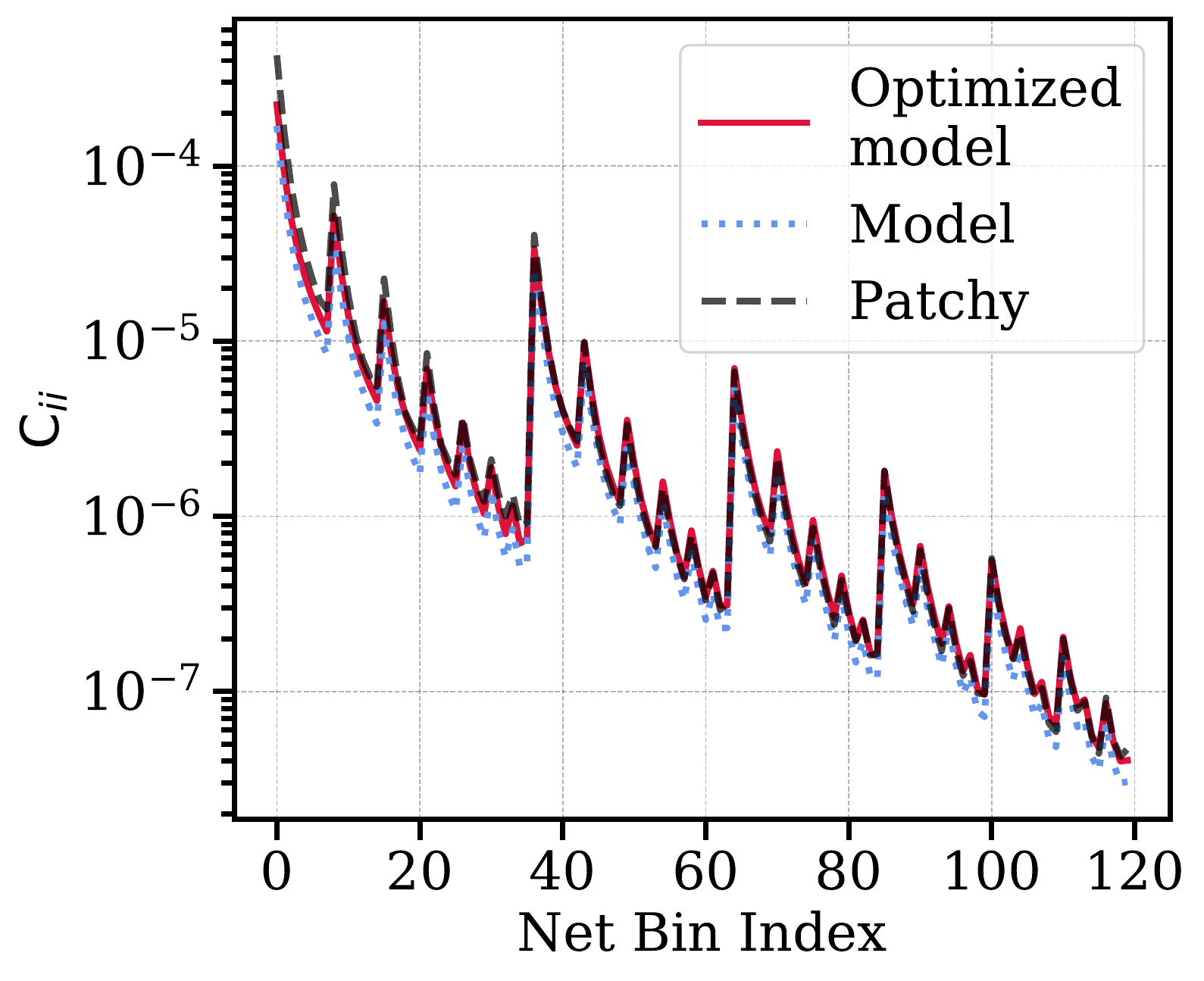}
         \caption{Comparison of diagonals}
         \label{fig:patchy_ngc_diag_000_000}
     \end{subfigure}
    \caption{As Fig.\,\ref{fig:compare1_logn_real_000_000} but for $999$ \patchy mocks. These include both RSD and survey geometry. %Comparison of correlation matrix for angular momenta $\{\Lambda, \Lambda'\}= \{000, 000\}$. Left panel shows the model prediction for the fully-coupled 4PCF correlation matrix, the insets show the Gaussian 4PCF model in redshift space. 
    %The middle panel shows the correlation matrix estimated from $999$ \patchy mocks, the inset shows the measurements of the full 4PCF from the \patchy mocks. 
    The third panel shows a comparison of the diagonal elements for the \patchy covariance (grey dashed curve), analytic covariance with and without optimization (red solid curve and black dotted curve, respectively).}
\label{fig:compare1_patchy_000_000}
\end{figure}

\begin{figure}
     \centering
     \begin{subfigure}[b]{0.45\textwidth}
         \centering
         \includegraphics[width=.7\textwidth]{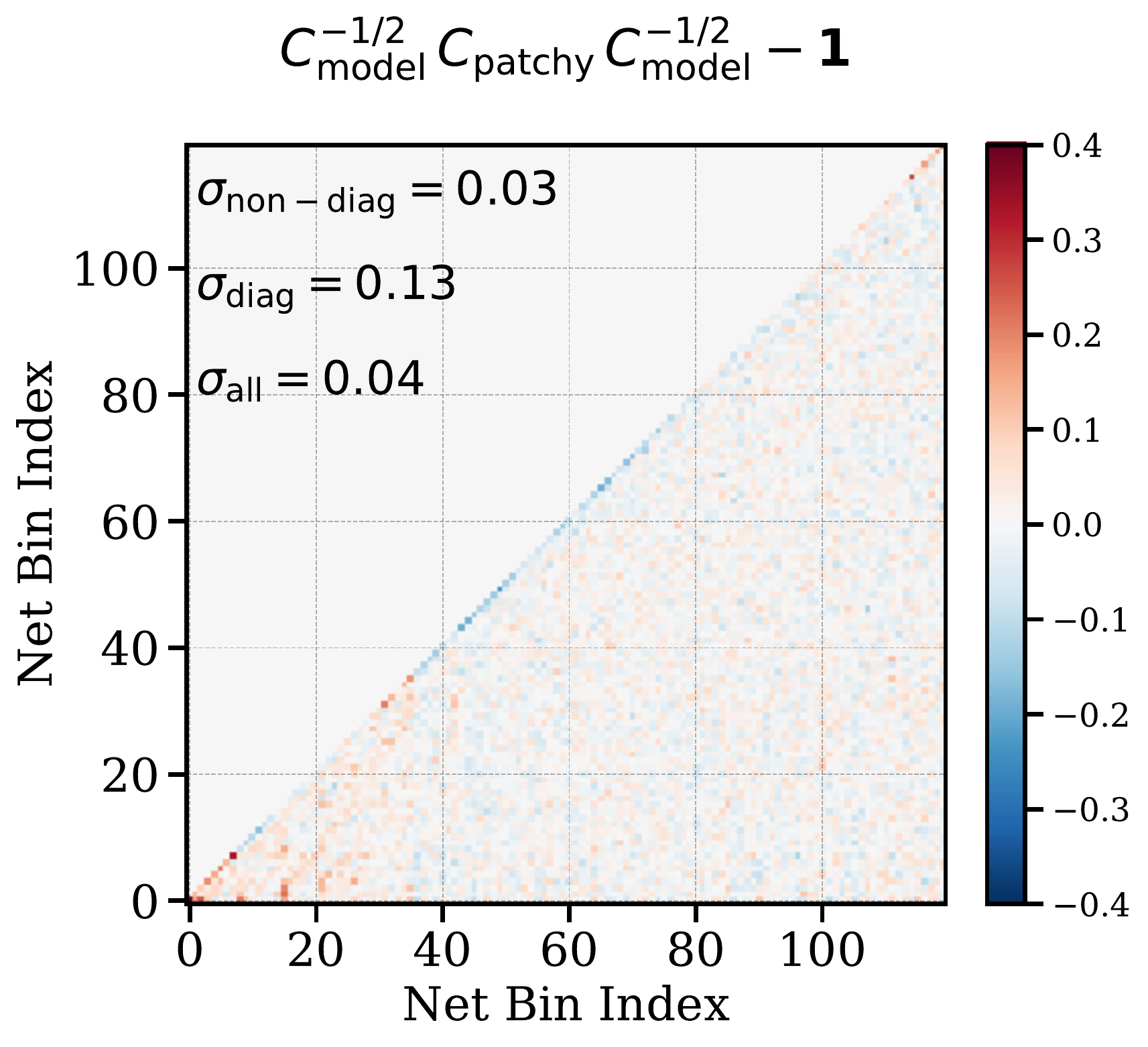}
         \caption{Half-inverse test $\{\Lambda, \Lambda'\}= \{000, 000\}$}
         \label{fig:half_inv_patchy_ngc_nbar2.0E-4_000_000}
     \end{subfigure}
    %  \hfill
     \begin{subfigure}[b]{0.45\textwidth}
         \centering
         \includegraphics[width=.7\textwidth]{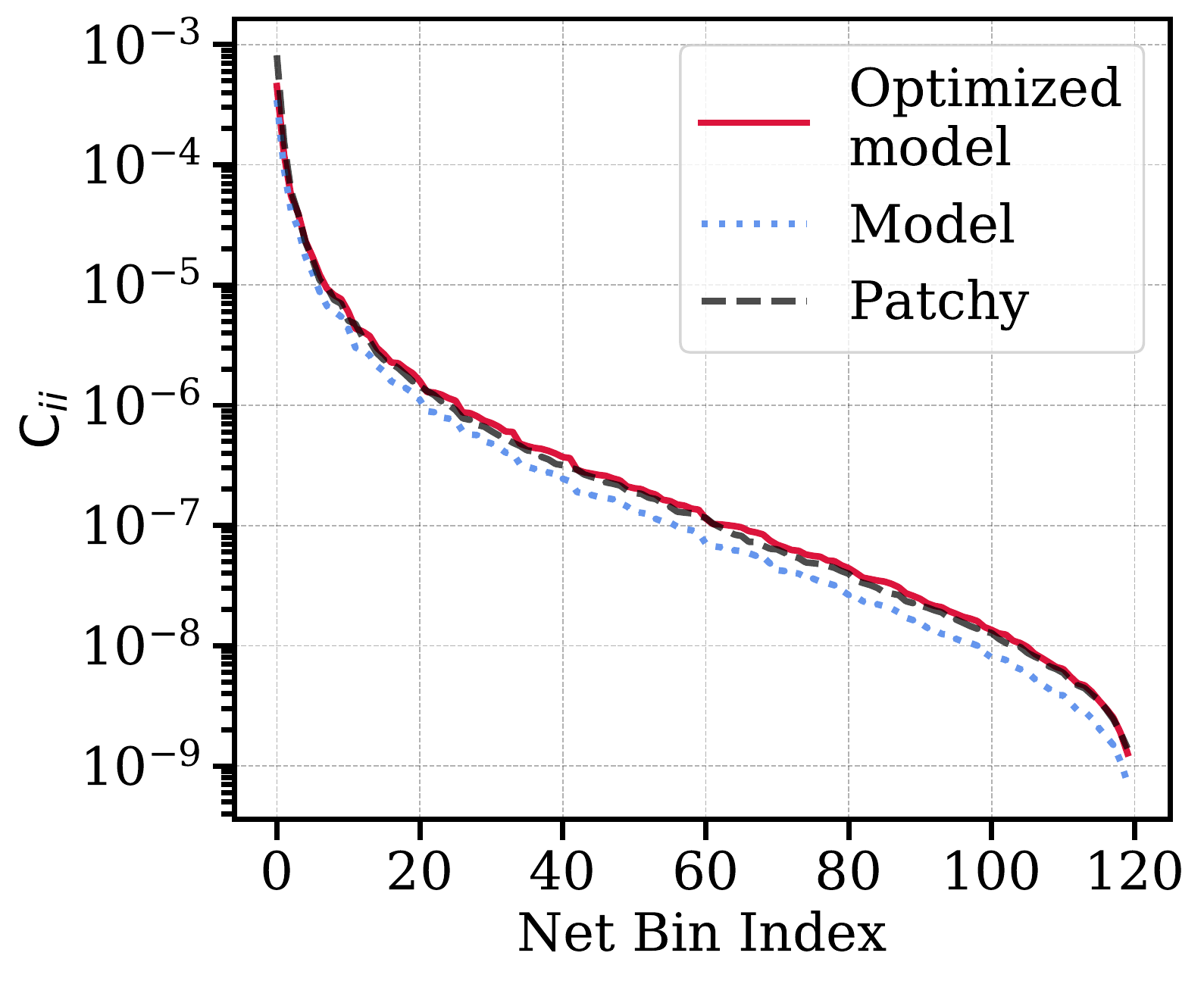}
         \caption{Half-inverse test $\{\Lambda, \Lambda'\}= \{000, 000\}$}
         \label{fig:eval_patchy_ngc_nbar2.0E-4_000_000}
     \end{subfigure}
    \caption{{\bf Left panel:} half-inverse test for the model applied to the \patchy NGC mocks for angular momenta $\{\Lambda, \Lambda'\}= \{000, 000\}$, as in Fig.\,\ref{fig:half_inv_logn_real_000_000}. {\bf Right panel:} comparison of the eigenvalues for the theoretical covariance before optimization (dotted black curve), after optimization (solid red curve), and from the \patchy mocks (dashed grey curve)}.
    \label{fig:compare2_patchy_000_000}
\end{figure}

The comparison of the correlation matrix for $\{\Lambda, \Lambda'\}= \{000, 000\}$ is shown in Fig.~\ref{fig:compare1_patchy_000_000}. The left and middle panels show the optimized correlation matrix from the model prediction and the covariance obtained from the \patchy NGC mocks, respectively. The right panel shows a comparison for the diagonal elements of the analytic covariance model with and without optimization (solid red curve and dotted black curve, respectively), and the \patchy mocks (dashed grey curve). Fig.~\ref{fig:compare2_patchy_000_000} shows the half-inverse test in the left panel, with the right panel giving the covariance matrix eigenvalues predicted by the analytic model before optimization (dotted black curve), after optimization (solid red curve), and estimated from the \patchy mocks (grey curve).
Before applying the optimization, there is a clear mismatch between the theoretical prediction and the mock measurement, both in terms of its diagonal elements and the eigenvalues. The mean of the half-inverse matrix gives $\langle \mathsf{S}\rangle=-9\times 10^{-4}$, while the mean of the diagonal is $-0.012$.
The tests using the \quijote mocks indicate no obvious deviations from RSD not nonlinearity, thus we expect the offset to arise due to the survey geometry. Fitting for the number density and effective volume, we find that one can moderately compensate for this effect.

\begin{figure}
     \centering
     \begin{subfigure}[b]{1\textwidth}
         \centering
         \includegraphics[width=.7\textwidth]{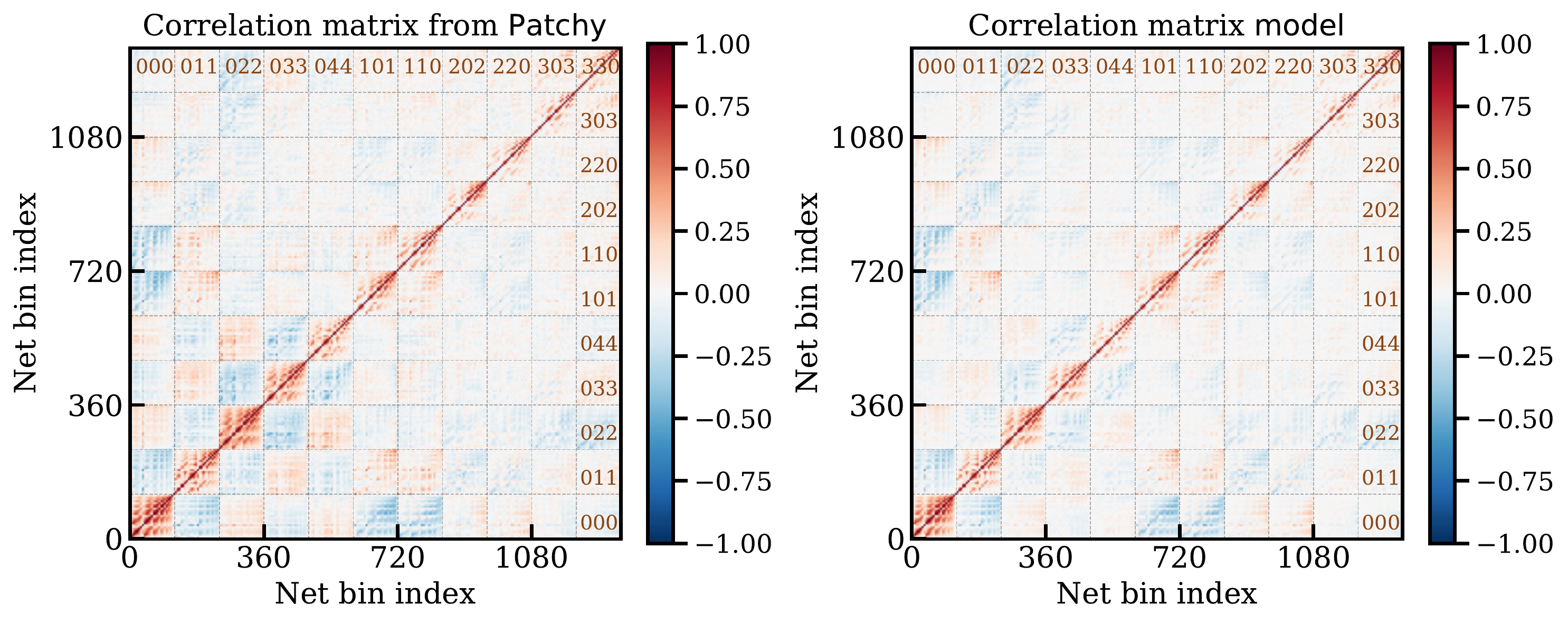}
     \end{subfigure}
    \caption{Comparison of correlation matrices estimated from \patchy NGC mocks {\bf (left)} and model {\bf (right)}. Unlike previous plots, we include 11 different choices of $\Lambda$, with each submatrix being the correlation between angular momentum sets $\{\Lambda, \Lambda'\} = \{\ell_1\, \ell_2\, \ell_3, \ell'_1\, \ell'_2\, \ell'_3\}$. The shot noise and volume entering the analytic covariance are optimized using 13 choices of $\Lambda$ (those involving $\ell_i = 0$, up to $\ell_{\rm max}=4$). 
    Overall, we find reasonably good agreement between the Gaussian model and the sample covariance. We see some differences in the off-diagonal terms, and these differences increase with rising angular momenta. The diagonal terms are relatively consistent with each other, mostly as a result of the parameter fitting.}% predicted diagonal terms are in general close to the mock covariance.}
    \label{fig:corr_matrix_comparison_patchy_ngc_nells11}
\end{figure}
To this end, we also perform a parameter fit including a total of 13 auto-covariance terms in which $\Lambda=\Lambda'$ (using those values of $\Lambda$ which include at least one zero). We find the optimized number density and volume $\bar{n}=2.0\times 10^{-4}\, (h^{-1}\,\mathrm{Mpc})^{-3}$ and $V^{-1}_{\rm eff}=1.57\, (h^{-1}\,\mathrm{Gpc})^3$. Fig.~\ref{fig:corr_matrix_comparison_patchy_ngc_nells11} shows a comparison of the correlation matrices estimated from the \patchy NGC mocks and model prediction; for visibility we show 11 terms. Despite an overall good agreement between the mock correlation matrix and the model one, we find that different angular momentum orders are affected by the non-Gaussianity and survey geometry in different ways. As such, the number density and effective volume optimized for a specific angular momentum combination is not necessarily the optimal combination for the others. This indicates a fundamental limitation of the fitting approximation.% and in this case a new fitting is recommend in this approximated approach.

\section{Summary}
\label{sec:summary}

Summary statistics, such as the $N$-point correlation functions, can effectively capture cosmological information from the spatial distribution of LSS. Throughout the past decades, significant work has been devoted to developing pipelines for the analysis of two-point statistics, focused primarily on the extraction of the BAO position and the growth parameter, $f\sigma_8$. %have been developed using the two-point statistics to perform BAO and RSD analysis in the past. 
The next generation of surveys, {\textit e.g.} the Dark Energy Spectroscopic Instrument~\citep{DesiCollaboration2016}, the Euclid satellite~\citep{Laureijs2011,Amendola2018}, and the Rubin Observatory~\citep{collaboration2009lsst} will map out much larger survey volumes with increased statistical power, facilitating analysis beyond the two-point function.

Higher-order statistics allow us to gain new insight into gravity-induced nonlinearities and neutrino masses, particularly in combination with two-point statistics. Further, they can be used to study scalar parity violation, which cannot be probed at all for NPCFs with $N\leq 3$. A particular challenge is that higher-order statistics usually imply high dimensionality; if one purses a simulation-based covariance estimation, a large number of mocks are required, which is computationally demanding.

In this paper we discuss an analytic approach to computing the NPCF covariance. In particular, we decompose the NPCF into the isotropic basis functions described in~\citet{Cahn202010}, and compute the covariance in this basis. Assuming the density field to be statistically isotropic (\textit{i.e.} ignoring RSD), this is a natural basis to use, since it has full 3D rotational symmetry%. the 3-dimensional coordinates and is an optimal basis for studying cosmology by assuming homogeneity of the Universe in the absence of the RSD on the very large scales \jiamin{still need to discuss `isotropy'}.

When constructing higher-order NPCFs, it is important to subtract any contributions which also appear in the lower-order statistics, \textit{i.e.} to use only the connected NPCF. As we have shown, the full NPCF covariance matrix can be written as a sum of two pieces, denoted as fully-coupled and partially-coupled, with only the former contributing to the covariances of connected NPCFs.
%Since most of the information in the higher-order statistics is replicated by the lower-order statistics, they lead to partially-coupled covariance in the presence of self-contracting pairs. Subtracting those self-contracting pairs (disconnected terms) allows us to concentrate on the connected NPCF estimators, which are most relevant in understanding the non-Gaussian signals.
%Therefore in this work we focus on the fully-coupled covariance by considering the connected terms in the NPCF estimator. 
We present a general formalism for the NPCF covariance under the assumption of Gaussianity, which we can further break down into basic elements as contractions between two overdensity fields. Each basic element consists of an $f$-integral (Eq.\,\ref{eqn:f3l-tensor}) with coefficients involving products of angular momenta and 3-$j$ symbols multiplied by a phase. We show that the general NPCF covariance can be built directly out of these basic elements by invoking properties of the isotropic basis functions. In the $N=4$ case, we explicitly derive the analytic form for the 4PCF covariance, introducing a diagrammatic representation to assist with understanding of the coupling structure. We also numerically implement the analytic formalism for this case.
%. Since the primary vertices $\bfr_0$ and $\bfr'_0$ are always associated with zero angular momenta. To better illustrate the derivation, we distinguish the contraction of the overdensity fields between primary vertices or primary vertices with endpoints. To this end, we also introduce diagrammatic representation to assist understanding of the coupling structure. We also numerically implemented the analytic formalism for the $N=4$ case.

We compare our theoretical model, which assumes Gaussianity, isotropy, and a uniform survey geometry, to simulations with various levels of realism, including the \logn mocks, which have high redshift and high shot noise, but suppressed gravitational non-linearity, and the \quijote simulations, which have low redshift and low shot noise, and include non-linear effects. One of the most interesting conclusions from these numerical tests is that, even though our na\"{i}ve Gaussian model takes neither RSD nor gravitational non-Gaussianities into account, it produces a reasonably accurate estimate of the \quijote covariances in real and redshift space. However, despite a good overall match for the \logn mocks, we do observe spurious residuals via the half-inverse test. In particular, we find a residual in the diagonal elements, which is likely due to beyond-Gaussian correlators induced by shot noise effects. Finally, we also test our model using the \patchy mocks. These have a realistic survey geometry, matching that of the BOSS DR12 CMASS sample. In this case, we found the survey geometry to have a major impact on our theoretical prediction. Since our analytic formalism does not include full treatment of the window function, we account for the geometry by fitting for the number density and the effective volume by maximizing a likelihood based on the KL-divergence. This is shown to roughly compensate for the window function. 
Our companion paper~\citep{Philcox202002} shows that the theoretical covariance can be used as an important tool to facilitate data compression \citep{Scoccimarro200012,Taylor201305}, allowing a detection of gravitationally-induced non-Gaussianity from the BOSS 4PCF.

This work represents an important step along the path to constraining cosmology using. higher-point functions. A number of extensions are possible, in particular, including modeling of window function effects, numerical implementation of the covariances including RSD, extension to higher-order statistics such as the 5PCF and 6PCF, and a more thorough study of the performance of the Gaussian model in the limit of high shot noise.

%%%%%%%%%%%%%%%%%%%%%%%%%%%%%%%%%%%%%%%%%%%%%%%%%%

\section*{Acknowledgments}

We thank all members of the Slepian research group for useful discussions. JH thanks Hao Ding for insightful discussions.
ZS thanks Simone Ferraro, Adam Ginsburg, Alex Krolewski, and Kristen Lavelle for useful discussions. OP acknowledges funding from the WFIRST program through NNG26PJ30C and NNN12AA01C, and thanks the University of Florida and the Simons Foundation for additional support. 

The authors are pleased to acknowledge that the work reported on in this paper was substantially performed using the Princeton Research Computing resources at Princeton University which is consortium of groups led by the Princeton Institute for Computational Science and Engineering (PICSciE) and Office of Information Technology's Research Computing.

%%%%%%%%%%%%%%%%% APPENDICES %%%%%%%%%%%%%%%%%%%%%

\appendix

\section{Explicit Results for the Generalized Gaunt Integrals with $N=2,3$ and $4$}
\label{appendix:generalized_gaunt_N234}
In \S\ref{subsec:calY-from-avg} we discussed the generalized Gaunt integral; here, we present explicit results for $n=2,3$ and $4$, following~\citep{Cahn202010}. This uses the definition of Eq.~\eqref{eqn:generalized-gaunt}, which includes the quantity $\calQ^{\Lambda\Lambda'\Lambda''}$. For $n=2$, given the definition of $\calQ$ in Eq.~\eqref{eqn:defcalQ}, we have $\Lambda\to (\ell,\ell),\; \Lambda'\to (\ell',\ell')$, and $\Lambda''\to (\ell'',\ell'')$. This leads to
\begeqar
\calG^{\Lambda\Lambda'\Lambda''}=(4\uppi)^{-1}\sqrt{(2\ell+1)(2\ell'+1)(2\ell''+1)}\six{\ell}{\ell'}{\ell''}000^2.
\label{eq:A2}
\endeqar
This is a rescaling of the well-known result~\citep{Adams1878} for the coefficient when a product of two Legendre polynomials is expanded into a sum over single Legendre polynomials.  

For $n=3$ the generalized Gaunt integral is given by 
\begeqar\label{eqn:generalized-gaunt-N3}
    \calG^{\Lambda\Lambda'\Lambda''}&=&(4\pi)^{-3/2}\calQ^{\Lambda\Lambda'\Lambda''}\prod_{i=1}^3\six{\ell_i}{\ell'_i}{\ell''_i}000 \sqrt{(2\ell_i+1)(2\ell'_i+1)(2\ell''_i+1)}\non
    &=& (4\pi)^{-3/2}\nine{\ell_1}{\ell'_1}{\ell''_1}{\ell_2}{\ell'_2}{\ell''_2}{\ell_3}{\ell'_3}{{\ell_3''}}\prod_{i=1}^3\six{\ell_i}{\ell'_i}{\ell''_i}000 \sqrt{(2\ell_i+1)(2\ell'_i+1)(2\ell''_i+1)},
\endeqar
where we have used the definition of $\calC^{\ell_i\ell'_i\ell''_i}_{000}$ (cf.~Eq.~\ref{eqn:clebsch_gorden_N3}) and $\calD^{\rm P}_{\ell_i\ell'_i\ell''_i}$ (cf.~Eq.~\ref{eqn:defcalD_var}), and the quantity $\calQ^{\Lambda\Lambda'\Lambda''}$ is given by a 9-$j$ symbol, after summing over $m_i$, $m'_i$, and $m''_i$ (for $i=1, 2, 3$).

% after summing over the $m$ the quantity in Eq.~\eqref{eqn:defcalQ}  reduces to a 9-$j$ symbol:
% \begeqar\label{eqn:wigner-9j}
% \calQ^{\Lambda\Lambda'\Lambda''}&=&\sum_{m_1m_2m_3}\;\;\sum_{m_1'm_2'm_3'}\;\;\sum_{m''_1m''_2m''_3}\tj{\ell_1}{\ell_2}{\ell_3}{m_1}{m_2}{m_3}\tj{\ell_1'}{\ell_2'}{\ell_3'}{m_1'}{m_2'}{m_3'}\tj{\ell''_1}{\ell''_2}{\ell''_3}{m''_1}{m''_2}{m''_3}\nonumber\\
%     &&\,\qquad\times\,\tj{\ell_{1}}{\ell_1'}{\ell''_1}{m_{1}}{m_1'}{m''_1}\tj{\ell_{2}}{\ell_2'}{\ell''_2}{m_{2}}{m_2'}{m''_2}\tj{\ell_{3}}{\ell_3'}{\ell''_3}{m_{3}}{m_3'}{m''_3}\equiv\,\nine{\ell_1}{\ell'_1}{\ell''_1}{\ell_2}{\ell'_2}{\ell''_2}{\ell_3}{\ell'_3}{{\ell_3''}},
% \endeqar

For $n=4$, expanding the $\calQ^{\Lambda\Lambda'\Lambda''}$ quantity leads to 10 Wigner 3-$j$ symbols, and consequently
the product of two 9-$j$ symbols. The detailed derivation of this is given in ~\citet{Cahn202010} (section 6.4 and equation 71), leading to the final result:
\begeqar\label{eqn:generalized-gaunt-N4}
\calG^{\Lambda\Lambda'\Lambda''}&=&(4\uppi)^{-2}
\sqrt{(2\ell_{12}+1)(2\ell'_{12}+1)(2\ell''_{12}+1)}\non
&&\times\prod_{i=1}^4\sqrt{(2\ell_i+1)(2\ell'_i+1)(2\ell''_i+1)}\six{\ell_i}{\ell'_i}{\ell''_i}000\non
&&\quad\times\nine{\ell_1}{\ell_{2}}{\ell_{12}}{\ell'_1}{\ell'_{2}}{\ell'_{12}}{\ell''_1}{\ell''_{2}}{{\ell''}_{12}}
\nine{\ell_{12}}{\ell_3}{\ell_4}{\ell'_{12}}{\ell'_3}{\ell'_4}{\ell''_{12}}{{\ell''}_3}{\ell''_4}.
\endeqar

\section{Derivation of the Basic Covariance Elements}
\label{appendix:basic_elements}
\subsection{Real Space}%Basic Elements in Real Space}
Here we derive the basic covariance elements presented in \S\ref{subsec:basic-cov-elements}. Without loss of generality we consider only the contraction between a single pair of endpoints, neglecting the subindices and denoting the positions as $\bfr$ and $\bfr'$. The coupling between two endpoints across the unprimed and primed families can be expanded as:
\begeqar\label{eqn:xi_r_rp_s_appendix}
    \langle \delta(\bfx+\mathbf{r}) \delta(\bfx+\bfs+\bfr') \rangle = \xi(|\bfr'+\bfs-\bfr|) &=& \int_{\bfk}e^{i\bfk\cdot(\bfr'+\bfs-\bfr)} P(k) \nonumber\\
    &=& (4\uppi)^3 \sum_{\ell m}\; \sum_{\ell' m'}\; \sum_{LM} i^{\ell'+L-\ell}   \int_{\bfk} P(k) \nonumber\\
    && \times\, j_{\ell'}(kr')j_{L}(ks)j_{\ell}(kr) 
        Y^*_{\ell' m'}(\bfkhat)Y_{\ell' m'}(\bfrhat') Y_{L M}(\bfkhat)Y^*_{L M}(\bfshat)Y^*_{\ell m}(\bfkhat) Y_{\ell m}(\bfrhat), 
\endeqar
where, as stated in \S\ref{sec:npcf}, we have assumed isotropy (\textit{i.e.} that $P(\bfk)=P(k)$) in the first equality. The second equality arises from applying the plane wave expansion three times. Performing the angular integral over $\bfkhat$ gives the Gaunt integral:
%\bob{Haven't we already used $\mathcal{G}$ for something slightly different?} \jiamin{we used the same symbol for the gaunt integral and the generalized integral also in Encore paper.}
\begeqar
\mathcal{G}_{\ell \ell' L}^{m m' M} \equiv \int {d\Omega_k} Y^*_{\ell m}(\bfkhat)Y^*_{\ell' m'}(\bfkhat)Y^*_{LM}(\bfkhat)  &=&\sqrt{\frac{(2\ell+1)(2\ell'+1)(2L+1)}{4\uppi}}\six{\ell}{\ell'}{L}{0}{0}{0} \six{\ell}{\ell'}{L}{m}{m'}{M}\non
&=& (4\uppi)^{-1/2} \mathcal{D}^{\rm P}_{\ell \ell' L} \mathcal{C}^{\ell\ell' L}_{000}\mathcal{C}^{\ell\ell' L}_{m m' M}.
\endeqar
Inserting the definition of the $f$-integral, Eq.~\eqref{eqn:xi_r_rp_s_appendix} becomes 
%\jiamin{The result differs to Bob by a phase, but is consistent with the covariance Case I}
%\bob{I think our phases are the same.  The factor $i^{xxx}$ is real so it is the same as $(-i)^{xxx}=i^{-xxx}$} \jiamin{Ok! I think I found out the reason: I added another phase in Eq.~\eqref{eqn:clebsch_gorden_N2N3} which was already included in CG coefficients.}
\begeqar\label{eqn:2pcf_r_rp_s_appendix}
    \langle \delta(\bfx+\mathbf{r}) \delta(\bfx+\mathbf{s}+\mathbf{r}') \rangle 
    &=& (4\uppi)^3 \sum_{\ell \ell' L}\; \sum_{m m' M} i^{-\ell+\ell'+L} (4\uppi)^{-1} f_{\ell \ell' L}(r, r', s) (4\uppi)^{-1/2} \mathcal{D}^{\rm P}_{\ell \ell' L} \mathcal{C}^{\ell\ell' L}_{000}\mathcal{C}^{\ell\ell' L}_{m m' M} \nonumber \\
    && \times\, Y_{\ell m}(\bfrhat) Y_{\ell' m'}(\bfrhat') Y_{L M}(\bfshat) \nonumber\\
    &=& (4\uppi)^{3/2} \sum_{\ell \ell' L} i^{-\ell+\ell'+L} f_{\ell \ell' L}(r, r', s) \;\calD^{\rm P}_{\ell \ell' L} \;\mathcal{C}^{\ell \ell' L}_{000} \;\mathcal{P}_{\ell \ell' L}(\bfrhat, \bfrhat', \bfshat).
\endeqar

Finally, we give expressions for the contraction of two overdensity fields from the same family. These self-coupling terms do not occur in the calculation of the covariance of the connected NPCF, but do appear if one considers a covariance which includes the disconnected piece (as in Appendix \ref{appendix:cov_pc}). 
%\bob{PLEASE NOTE MY CHANGE IN THE TEXT HERE} \jiamin{Ok, reads good}
In this case, $\bfr_i$ and $\bfr_j$ denote two endpoints from the same family. As before, we apply the plane wave expansion to the exponentials in Eq.~\eqref{eqn:iso_pk_Fourier}, then integrate over $\bfkhat$ to find:
\begeqar\label{eqn:2pcf_r_r_appendix}
\left<\delta(\bfx+\bfr_i)\delta(\bfx+\bfr_j)\right>=\xi(|\bfr_i-\bfr_j|)&=&\int\frac{k^2dk}{2\pi^2}P(k)\sum_\ell j_\ell(kr_i) j_\ell(kr_j)(2\ell+1) \calL_\ell(\bfrhat_i\cdot\bfrhat_j)\non
&=& (4\uppi)^{3/2}\sum_\ell (-1)^\ell\sqrt{2\ell+1}\;f_{\ell\ell0}(r_i,r_j,0)\;\calY_{\ell\ell0}(\bfrhat_i,\bfrhat_j, 0).
\endeqar
In the second line, we have written our result in terms of the $N=3$ isotropic functions to maintain a consistent structure for all the basic elements. If one of the two overdensity fields is a primary, the expectation value is simply a 2PCF:
\begeqar
\left<\delta(\bfx+\bfr_0)\delta(\bfx+\bfr_i)\right>|_{\bfr_0\to 0}=\xi(|\bfr_i-\bfr_0|)|_{\bfr_0\to 0}=(4\uppi)^{3/2}f_{000}(r,0,0)\calY_{000}(\bfrhat,0,0).
\label{eqn:2pcf_r} 
\endeqar

\subsection{Redshift Space}%Basic Elements in Redshift Space}
Below, we derive the basic elements in redshift space, as a preparation for the fully-coupled covariance with RSD discussed in Appendix \ref{appendix:cov_fc_rsd_iso}. 
We first expand the power spectrum in terms of Legendre polynomials:
\begeqar\label{eq: P-legendre-expan}
    P(\vk) = \sum_{\lambda} P_\lambda(k)L_\lambda(\hat{\vk}\cdot\hn) = \sum_{\lambda\,\mu} \frac{4\pi}{2\lambda+1}P_\lambda(k)Y^*_{\lambda\mu}(\hat{\vk})Y_{\lambda\mu}(\hn)
\endeqar
where $P_\lambda(k)$ is the $\lambda$-th Legendre multipole of the power spectrum (where $\lambda$ is even) and $\hn$ is the line of sight. 

The expectation value of the product of two overdensity fields now reads
\begeqar
&&\av{\delta(\bfx+\bfr)\delta(\bfx+\bfr'+\bfs)}=\int_{\bfk} e^{i\bfk\cdot(\bfr'+\bfs-\bfr)} P(\bfk)\non
&=& \int_{\bfk} e^{i\bfk\cdot(\bfr'+\bfs-\bfr)} \sum_{\lambda\mu}\frac{4\uppi}{2\lambda+1} P_{\lambda}(k)Y^*_{\lambda\mu}(\hat{\vk})Y_{\lambda\mu}(\hn)\non
&=& \int \frac{d\hat{\vk}}{4\uppi}\int \frac{k^2 dk}{2\uppi^2} (4\uppi)^3 \sum_{\ell\ell'L}\sum_{mm'M} i^{\ell'+L-\ell} j_{\ell'}(kr')j_{\ell''}(ks)j_{\ell}(kr) Y^*_{\ell' m'}(\hat{\vk})Y^*_{LM}(\hat{\vk})Y^*_{\ell m}(\hat{\vk})\non
&&\,\times\,Y_{\ell' m'}(\bfrhat')Y_{LM}(\bfshat)Y_{\ell m}(\bfrhat)
\sum_{\lambda\mu}\frac{4\uppi}{2\lambda+1} P_{\lambda}(k)Y^*_{\lambda\mu}(\hat{\vk})Y_{\lambda\mu}(\hn).
\endeqar
We can perform an angular integral over $\hat{\vk}$:
\begeqar
\int d\hat{\vk}\; Y^*_{\ell m}(\hat{\vk})Y^*_{\ell' m'}(\hat{\vk})Y^*_{LM}(\hat{\vk}) Y^*_{\lambda\mu}(\hat{\vk}) &=& \sum_{\bar{L}} (-1)^{\bar{M}} \calG^{\ell\ell'\bar{L}}_{mm'\bar{M}}\calG^{\bar{L}L\lambda}_{-\bar{M}M\mu}\non
&=&{(4\uppi)}^{-1}\sum_{\bar{L}} (-1)^{\bar{M}} \calD^{\rm P}_{\ell\ell'L\lambda}(2\bar{L}+1) \calC^{\ell\ell'\bar{L}}_{mm'\bar{M}} \calC^{\bar{L}L\lambda}_{-\bar{M}M\mu}
\calC^{\ell\ell'\bar{L}}_{000} \calC^{\bar{L}L\lambda}_{000}\non
&=& {(4\uppi)}^{-1} \calD^{\rm P}_{\ell\ell'L\lambda} \calC^{\ell\ell'L\lambda}_{mm'm''\mu} 
\calC^{\ell\ell'L\lambda}_{0000};
\endeqar
because of the additional l.o.s. direction $\hn$, we need to consider isotropic functions with four arguments:
\begeqar
\calY_{\ell\ell'L\lambda}(\bfrhat,\bfrhat',\bfshat,\hn) = \sum_{mm'M\mu}\calC^{\ell\ell'L\lambda}_{mm'M\mu} Y_{\ell m}(\bfrhat)Y_{\ell' m'}(\bfrhat')Y_{L M}(\bfshat)Y_{\lambda\mu}(\hn).
\endeqar
To incorporate the power spectrum multipole decomposition, we extend the definition of the $f$-integral as follows:
\begeqar
    f_{\ell_1\ell_2\ell_3}^{\lambda}(r_1,r_2,r_3) &=&  \int \frac{k^2dk}{2\pi^2}P_{\lambda}(k)j_{\ell_1}(kr_1)j_{\ell_2}(kr_2)j_{\ell_3}(kr_3).
\endeqar
The redshift space basic covariance element can thus be written:
%the basic element expressed in redshift space is given by:
\begeqar\label{eqn:2pcf_r_rp_s_rsd}
&&\av{\delta(\bfx+\bfr)\delta(\bfx+\bfr'+\bfs)}=(4\uppi)^2\sum_{\ell\ell'L\lambda} i^{-\ell+\ell'+L} \frac{1}{2\lambda+1} \calD^{\rm P}_{\ell\ell'L\lambda} 
\calC^{\ell\ell'L\lambda}_{0000}
f_{\ell\ell'L}^{\lambda}(r,r',s) \calY_{\ell\ell'L\lambda}(\bfrhat,\bfrhat',\bfshat,\hn).
\endeqar

\section{Partially-coupled 4PCF covariance}
\label{appendix:cov_pc}

\subsection{Fully-Coupled and Partially-Coupled Covariances}%Partially-coupled Covariance and its Scaling}
\label{appendix:intro_cov_pc}
%\jiamin{let me give it a try to add a section explaining the scaling of the partially-coupled covariance -- try to keep the following as concise as possible, the major discussion disconnected estimator will also be discussed in the data paper}

In \S\ref{subsec:cov_fc} we presented the fully-coupled covariance, which is the part of relevance for the connected NPCF estimator. As before, the connected estimator is obtained by subtracting the disconnected piece from full estimator as in Eq.~\eqref{eqn:4pcf-c-dc}. This feature is now included in the \textsc{encore} code, and is discussed at length in our companion paper~\citet{Philcox2021boss4pcf}. For completeness however, we will discuss in this section how one may estimate the partially-coupled covariance.% can be handled and the approximation we made. 

We first sketch our reasoning for ignoring the partially-coupled terms in the connected 4PCF covariance. Following the definition of our estimator, the fully-coupled covariance can be written as
\begeqar
\cov^{\rm fc}(\bfR, \bfR') &\equiv& \cov(\hat{\zeta}_{\rm c}, \hat{\zeta}_{\rm c})\non 
&=& \cov(\hat{\zeta},\hat{\zeta})-\cov(\hat{\zeta}_{\rm dc}, \hat{\zeta}) - \cov(\hat{\zeta}, \hat{\zeta}_{\rm dc}) + \cov(\hat{\zeta}_{\rm dc}, \hat{\zeta}_{\rm dc}),
\endeqar
where the the first term in the second equality, the covariance of the full estimator, is simply the covariance obtained from all combinations of eight overdensity fields. %is precisely a term we have direct access to.
We use $\av{\delta\delta\delta\delta}$ to denote the full estimator; given the symmetry, any one of the overdensity fields can be thought of as a primary vertex, with the position of its neighbours fixed relative to the primary. As before, the covariance of the full estimator consists of both fully- and partially-coupled parts. Below, we give an example of a contraction that leads to a partially-coupled term (here with angle brackets representing spatial integrals rather than statistical expectations):
\begin{eqnarray*}
% &&&&&&&&&
% full-full
% &&&&&&&&&
% contraction 1
\contraction[0.8ex]{\cov(\hat{\zeta}, \hat{\zeta}) \to\langle}{\delta}{\,}{\delta}
% contraction 2
\contraction[1.2ex]{\cov(\hat{\zeta}, \hat{\zeta}) \to\langle\delta\, \delta \,}{\delta}{\,\delta\rangle \langle \delta'\delta'}{\delta'}
% contraction 3
\contraction[1.6ex]{\cov(\hat{\zeta}, \hat{\zeta}) \to\langle\delta\, \delta \,\delta\,}{\delta}{\rangle \langle \delta'\delta'\delta'}{\delta'}
% contraction 4
\contraction[0.4ex]{\cov(\hat{\zeta}, \hat{\zeta}) \to\langle\delta\, \delta \,\delta\,\delta\rangle \langle}{\delta'}{}{\delta'}
% expression
\cov(\hat{\zeta}, \hat{\zeta}) \to \langle\delta\, \delta \,\delta\,\delta\rangle \langle \delta' \delta' \delta' \delta'\rangle.
\end{eqnarray*}
The disconnected estimator is represented by $\av{\delta\delta}\av{\delta\delta}$. Again, we know the relative position between overdensity fields appears within a $\av{\cdots}$ integral, but the relative position between two $\av{\cdots}$ is free. This leads us to consider only the self-coupling contractions within an integral such as 
$\contraction[0.8ex]{\langle}{\delta}{\,}{\delta}
\langle\delta\, \delta\rangle\langle\delta\,\delta\rangle$; this contraction is, by definition, a 2PCF. Contractions such as $\contraction[0.8ex]{\langle\delta}{\delta}{\rangle\langle}{\delta}
\langle\delta\, \delta\rangle\langle\delta\,\delta\rangle$ must be integrated over the unfixed pair separation vector, resulting in an additional volume factor $V^{-1}$, which leads to a strong suppression. Below, we list the contractions that contribute to the partially-coupled covariance at leading order:
\begin{eqnarray*}
% &&&&&&&&&
% dc-full
% &&&&&&&&&
% contraction 1
\contraction[0.8ex]{\cov(\hat{\zeta}_{\rm dc}, \hat{\zeta})\to\langle}{\delta}{\,}{\delta}
% contraction 2
\contraction[1.2ex]{\cov(\hat{\zeta}_{\rm dc}, \hat{\zeta})\to\langle\delta\, \delta\rangle\langle}{\delta}{\,\delta\rangle \langle \delta'\delta'}{\delta'}
% contraction 3
\contraction[1.6ex]{\cov(\hat{\zeta}_{\rm dc}, \hat{\zeta})\to\langle\delta\, \delta \rangle\langle\delta\,}{\delta}{\rangle \langle \delta'\delta'\delta'}{\delta'}
% contraction 4
\contraction[0.4ex]{\cov(\hat{\zeta}_{\rm dc}, \hat{\zeta})\to\langle\delta\, \delta \rangle\langle\delta\,\delta\rangle \langle}{\delta'}{}{\delta'}
% expression
\cov(\hat{\zeta}_{\rm dc}, \hat{\zeta})\to\langle\delta\, \delta\rangle\langle\delta\,\delta\rangle \langle \delta' \delta' \delta' \delta'\rangle\\
% &&&&&&&&&
% full-dc
% &&&&&&&&&
% contraction 1
\contraction[0.8ex]{\cov(\hat{\zeta}, \hat{\zeta}_{\rm dc})\to\langle}{\delta}{\,}{\delta}
% contraction 2
\contraction[1.2ex]{\cov(\hat{\zeta}, \hat{\zeta}_{\rm dc})\to\langle\delta\, \delta\,}{\delta}{\,\delta\rangle \langle \delta'\rangle\langle\delta'}{\delta'}
% contraction 3
\contraction[1.6ex]{\cov(\hat{\zeta}, \hat{\zeta}_{\rm dc})\to\langle\delta\, \delta\,\delta\,}{\delta}{\rangle \langle \delta'\delta'\rangle\langle\delta'}{\delta'}
% contraction 4
\contraction[0.4ex]{\cov(\hat{\zeta}, \hat{\zeta}_{\rm dc})\to\langle\delta\, \delta\,\delta\,\delta\rangle\langle}{\delta'}{}{\delta'}
% expression
\cov(\hat{\zeta}, \hat{\zeta}_{\rm dc})\to\langle\delta\, \delta\,\delta\,\delta\rangle\langle \delta' \delta' \rangle \langle\delta' \delta'\rangle\\
% &&&&&&&&&
% dc-dc
% &&&&&&&&&
% contraction 1
\contraction[0.8ex]{\cov(\hat{\zeta}_{\rm dc}, \hat{\zeta}_{\rm dc})\to\langle}{\delta}{\,}{\delta}
% contraction 2
\contraction[1.2ex]{\cov(\hat{\zeta}_{\rm dc}, \hat{\zeta}_{\rm dc})\to\langle\delta\, \delta\rangle\langle}{\delta}{\,\delta\rangle \langle \delta'\rangle\langle\delta'}{\delta'}
% contraction 3
\contraction[1.6ex]{\cov(\hat{\zeta}_{\rm dc}, \hat{\zeta}_{\rm dc})\to\langle\delta\, \delta \rangle\langle\delta\,}{\delta}{\rangle \langle \delta'\delta'\rangle\langle\delta'}{\delta'}
% contraction 4
\contraction[0.4ex]{\cov(\hat{\zeta}_{\rm dc}, \hat{\zeta}_{\rm dc})\to\langle\delta\, \delta \rangle\langle\delta\,\delta\rangle \langle}{\delta'}{}{\delta'}
% expression
\cov(\hat{\zeta}_{\rm dc}, \hat{\zeta}_{\rm dc})\to \langle\delta\, \delta\rangle\langle\delta\,\delta\rangle \langle \delta' \delta' \rangle\langle\delta' \delta'\rangle\nonumber.
\end{eqnarray*}
After counting the permutations, we find $72$ terms in each case, all of which cancel. This leads only corrections of $\calO((r_{\rm c}^3/V)^2)$ and higher, where $r_{\rm c}$$\sim$$100\, h^{-1} {\rm Mpc}$ is the correlation length. This correction is typically $\sim$$0.1\%$ and hence can be neglected when comparing to the measurements from the mock simulations with box length of $L_{\rm box}$$\sim$$\calO(1)\, h^{-1} {\rm Gpc}$. We thus conclude that the fully coupled covariance does represent that of the connected 4PCF in the large-volume limit.

\subsection{Analytic Form}%Partially-coupled 4PCF Covariance arises from the Full Estimator}
\begin{figure}
    \centering
    \includegraphics[width=.9\textwidth]{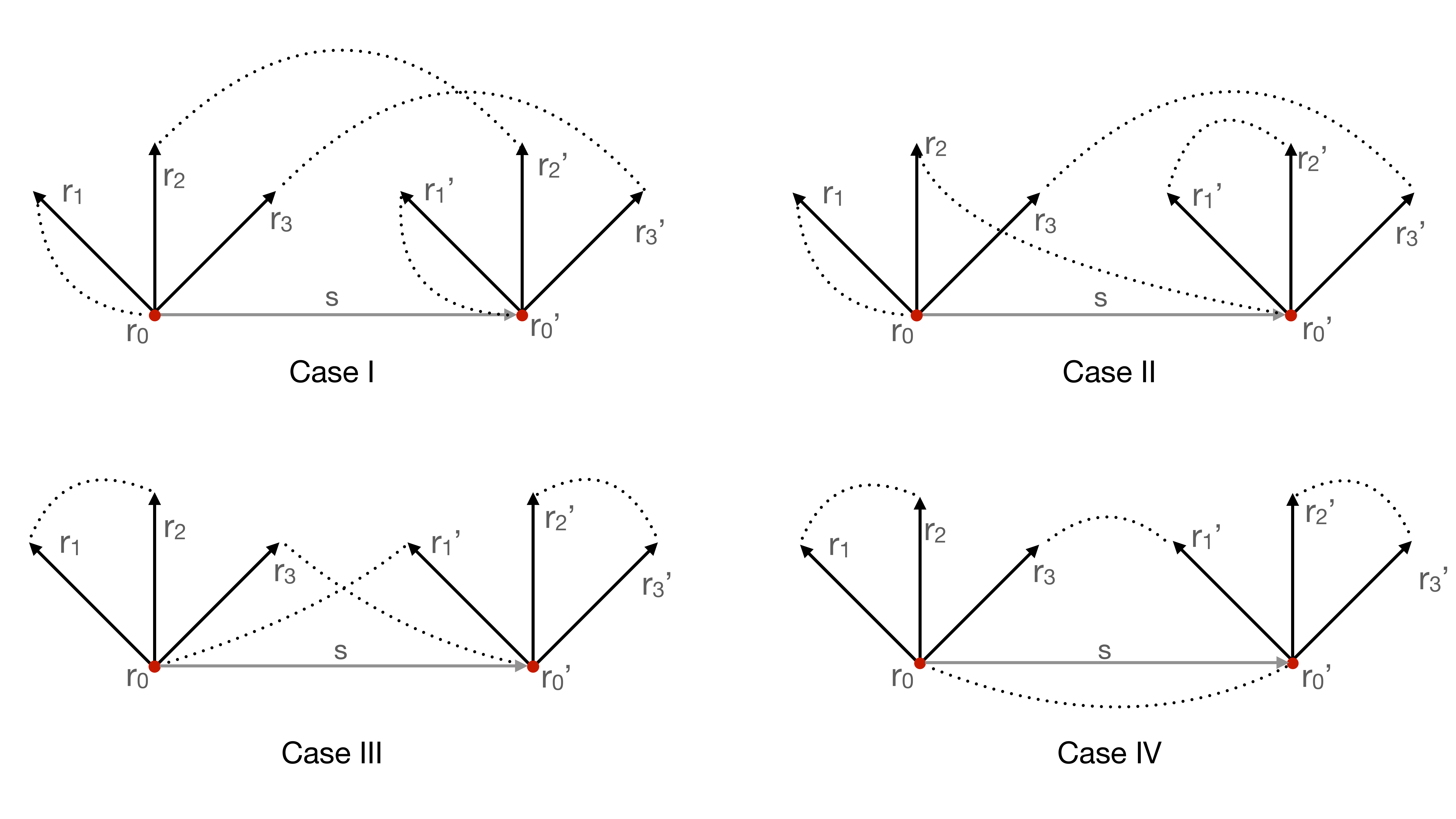}
    \caption{Diagrams for the partially-coupled covariance. This Figure is analogous to Fig.\,\ref{fig:tree_connected_2cases}, but gives the terms necessary to model the \textit{disconnected} 4PCF covariance.}
    \label{fig:tree_disconnected_4cases}
\end{figure}

For completeness, we also derive analytic expressions for the partially-coupled covariance. % in the covariance for the full estimator.
These contributions are composed of similar structures to the basic elements shown in \S\ref{subsec:basic-cov-elements} and can be divided into four pieces as shown in Fig.~\ref{fig:tree_disconnected_4cases}. 
All terms involve a self-coupling, \textit{i.e.} the contraction of overdensity fields within a primed or unprimed family. As a result, the basis function will end up with one of the angular momenta being zero, with the other two equal. This implies that the partially-coupled covariance can be fully characterized just by $\ell$ and $\ell'$. 
The fundamental idea of the derivation is similar to that underlying the fully-coupled covariance derivation. First, identify the basic elements that contribute to the given cases. Second, apply a rotational average over the three direction vectors $\bfrhat$, $\bfrhat'$, and $\bfshat$ and reorder the permuted coordinates into canonical ordering. Third, project the covariance onto the isotropic basis, picking out the terms proportional to $\calY_{\Lambda}(\bfRhat)$ and $\calY_{\Lambda'}(\bfRhat')$. Here we necessarily need to introduce both permutations $G$ and $H$ because self-contraction breaks the symmetry of the coupling structure. As before, we restrict $G$ to cyclic permutations, allowing $H$ to explore all possibilities.

\paragraph*{Case I}
The partially-coupled covariance in this case contains the self-contraction between primary vertices, ${\bf r}_0$ and ${\bf r}_0'$, and endpoints of their own family (see Fig.~\ref{fig:tree_disconnected_4cases}). This can be expressed as
\begeqar
I_{\rm I}(\bfR,\bfR';\bfs)&=&\left<\delta(\bfx+\bfr_0)\delta(\bfx+\bfr_{G1})\right> \left<\delta(\bfx+\bfs+\bfr'_0)\delta(\bfx+\bfs+\bfr'_{H1})\right>|_{\bfr_{0}=0, \bfr_0' = 0}\\
&&\,\times\,\left<\delta(\bfx+\bfr_{G2})\delta(\bfx+\bfs+\bfr'_{H2})\right>\left<\delta(\bfx+\bfr_{G3})\delta(\bfx+\bfs+\bfr'_{H3})\right>\nonumber.
\endeqar
Inserting the definition of the basic elements defined in \S\ref{subsec:basic-cov-elements}, we find
\begeqar
&&I_{\rm I}(\bfR,\bfR';\bfs)=\non
&&\quad \sum_{G,H}\, (4\uppi)^{3/2}f_{000}(r_{G1},0,0)\calY_{000}(\bfrhat_{G1},0,0)\,(4\uppi)^{3/2}f_{000}(0,r'_{H1},0)\calY_{000}(0,\bfrhat'_{H1},0)\non
&&\,\quad\times\,(4\uppi)^{3/2}\sum_{\ell_{G2}\ell'_{H2}L_2} i^{-\ell_{G2}+\ell'_{H2}+L_2}
f_{\ell_{G2}\ell'_{H2}L_2}(r_{G2},r'_{H2},s)\calD^{\rm P}_{\ell_{G2}\ell'_{H2}L_2}\calC^{\ell_{G2}\ell'_{H2}L_2}_0\calY_{\ell_{G2}\ell'_{H2}L_2}(\bfrhat_{G2},\bfrhat'_{H2},\bfshat)\non
&&\qquad\qquad\times (4\uppi)^{3/2}\sum_{\ell_{G3}\ell'_{H3}L_3} i^{-\ell_{G3}+\ell'_{H3}+L_3}
f_{\ell_{G3}\ell'_{H3}L_3}(r_{G3},r'_{H3},s)\calD^{\rm P}_{\ell_{G3}\ell'_{H3}L_3}\calC^{\ell_{G3}\ell'_{H3}L_3}_{000}\calY_{\ell_{G3}\ell'_{H3}L_3}(\bfrhat_{G3},\bfrhat'_{H3},\bfshat).
\endeqar
Given that the sum of the orbital angular momentum must be an even number, $\ell_{G2}=\ell'_{H2}$ and $\ell_{G3}=\ell'_{H3}$, thus the sum reduces to one over $\ell,\ell',\ell''$ with $\calL=(\ell,\ell)$, $\calL'=(\ell',\ell')$, $\calL''=(L,L)$. As a reminder, the coefficients $\calC$ and $\calD^{\rm P}$ are given in Eq.~\eqref{eqn:clebsch_gorden_N3} and Eq.~\eqref{eqn:defcalD_var}, respectively.
Integrating over $\vs$ we find
\begeqar
&&\int\frac{d^3\bfs}{V}I_{\rm I}(\bfR,\bfR';\bfs)= \sum_{G,H}\,\sum_{\calL_G, \calL'_H}\int\frac{s^2 ds}{V}\xi(r_{G1})\xi(r'_{H1}) (4\uppi)^4 \sum_{\ell\ell'L}
(-1)^{\ell+\ell'+L} f_{\ell\ell'L}(r_{G2},r'_{H2},s)f_{\ell\ell'L}(r_{G3},r'_{H3},s)\non
&&\qquad\qquad\times ({\calD^{\rm P}_{\ell\ell'L}})^2({{\calC^{\ell\ell'L}_{000}}})^2
\calQ^{(\ell\ell)(\ell'\ell')(LL)}\calD^{\rm P}_{LL}\calC^{LL}_{00}\calY_{\calL_G}(\bfrhat_{G1},\bfrhat_{G2},\bfrhat_{G3})\calY_{\calL'_H}(\bfrhat'_{H1},\bfrhat'_{H2},\bfrhat'_{H3}),
\endeqar
where the rotational average over $d\calS$ gives a factor of $4\uppi$, following our normalization convention. $\calL_{G}$ has one angular momentum of zero with the other two equal; the same goes for $\calL'_{H}$. Expressing the two-argument isotropic basis functions in terms of those with three arguments, for example, $\calY_{\ell\ell}(\bfrhat_{G2},\bfrhat_{G3})=(4\uppi)^{1/2}\calY_{0\ell\ell}(\bfrhat_{G1},\bfrhat_{G2},\bfrhat_{G3})$, we obtain an additional $4\uppi$. We now insert the definition of the generalized Gaunt integral for $N=2$ (cf. Eq.~\ref{eq:A2}), giving
\begeqar
\int\frac{d^3\bfs}{V}I_{\rm I}(\bfR,\bfR';\bfs) &=& \sum_{G,H}\,\sum_{\calL_G, \calL'_H}\int\frac{s^2 ds}{V} \xi(r_{G1})\xi(r'_{H1}) (4\uppi)^4 \sum_{\ell\ell'L}
(-1)^{\ell+\ell'} f_{\ell\ell'L}(r_{G2},r'_{H2},s)f_{\ell\ell'L}(r_{G3},r'_{H3},s) \nonumber\\ &&\,\times\,\sqrt{(2\ell+1)(2\ell'+1)}(2L+1)\six{\ell}{\ell'}{L}000 ^2 \calY_{\calL_{G}}(\bfrhat_{G1},\bfrhat_{G2},
\bfrhat_{G3})\calY_{\calL'_{H}}(\bfrhat'_{H1},\bfrhat'_{H2},\bfrhat'_{H3}).
\endeqar
Using Eq.~\eqref{eqn:generalized-kronecker-delta}, we may restore the arguments to canonical order:
\begeqar
\calY_{\calL_G}(\bfRhat_{G})= \sum_{J} \calB^{G^{-1}}_{\calL_G,J} \calY_{J}(\bfRhat), \qquad \calY_{\calL'_H}(\bfRhat'_{H})= \sum_{J'} \calB^{H^{-1}}_{\calL'_H,J'} \calY_{J'}(\bfRhat').
\endeqar
In this case, $\calB^{G^{-1}}_{\calL_G,J}$ and $\calB^{H^{-1}}_{\calL'_H,J'}$ are given by products of Kronecker deltas since one of the angular momenta is zero.
Since the partially-coupled covariance always leads to products of two $f$-integrals, it is useful to introduce the $g$-integral, defined by:
\begeqar\label{eqn:def-gllll}
&&\int s^2ds f_{\ell\ell\lambda}(r_1,r_2,s) f_{\ell'\ell'\lambda}(r'_1,r'_2,s)\non
&&\qquad=\int\frac{k^2dk}{(2\pi)^3}j_{\ell}(kr_1)j_{\ell}(kr_2)j_{\ell'}(kr'_1)j_{\ell'}(kr'_2)P^2(k)\equiv g_{\ell\ell\ell'\ell'}(r_1,r_2,r'_1,r'_2).
\endeqar
It is worth noting that, unlike the $f$-integral, the $g$-integral has dimensions of volume. The coefficient $(2\uppi)^{-3}$ appears due to the definition of the $f$-integral, together with the coefficient in the identity for the integral of two sBFs:
\begeqar\label{eqn:integral-2sbfs}
\int s^2ds j_\lambda(sa)j_\lambda(sb)=\frac \pi{2ab}\delta_{\rm D}(a-b).
\endeqar
Together with the relation:
\begeqar
\sum_L (2L+1)\six\ell{\ell'}{L}000 ^2=1,
\endeqar
we find the final expression for the partially-coupled covariance Case I:
\begeqar\label{eqn:cov_pc_caseI}
    \cov^{\rm{(pc)}, I}_{\Lambda,\Lambda'}=\sum_{G,H}\,\sum_{\calL_G, \calL'_H}\frac{(4\uppi)^4}V(-1)^{\ell+\ell'} \sqrt{(2\ell+1)(2\ell'+1)}\xi(r_{G1})\xi(r'_{H1})
    g_{\ell\ell\ell'\ell'}(r_{G2},r_{G3},r'_{H2},r'_{H3})\calB^{G^{-1}}_{\calL_G,\Lambda}\calB^{H^{-1}}_{\calL'_H,\Lambda'}.
\endeqar
Here we keep the inverse reordering coefficient to make clear that the partially-coupled covariance only contributes to the collection of the three angular momenta with the following form $\{\Lambda,\Lambda'\}=\{0 \ell \ell, 0\ell'\ell'\} + \rm{8\,perms.}$. %depending on where the zeros appears, \textit{i.e.} there must be one zero which forces the rest two angular momenta to be the same.

\paragraph*{Case II} 
In this case, only one of the primary vertices is connected intra-family-wise (as shown in Fig.~\ref{fig:tree_disconnected_4cases}). It can happen that the primary vertex of the primed tetrahedron is coupled to an unprimed vertex, or the other way around. By symmetry, we need only discuss one of the two possibilities. The contraction of the eight overdensity fields can be expressed as
\begeqar
I_{\rm II}(\bfR,\bfR';\bfs)&=&\left<\delta(\bfx+\bfr_0)\delta(\bfx+\bfr_{G1})\right>\left<\delta(\bfx+\bfr_{G2})\delta(\bfx+\bfs+\bfr'_{0})\right>|_{r_{0}=r_0'=0}\non
&&\,\times\,\left<\delta(\bfx+\bfs+\bfr'_{H1})\delta(\bfx+\bfs+\bfr'_{H2})\right>\left<\delta(\bfx+\bfr_{G3})\delta(\bfx+\bfs+\bfr'_{H3})\right>.
\endeqar
In terms of the basic elements, $I_{\rm II}(\bfR,\bfR';\bfs)$ becomes
\begeqar
&& I_{\rm II}(\bfR,\bfR';\bfs)\non
&&\quad = \sum_{G,H}\,(4\uppi)^{3/2}f_{000}(r_{G1},0,0)\calY_{000}(\bfrhat_{G1},0,0)\non
&&\quad\quad\times(4\uppi)^{3/2}\sum_{\ell'_{H1}} (-1)^{\ell'_{H1}}f_{\ell'_{H1}\ell'_{H1}0}(r'_{H1},r'_{H2},0)\sqrt{2\ell'_{H1}+1} \calY_{\ell'_{H1}\ell'_{H1}0}(\bfrhat'_{H1},\bfrhat'_{H2},0)\non
&&\qquad\quad\times
(4\uppi)^{3/2}\sum_{\ell_{G2}}
(-1)^{\ell_{G2}}f_{\ell_{G2}0\ell_{G2}}(r_{G2},0,s)\sqrt{2\ell_{G2}+1}\calY_{\ell_{G2} 0\ell_{G2}}(\bfrhat_{G2},0,\bfshat)\non
&&\qquad\qquad\times(4\uppi)^{3/2}\sum_{\ell_{G3}\ell'_{H3}L_3} i^{-\ell_{G3}+\ell'_{H3}+L_3}
f_{\ell_{G3}\ell'_{H3}L_3}(r_{G3},r'_{H3},s)\calD^{\rm P}_{\ell_{G3}\ell'_{H3}L_3}\calC^{\ell_{G3}\ell'_{H3}L_3}_{000}\calY_{\ell_{G3}\ell'_{H3}L_3}(\bfrhat_{G3},\bfrhat'_{H3},\bfshat).
\endeqar
Averaging over $d\calS$ involves only two angular momenta, $\ell_{G2}$ and $L_3$, enforcing $\ell_{G2}=L_3\equiv\ell$. Similarly, averaging over $d\calR$ involves just $\ell_{G2}$ and $\ell_{G3}$ and sets $\ell_{G2}=\ell_{G3}\equiv\ell$. Finally, since $\bfrhat'_{H1}$ and $\bfrhat'_{H2}$ are already combined into an isotropic function, the integration over $d\calR'$ effectively involves only $\bfrhat'_{H3}$ and will result in $\ell'_{H3}=0$. The imaginary phase also becomes unity.

Using the definition given in Eq. (\ref{eqn:def-gllll}), we have
\begeqar
\int s^2 ds f_{\ell 0\ell}(r_{G2},0,s)f_{\ell 0\ell}(r_{G3},r'_{H3},s)=g_{\ell0\ell0 }(r_{G2}, 0,r_{G3},r'_{H3});
\endeqar
in this case, the $g$-integral can be reduced to an $f$-integral. 
%Projecting the covariance on to the basis proportional to $\calY_{0\ell\ell}(0,\bfrhat_{G2},\bfrhat_{G3})\calY_{\ell'\ell' 0}(\bfrhat'_{H1},\bfrhat'_{H2},0)$ 
The final form of Case II reads:
\begeqar\label{eqn:cov_pc_caseII}
\cov^{\rm{(pc)}, II}_{\Lambda,\Lambda'}
 = \sum_{G,H}\,\sum_{\calL_G, \calL'_H} \frac{(4\uppi)^4}V(-1)^{\ell+\ell'} \xi(r_{G1})\sqrt{(2\ell+1)(2\ell'+1)} f_{\ell'\ell' 0}(r'_{H1},r'_{H2},0)g_{\ell0\ell0 }(r_{G2}, 0,r_{G3},r'_{H3})\calB^{G^{-1}}_{\calL_G,\Lambda}\calB^{H^{-1}}_{\calL'_H,\Lambda'}.
\endeqar
The case in which an unprimed primary vertex $\delta(\bfr_0)$ couples to an endpoint from the primed family $\delta(\bfr'_i)$ follows similarly.
%In case II, there is another mirrored scenario where the primary vertex $\delta(\bfr_0)$ from the unprimed family coupled to the one from the endpoints in the primed family $\delta(\bfr'_i)$, which ends up in an expression with similar fashion and we do not replicate it here.

\paragraph*{Case III} 
The next form to consider occurs when both the primed vertices are coupled to a vertex from the opposite family. In this case:
\begeqar
I_{\rm III}(\bfR,\bfR';\bfs) &=&\left<\delta(\bfx+\bfr_{G1})\delta(\bfx+\bfr_{G2})\right>\left<\delta(\bfx+\bfs+\bfr'_{H2})\delta(\bfx+\bfs+\bfr'_{H3})\right>\non
&&\,\times\,\left<\delta(\bfx+\bfr_0)\delta(\bfx+\bfs+\bfr'_{H1})\right>\left<\delta(\bfx+\bfr_{G3})\delta(\bfx+\bfs+\bfr'_0)\right>|_{r_{0}=r_0'=0}.
\endeqar
Na\"ively, this case also involves an isotropic function of the form $\calY_{0\ell\ell}$; however, the rotational average over the endpoint vectors forces their paired angular momenta to be zero.

Inserting our basic elements, we have:
\begeqar
I_{\rm III}(\bfR,\bfR';\bfs) &=& 
\sum_{G,H}\, (4\uppi)^{3/2}\sum_{\ell_{G1}}
 f_{\ell_{G1}\ell_{G1}0}(r_{G1},r_{G2},0)(-1)^{\ell_{G1}}\sqrt{2\ell_{G1}+1}\calY_{\ell_{G1}\ell_{G1} 0}(\bfrhat_{G1},\bfrhat_{G2},0) \non
&&\,\times\, (4\uppi)^{3/2}\sum_{\ell'_{H1}}
 f_{\ell'_{H1}\ell'_{H1}0}(r'_{H1},r'_{H2},0)(-1)^{\ell'_{H1}}\sqrt{2\ell'_{H1}+1}\calY_{\ell'_{H1}\ell'_{H1}0}(\bfrhat'_{H1},\bfrhat'_{H2},0)\non
&&\,\times\, (4\uppi)^{3/2} \sum_{\ell'_{H3}}
 f_{0\ell'_{H3}\ell'_{H3}}(0,r'_{H3},s)(-1)^{\ell'_{H3}}\sqrt{2\ell'_{H3}+1}\calY_{0\ell'_{H3}\ell'_{H3}}(0,\bfrhat'_{H3},\bfs)\non
&&\,\times\, (4\uppi)^{3/2}\sum_{\ell_{G3}}
 f_{0\ell_{G3}\ell_{G3}}(0,r_{G3},s)(-1)^{\ell_{G3}}\sqrt{2\ell_{G3}+1}\calY_{0\ell_{G3}\ell_{G3}}(0,\bfrhat_{G3},\bfs).
\endeqar
In this case, the rotation average over $d\calR$ will leave only the $\ell_{G3}=0$ term since $\bfrhat_{G1}$ and $\bfrhat_{G2}$ are already combined into an isotropic function. Similarly, averaging over $d\calR'$ will force $\ell'_{H3}=0$, allowing us to simplify $\ell_{G1}\equiv\ell$ and $\ell'_{H1}\equiv\ell'$. Therefore, the two $f$-integrals associated with $\ell_{G3}$ and $\ell'_{H3}$ are given by
\begeqar
 &&\int s^2 ds\,f_{000}(0,r_{G3},s) f_{000}(0,r'_{H3},s) =g_{0000}(0,r_{G3},0,r'_{H3}),  
\endeqar
where we have used the identity for the integral of a product of two sBFs given in Eq.~\eqref{eqn:integral-2sbfs}.
%insert this into the definition of case III and  projecting out the contribution to covariance for $\calY_{\ell\ell 0}(\bfrhat_{G1},\bfrhat_{G2},\bfrhat_{G3})\calY_{\ell'\ell' 0}(\bfrhat'_{H1},\bfrhat'_{H2},\bfrhat'_{H3})$, we arrive at 
The final form of Case III reads:
\begeqar\label{eqn:cov_pc_caseIII}
&&\cov^{\rm{(pc)}, III}_{\Lambda,\Lambda'}=\\
&&\quad\sum_{G,H}\,\sum_{\calL_G, \calL'_H}\frac{(4\pi)^4}V
 (-1)^{\ell+\ell'}\sqrt{(2\ell+1)(2\ell'+1)}f_{\ell\ell0}(r_{G1},r_{G2},0)f_{\ell'\ell'0}(r'_{H1},r'_{H2},0)g_{0000}(0,r_{G3},0,r'_{H3})\calB^{G^{-1}}_{\calL_G,\Lambda}\calB^{H^{-1}}_{\calL'_H,\Lambda'}\nonumber
 \endeqar

\paragraph*{Case IV} 
Finally, consider the direct contraction between two primary vertices, accompanied by the contraction of two endpoints from each family
\begeqar
I_{\rm IV}(\bfR,\bfR';\bfs)&=&\left<\delta(\bfx+\bfr_0)\delta(\bfx+\bfs+\bfr'_0)\right>|_{r_{0}=r_0'=0}\non
&&\,\times\,\left<\delta(\bfx+\bfr_{G1})\delta(\bfx+\bfr_{G2})\right>\left<\delta(\bfx+\bfs+\bfr'_{H2})\delta(\bfx+\bfs+\bfr'_{H3})\right>\left<\delta(\bfx+\bfr_{G3})\delta(\bfx+\bfs+\bfr'_{H1})\right>.
\endeqar
As before, inserting the basic elements leads to
\begeqar
I_{\rm IV}(\bfR,\bfR';\bfs)
&=& \sum_{G,H}\,(4\uppi)^{3/2} f_{000}(0,0,s)\calY_{000}(0,0,\bfshat)\non
&&\,\times\,(4\uppi)^{3/2}\sum_{\ell_{G1}} (-1)^{\ell_{G1}}f_{\ell_{G1}\ell_{G1}0}(r_{G1},r_{G2},0)\sqrt{2\ell_{G1}+1}\calY_{\ell_{G1}\ell_{G1}0}(\bfrhat_{G1},\bfrhat_{G2},0)\non
&&\,\times\, (4\uppi)^{3/2} \sum_{\ell'_{H1}} (-1)^{\ell'_{H2}}f_{0\ell'_{H2}\ell'_{H2}}(0,r'_{H2},r'_{H3})\sqrt{2\ell'_{H2}+1}\calY_{0\ell'_{H2}\ell'_{H2}}(0,\bfrhat'_{H2},\bfrhat'_{H3})\non
&&\,\times\,(4\uppi)^{3/2}\sum_{\ell_{G3}\ell'_{H1}L_3} i^{-\ell_{G3}+\ell'_{H1}+L_3} f_{\ell_{G3}\ell'_{H1}L_3}(r_{G3},r'_{H1},s)\calD^{\rm P}_{\ell_{G3}\ell'_{H1}L_3}\calY_{\ell_{G3}\ell'_{H1}L_3}(\bfrhat_{G3},\bfrhat'_{H1},\bfshat),
\endeqar
simplifying $\ell_{G1}\equiv\ell$ and $\ell'_{H2}\equiv\ell'$. We can see that the rotational average over $d\calS$ forces $L_3=0$ and thus $\ell_{G3}=\ell'_{H1}$. Moreover, since $\bfrhat_{G1}$ and $\bfrhat_{G2}$ are already in an isotropic configuration in $\calY_{\ell\ell0}(\bfrhat_{G1},\bfrhat_{G2},\bfrhat_{G3})$, the only allowed values of $\ell_{G3}$ and $\ell'_{H1}$ are zero. It follows that the isotropic functions reduce to constants: $\calY_{\ell_{G3}\ell'_{H1}L_3}(\bfrhat_{G3},\bfrhat'_{H1},\bfshat)=(4\uppi)^{-3/2}$ and $f_{\ell_{G3}\ell'_{H1}L_3}(r_{G3},r'_{H1},s) =f_{000}(r_{G3},r'_{H1},s)$. Integrating over $s$ and using Eq.~\eqref{eqn:def-gllll} we find
\begeqar
\int s^2 ds\, f_{000}(0,0,s) f_{000}(r_{G3},r'_{H1},s) = g_{0000}(0,0,r_{G3}, r'_{H1}),
\endeqar
% Now the contribution to the covariance proportional to
% $\calY_{\ell\ell0}(\bfrhat_{G1},\bfrhat_{G2},\bfrhat_{G3})\calY_{0\ell'\ell'}(\bfrhat'_{H1},\bfrhat'_{H2},\bfrhat'_{H3})$ is
The final form of Case IV is given by
\begeqar\label{eqn:cov_pc_caseIV}
    &&\cov^{\rm{(pc)}, IV}_{\Lambda,\Lambda'}=\\
    &&\quad\sum_{G,H}\,\sum_{\calL_G, \calL'_H}\frac{(4\uppi)^4}V (-1)^{\ell+\ell'} \sqrt{(2\ell+1)(2\ell'+1)} f_{\ell\ell0}(r_{G1},r_{G2},0)f_{0\ell'\ell'}(0,r'_{H2},r'_{H3}) g_{0000}(0,0,r_{G3},r'_{H1})\calB^{G^{-1}}_{\calL_G,\Lambda}\calB^{H^{-1}}_{\calL'_H,\Lambda'}.\nonumber
\endeqar

\section{Analytic Solution for Integral of Product of Three Spherical Bessel Functions}
\label{appendix:fabrikant_rbin}

When radial binning is included, the $f$-integral is evaluated with the bin-averaged sBFs:
\begeqar\label{eqn:f3l_bin_avg}
f_{\ell_1\ell_2\ell_3}(r_1, r_2, r_3)=\int \frac{k^2 dk}{2\uppi^2} P(k) \bar{j}_{\ell_1}(k;r_1)\bar{j}_{\ell_2}(k;r_2){j}_{\ell_3}(k;r_3),
\endeqar
where the bin-averaged sBFs are defined as:
\begeqar\label{eqn:sbf_bin_avg}
\bar{j}_{\ell_i}(k;r_i)=\frac{\int r^2 dr\, j_{\ell_i}(kr_i)\Theta(r_i)}{\int r^2 dr\, \Theta(r_i)}.
\endeqar
Here $\Theta(r_i)$ is a binning function equal to unity within bin $r_i$ and zero elsewhere.

In order to check the evaluation and implementation of the $f$-integral, we compare the numerical result to an analytic form with bin-averaged sBFs derived using~\citet{Fabrikant2013}, Eq. (24):
\begeqar\label{eq: Ipqm-tmp}
    I_{e}(p, q, m, n, \ell ; a, b, c)=\int_{0}^{\infty} \exp (-p k) k^{q} j_{m}(a k) j_{n}(b k) j_{\ell}(c k) d k.
\endeqar

\begin{figure}
    \centering
    \includegraphics[width=.4\textwidth]{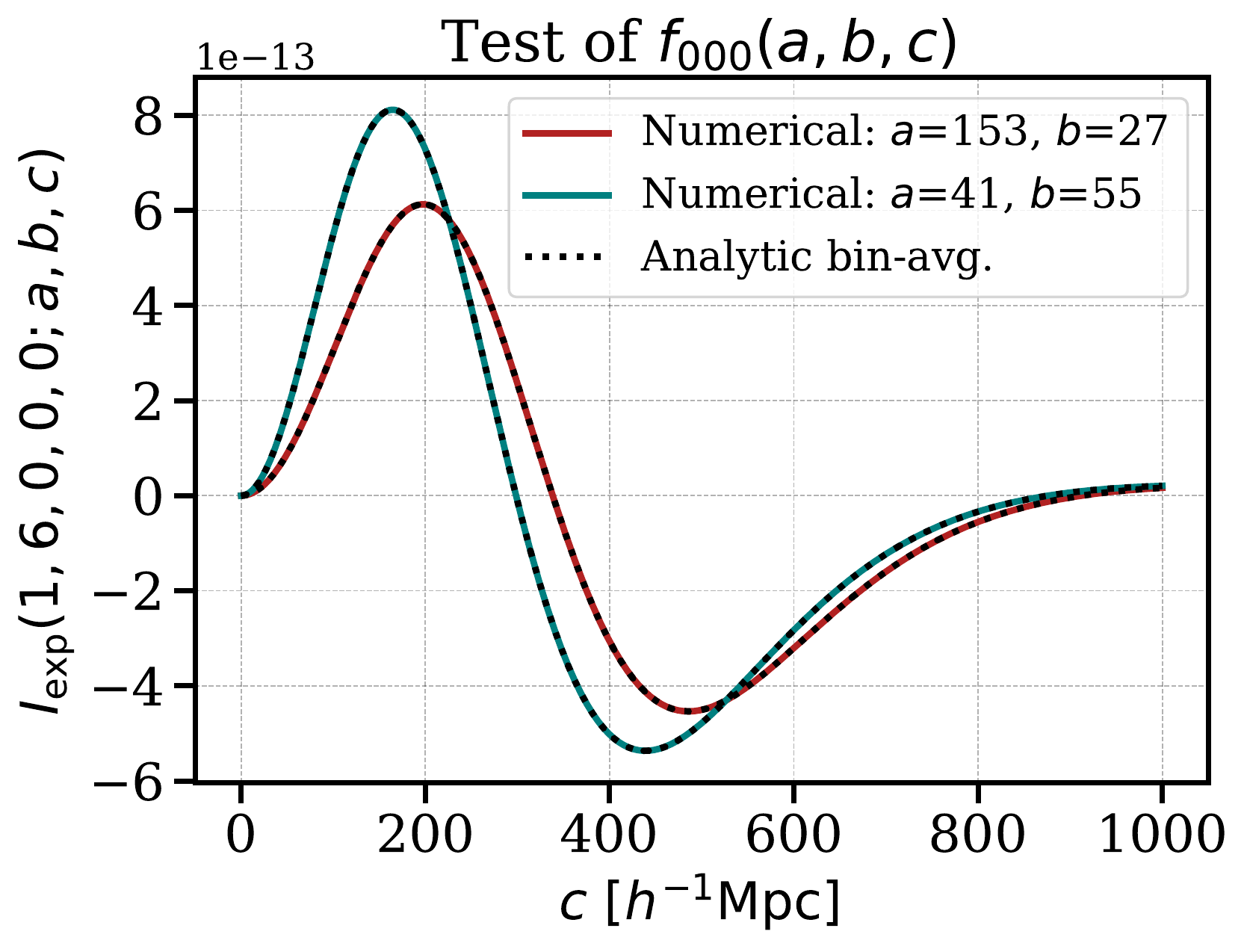}
    \caption{Comparison of the numerical and analytic implementations of the bin-averaged $f$-integral, setting $\ell_1=\ell_2=\ell_3=0$, and using a damped power-law power spectrum. We evaluate the integral at radial bin centers $a$ and $b$ as given in the legend, and their units are $h^{-1} {\rm Mpc}$.}
    \label{fig:fabrikant_bin_avg_rbin153x27_41x55}
\end{figure}

For this test, we make use of Eq. (26) in ~\citet{Fabrikant2013}, which provides an explicit solution for Eq.~\eqref{eq: Ipqm-tmp} with $p=1$, $q=2$, and $m=n=\ell=0$, and thus of $f_{000}(a,b,c)$, when the power spectrum is replaced by a power law damped by an exponential:
\begeqar
    &&I_{\rm exp}(1, 2, 0, 0, 0 ; a, b, c) \non
    % &=& \frac{1}{4abc}\left(-\tan ^{-1}[(a+b+c)/p] +\tan ^{-1}[(-a+b+c)/p]+\tan ^{-1}[(a-b+c)/p]+\tan ^{-1}[(a+b-c)/p]\right)\non
    &=& \frac{1}{4abc}\left(-T_{+++}^{abc} +T_{-++}^{abc}+T_{+-+}^{abc}+T_{++-}^{abc}\right).
    \label{eqn:integral_sbf_q2}
\endeqar
Here, we have introduced the notation that $T_{\pm\pm\pm}^{abc}\equiv\tan^{-1}[(\pm a\pm b \pm c)/p]$.
In practice the sBFs with arguments $a$ and $b$ are bin-averaged, and can be written as
\begeqar\label{eqn:jbar_0}
    \bar{j}_0(ak) = \frac{3}{k(a_{\rm max}^3-a_{\rm min}^3)}\, \left[a^2_{\rm max}j_1(a_{\rm max}k)-a^2_{\rm min}j_1(a_{\rm min}k)\right],
\endeqar
where the recurrence relation (Rayleigh's formula) gives
\begeqar
   j_1(xk) =  - \frac{1}{k}\frac{d}{dx} j_0(xk)
\endeqar
Replacing the sBF with the bin-averaged one given by Eq.~\eqref{eqn:jbar_0} and inserting the result into Eq.~\eqref{eqn:integral_sbf_q2} (setting $q=6$ in order to use the analytic solution), we have
\begeqar
    &&I_{\rm exp}(1, 6, 0, 0, 0 ; a, b, c)\non
    &=&\int_{0}^{\infty} \exp (- k) k^{6}\, \bar{j}_{0}(a k) \bar{j}_{0}(b k) j_{0}(c k) d k\non
    % =&\frac{3}{a_{\rm max}^3-a_{\rm min}^3}\frac{3}{b_{\rm max}^3-b_{\rm min}^3} \int_{0}^{\infty} \exp (- k) k^{6}\,j_{0}(c k)\\ 
    % &\frac{a^2_{\rm max}j_1(a_{\rm max}k)-a^2_{\rm min}j_1(a_{\rm min}k)}{k} \frac{b^2_{\rm max}j_1(b_{\rm max}k)-b^2_{\rm min}j_1(b_{\rm min}k)}{k}  d k\\
    % =& \frac{3}{a_{\rm max}^3-a_{\rm min}^3}\frac{3}{b_{\rm max}^3-b_{\rm min}^3} \int_{0}^{\infty} \exp (- k) k^{6}\,j_{0}(c k)\\
    % &\frac{1}{k^2}\left(a^2_{\rm max}b^2_{\rm max} j_1(a_{\rm max}k) j_1(b_{\rm max}k)-a^2_{\rm min}b^2_{\rm max} j_1(a_{\rm min}k) j_1(b_{\rm max}k)\right.\\
    % & \left. -a^2_{\rm max}b^2_{\rm min} j_1(a_{\rm max}k) j_1(b_{\rm min}k)+a^2_{\rm min}b^2_{\rm min} j_1(a_{\rm min}k) j_1(b_{\rm min}k)\right)\\
    % =& \frac{3}{a_{\rm max}^3-a_{\rm min}^3}\frac{3}{b_{\rm max}^3-b_{\rm min}^3} \int_{0}^{\infty} \exp (- k) k^{6}\,j_{0}(c k)\\    
    % &\frac{1}{k^2}\left(\frac{a^2_{\rm max}b^2_{\rm max}}{k^2}\frac{d}{d a_{\rm max}}\frac{d}{d b_{\rm max}} j_0(a_{\rm max}k) j_0(b_{\rm max}k)
    % -\frac{a^2_{\rm min}b^2_{\rm max}}{k^2}\frac{d}{d a_{\rm min}}\frac{d}{d b_{\rm max}} j_0(a_{\rm min}k) j_0(b_{\rm max}k)\right.\\
    % & \left. \frac{-a^2_{\rm max}b^2_{\rm min}}{k^2}\frac{d}{d a_{\rm min}}\frac{d}{d b_{\rm max}} j_0(a_{\rm max}k) j_0(b_{\rm min}k)
    % +\frac{a^2_{\rm min}b^2_{\rm min}}{k^2}\frac{d}{d a_{\rm min}}\frac{d}{d b_{\rm min}} j_0(a_{\rm min}k) j_0(b_{\rm min}k)\right)\\
    &=& \frac{3}{a_{\rm max}^3-a_{\rm min}^3}\frac{3}{b_{\rm max}^3-b_{\rm min}^3} \left[{a^2_{\rm max}b^2_{\rm max}}\frac{d}{d a_{\rm max}}\frac{d}{d b_{\rm max}} I_{\rm exp}(1, 2, 0, 0, 0 ; a_{\rm max}, b_{\rm max}, c)\right.\non 
    &&\left. \hspace{11em} - {a^2_{\rm min}b^2_{\rm max}}\frac{d}{d a_{\rm min}}\frac{d}{d b_{\rm max}} I_{\rm exp}(1, 2, 0, 0, 0 ; a_{\rm min}, b_{\rm max}, c) \right.\non
    &&\left. \hspace{11em} - {a^2_{\rm max}b^2_{\rm min}}\frac{d}{d a_{\rm max}}\frac{d}{d b_{\rm min}} I_{\rm exp}(1, 2, 0, 0, 0 ; a_{\rm max}, b_{\rm min}, c) \right. \non
    &&\left. \hspace{11em} +{a^2_{\rm min}b^2_{\rm min}}\frac{d}{d a_{\rm min}}\frac{d}{d b_{\rm min}} I_{\rm exp}(1, 2, 0, 0, 0 ; a_{\rm min}, b_{\rm min}, c)\right].
    \label{eqn:Ie_1}
\endeqar
In the above equation we obtain four types of terms, differing by their lower or upper bounds in $a$ or $b$. Next, we focus on the general form $\frac{d}{d a}\frac{d}{d b} I_{\rm exp}(...)$:
\begeqar
\frac{d}{d a}\frac{d}{d b} I_{\rm exp}(...) = \frac{d}{d a}\frac{d}{d b}\left[ \frac{1}{4abc}\left(-T_{+++}^{abc} +T_{-++}^{abc}+T_{+-+}^{abc}+T_{++-}^{abc}\right)\right].
\endeqar
Due to the symmetry of these expressions, in what follows we may focus on just the first term $T_{+++}^{abc}$. 
\begeqar
    &&\frac{d}{d a}\frac{d}{d b}\left( \frac{1}{4abc} T_{+++}^{abc}\right)\non
    &=&\frac{d}{d a}\frac{d}{d b}\left( \frac{1}{4abc}\tan ^{-1}[(c+b+a)/p]\right)\non
    % &=& \frac{d}{d a}\left(\frac{1}{4abcp}\frac{1}{(a+b+c)^2/p^2+1}-\frac{1}{4ab^2c}\tan ^{-1}[(c+b+a)/p]\right)\non
    % &=& -\frac{1}{4a^2bcp}\frac{1}{(a+b+c)^2/p^2+1}-\frac{1}{4abcp^3}\frac{2(a+b+c)}{((a+b+c)^2/p^2+1)^2} + \frac{1}{4a^2b^2c}\tan ^{-1}[(c+b+a)/p]-\frac{1}{4ab^2cp}\frac{1}{(a+b+c)^2/p^2+1}\non
    &=& \frac{1}{4a^2b^2c}\left( \tan ^{-1}[(c+b+a)/p]- \frac{a+b}{p}\frac{1}{(a+b+c)^2/p^2+1}-\frac{ab}{p^3}\frac{2(a+b+c)}{((a+b+c)^2/p^2+1)^2}\right),
    \label{eqn:dd_arctan}
\endeqar
This form remains the same for the rest of the $T_{\pm\pm\pm}^{abc}$ terms, except for the signs. Inserting Eq.~\eqref{eqn:dd_arctan} into Eq.~\eqref{eqn:Ie_1} we obtain the final result shown in Fig.~\ref{fig:fabrikant_bin_avg_rbin153x27_41x55} (dotted black curves) after integrating over $c$. As an example, we evaluate the integral for two cases $a=153\, h^{-1} {\rm Mpc}$, $b=27\, h^{-1} {\rm Mpc}$ and $a=41\, h^{-1} {\rm Mpc}$, $b=55\, h^{-1} {\rm Mpc}$. In both cases the numerical implementation and the analytic solution display excellent agreement. 

%% begin: some redundant derivation in Fabrikant 2015
\iffalse
\begeqar
    &I_{e}(1, 6, 0, 0, \ell ; a, b, c)\\
    =&\frac{3}{a_{\rm max}^3-a_{\rm min}^3}\frac{3}{b_{\rm max}^3-b_{\rm min}^3}\frac{1}{4cp} \left( \right.\\
    -&\left.\left(p\tan ^{-1}[(c+b_{\rm max}+a_{\rm max})/p]- \frac{a_{\rm max}+b_{\rm max}}{(a_{\rm max}+b_{\rm max}+c)^2/p^2+1}-\frac{a_{\rm max}b_{\rm max}}{p^2}\frac{2(a_{\rm max}+b_{\rm max}+c)}{((a_{\rm max}+b_{\rm max}+c)^2/p^2+1)^2}\right)\right.\\
    -&\left.\left(p\tan ^{-1}[(c+b_{\rm max}-a_{\rm min})/p]- \frac{a_{\rm min}+b_{\rm max}}{(-a_{\rm min}+b_{\rm max}+c)^2/p^2+1}-\frac{a_{\rm min}b_{\rm max}}{p^2}\frac{2(-a_{\rm min}+b_{\rm max}+c)}{((-a_{\rm min}+b_{\rm max}+c)^2/p^2+1)^2}\right)\right.\\
    -&\left.\left(p\tan ^{-1}[(c-b_{\rm min}+a_{\rm max})/p]- \frac{a_{\rm max}+b_{\rm min}}{(a_{\rm max}-b_{\rm min}+c)^2/p^2+1}-\frac{a_{\rm max}b_{\rm min}}{p^2}\frac{2(a_{\rm max}-b_{\rm min}+c)}{((a_{\rm max}-b_{\rm min}+c)^2/p^2+1)^2}\right)\right.\\
    +&\left.\left(p\tan ^{-1}[(-c+b_{\rm min}+a_{\rm min})/p]- \frac{a_{\rm min}+b_{\rm min}}{(a_{\rm min}+b_{\rm min}-c)^2/p^2+1}-\frac{a_{\rm min}}{p^2}b_{\rm min}\frac{2(a_{\rm min}+b_{\rm min}-c)}{((a_{\rm min}+b_{\rm min}-c)^2/p^2+1)^2}\right)
    \right)
\endeqar
\fi

%% end: some redundant derivation in Fabrikant 2015

\section{Gaussian NPCF Covariances including RSD}%Fully-coupled Gaussian NPCF Covariance with RSD in isotropic basis}
\label{appendix:cov_fc_rsd_iso}

Here we extend our general expression for the real-space covariance to include RSD.
As a preparation for the derivation, we extend the $\calQ$ quantity to involve four angular momenta:
\begeqar
\calQ^{\Lambda\Lambda'\Lambda''\Lambda'''} = \prod_{i=1}^{N} \sum_{m_im'_i M_i\mu_i} \calC^{\ell_i\ell'_iL_i\lambda_i}_{m_im'_iM_i\mu_i} \calC^{\Lambda}_{\rm M}\calC^{\Lambda'}_{\rm M'}\calC^{\Lambda''}_{\rm M''}\calC^{\Lambda'''}_{\rm M'''},
\endeqar
where the $C^{\Lambda}_{\rm M}$ coefficient is defined in Eq.~\eqref{eqn:C_threej} with
\begeqar
\calH^{\Lambda\Lambda'\Lambda''\Lambda'''} = (4\uppi)^{-N/2}\left[\prod_{i=1}^N\calD^{\rm P}_{\ell_i\ell'_iL_i\lambda_i}\calC^{\ell_i\ell'_iL_i\lambda_i}_{0000}\right]\calQ^{\Lambda\Lambda'\Lambda''\Lambda'''}.
\endeqar
Furthermore, averaging over isotropic functions of four arguments gives
\begeqar\label{eq:averaging_4ell}
    &&\int d\calR\, d\calR'\, d\calS\, d\calN \prod_{i=1}^N \calY_{\ell_i\ell'_i\ell''_i\lambda_i}(\bfrhat_i,\bfrhat'_i,\bfshat,\hn)\non
    &&\qquad\quad = (4\uppi)^{-N}\sum_{\calL\calL'\Lambda''\Lambda'''}\calQ^{\calL\calL'\Lambda''\Lambda'''}\calD^{\rm P}_{\Lambda''}\calC^{\Lambda''}_\bfzero\calD^{\rm P}_{\Lambda'''}\calC^{\Lambda'''}_\bfzero\calY_\calL(\bfRhat^{(N)})\calY_{\calL'}({\bfRhat'}{^{(N)}}).
\endeqar
For the fully-coupled covariance including RSD we start from Eq.~\eqref{eqn:general-contractions} and Eq.~\eqref{eqn:2pcf_r_rp_s_rsd}:
\begeqar
&&\sum_{\Lambda,\Lambda'}\calE(\Lambda')\cov_{\Lambda,\Lambda'}(R,R')\calY_\Lambda(\bfRhat) \calY_{\Lambda'}(\bfRhat')\non
 &&\quad= \int \frac {d^3\bfs}{V} (4\uppi)^{2N} \sum_{G} \prod_{i=0}^{N-1} \sum_{\ell_{Gi}\ell'_i L_i\lambda_i}\frac{1}{2\lambda_i+1} i^{-\ell_{Gi}+\ell'_i+L_i}\non 
&&\quad\qquad\times\, \calD^{\rm P}_{\ell_{Gi}\ell'_iL_i\lambda_i} 
\calC^{\ell_{Gi}\ell'_iL_i\lambda_i}_{0000} f^{\lambda_i}_{\ell_{Gi}\ell'_iL_i}(r_{Gi},r'_i,s)\calY_{\ell_{Gi}\ell'_i L_i\lambda_i}(\bfrhat_{Gi},\bfrhat'_i,\bfshat,\hn)|_{r_{0}=r_0'=0}.
\endeqar
Next, we apply the rotational average over $\bfrhat$, $\bfrhat'$, $\bfshat$, and $\hn$\footnote{The rotational average over $\hn$ is justified as the isotropic 4PCF must be invariant under rotations.}:
\begeqar
&&\sum_{\Lambda,\Lambda'}\calE(\Lambda')\cov_{\Lambda,\Lambda'}(R,R')\calY_\Lambda(\bfRhat) \calY_{\Lambda'}(\bfRhat')\non
&=& \int \frac {s^2 ds}{V} (4\uppi)^{3N/2} \sum_{G} \prod_{i=0}^{N-1} \sum_{\ell_{Gi}\ell'_i L_i\lambda_i}\frac{1}{2\lambda_i+1}\, i^{-\ell_{Gi}+\ell'_i+L_i}\,
\calH^{\calL_G\calL'\Lambda''\Lambda'''}\non
&&\,\times\,f^{\lambda_i}_{\ell_{Gi}\ell'_i L_i}(r_{Gi},r'_i,s)\,\calD^{\rm P}_{\Lambda''}\,\calC^{\Lambda''}_\bfzero\,\calD^{\rm P}_{\Lambda'''}\,\calC^{\Lambda'''}_\bfzero\calY_{\calL_{G}}(\bfRhat^{(N)}_{G})\calY_{\calL'}({\bfRhat'}{^{(N)}})|_{r_{0}=r_0'=0};
\endeqar
as before, going from $\bfRhat^{(N)}_{G} \rightarrow \bfRhat_{G}$ leads to a factor of $(4\uppi)^{-1/2}$, which is cancelled with the normalization factor arising from $d\calS$. Next, we use the reordering coefficient to restore the canonical ordering of the arguments, and project both sides onto the isotropic basis $\calY_\Lambda(\bfRhat)$ and $\calY_{\Lambda'}(\bfRhat')$. This yields the final form:
\begeqar
&&\cov_{\Lambda,\Lambda'}(R,R')\non
&=&  (4\uppi)^{3N/2} \int \frac {s^2 ds}{V} \sum_{G} \sum_{\calL_{G} \Lambda''\Lambda'''} \prod_{i=0}^{N-1} \left[\frac{1}{2\lambda_i+1} f^{\lambda_i}_{\ell_{Gi}\ell'_i L_i}(r_{Gi},r'_i,s)\right] (-1)^{\left[-\Sigma(\Lambda)-\Sigma(\Lambda')+\Sigma(\Lambda'')\right]/2} \non
&&\,\times\,\calB^{G^{-1}}_{\calL_G,\Lambda}\,\calH^{\calL_G\Lambda'\Lambda''\Lambda'''}\calD^{\rm P}_{\Lambda''}\calC^{\Lambda''}_\bfzero\calD^{\rm P}_{\Lambda'''}\calC^{\Lambda'''}_\bfzero|_{r_{G0}=r_0'=0}.
\endeqar

\bibliographystyle{mnras}
\bibliography{ref} 

\begin{thebibliography}{}
\makeatletter
\relax
\def\mn@urlcharsother{\let\do\@makeother \do\$\do\&\do\#\do\^\do\_\do\%\do\~}
\def\mn@doi{\begingroup\mn@urlcharsother \@ifnextchar [ {\mn@doi@}
  {\mn@doi@[]}}
\def\mn@doi@[#1]#2{\def\@tempa{#1}\ifx\@tempa\@empty \href
  {http://dx.doi.org/#2} {doi:#2}\else \href {http://dx.doi.org/#2} {#1}\fi
  \endgroup}
\def\mn@eprint#1#2{\mn@eprint@#1:#2::\@nil}
\def\mn@eprint@arXiv#1{\href {http://arxiv.org/abs/#1} {{\tt arXiv:#1}}}
\def\mn@eprint@dblp#1{\href {http://dblp.uni-trier.de/rec/bibtex/#1.xml}
  {dblp:#1}}
\def\mn@eprint@#1:#2:#3:#4\@nil{\def\@tempa {#1}\def\@tempb {#2}\def\@tempc
  {#3}\ifx \@tempc \@empty \let \@tempc \@tempb \let \@tempb \@tempa \fi \ifx
  \@tempb \@empty \def\@tempb {arXiv}\fi \@ifundefined
  {mn@eprint@\@tempb}{\@tempb:\@tempc}{\expandafter \expandafter \csname
  mn@eprint@\@tempb\endcsname \expandafter{\@tempc}}}

\bibitem[\protect\citeauthoryear{{Adams}}{{Adams}}{1878}]{Adams1878}
{Adams} J.~C.,  1878, Proceedings of the Royal Society of London Series I,
  \href {https://ui.adsabs.harvard.edu/abs/1878RSPS...27...63A} {27, 63}

\bibitem[\protect\citeauthoryear{{Agarwal}, {Desjacques}, {Jeong}  \&
  {Schmidt}}{{Agarwal} et~al.}{2021}]{Agarwal202103}
{Agarwal} N.,  {Desjacques} V.,  {Jeong} D.,   {Schmidt} F.,  2021, \mn@doi
  [\jcap] {10.1088/1475-7516/2021/03/021}, \href
  {https://ui.adsabs.harvard.edu/abs/2021JCAP...03..021A} {2021, 021}

\bibitem[\protect\citeauthoryear{Alam et~al.,}{Alam et~al.}{2015}]{Alam201507}
Alam S.,  et~al., 2015, \mn@doi [The Astrophysical Journal Supplement Series]
  {10.1088/0067-0049/219/1/12}, 219, 12

\bibitem[\protect\citeauthoryear{{Alam} et~al.,}{{Alam}
  et~al.}{2020}]{Alam202011}
{Alam} S.,  et~al., 2020, arXiv e-prints, \href
  {https://ui.adsabs.harvard.edu/abs/2020arXiv201105771A} {p. arXiv:2011.05771}

\bibitem[\protect\citeauthoryear{{Amendola} et~al.,}{{Amendola}
  et~al.}{2018}]{Amendola2018}
{Amendola} L.,  et~al., 2018, \mn@doi [Living Reviews in Relativity]
  {10.1007/s41114-017-0010-3}, \href
  {https://ui.adsabs.harvard.edu/abs/2018LRR....21....2A} {21, 2}

\bibitem[\protect\citeauthoryear{Anderson}{Anderson}{2009}]{Anderson2009}
Anderson T.,  2009, An Introduction to Multivariate Statistical Analysis, 3rd
  Ed..
Wiley India Pvt. Limited, \url {https://books.google.com/books?id=1iF0CgAAQBAJ}

\bibitem[\protect\citeauthoryear{Aviles, Banerjee, Niz  \& Slepian}{Aviles
  et~al.}{2021}]{Aviles202106}
Aviles A.,  Banerjee A.,  Niz G.,   Slepian Z.,  2021, Clustering in Massive
  Neutrino Cosmologies via Eulerian Perturbation Theory (\mn@eprint {arXiv}
  {2106.13771})

\bibitem[\protect\citeauthoryear{{BOSS Collaboration} et~al.,}{{BOSS
  Collaboration} et~al.}{2017}]{BOSSCollaboration2017}
{BOSS Collaboration} et~al., 2017, \mn@doi [\mnras] {10.1093/mnras/stx721},
  \href {https://ui.adsabs.harvard.edu/abs/2017MNRAS.470.2617A} {470, 2617}

\bibitem[\protect\citeauthoryear{Barreira}{Barreira}{2019}]{Barreira201903}
Barreira A.,  2019, \mn@doi [Journal of Cosmology and Astroparticle Physics]
  {10.1088/1475-7516/2019/03/008}, 2019, 008–008

\bibitem[\protect\citeauthoryear{{Bartolo}, {Bellini}, {Bertacca}  \&
  {Matarrese}}{{Bartolo} et~al.}{2013}]{Bartolo2013J03}
{Bartolo} N.,  {Bellini} E.,  {Bertacca} D.,   {Matarrese} S.,  2013, \mn@doi
  [\jcap] {10.1088/1475-7516/2013/03/034}, \href
  {https://ui.adsabs.harvard.edu/abs/2013JCAP...03..034B} {2013, 034}

\bibitem[\protect\citeauthoryear{Baumann, Nicolis, Senatore  \&
  Zaldarriaga}{Baumann et~al.}{2012}]{Baumann201207}
Baumann D.,  Nicolis A.,  Senatore L.,   Zaldarriaga M.,  2012, \mn@doi
  [Journal of Cosmology and Astroparticle Physics]
  {10.1088/1475-7516/2012/07/051}, 2012, 051–051

\bibitem[\protect\citeauthoryear{Cahn \& Slepian}{Cahn \&
  Slepian}{2020}]{Cahn202010}
Cahn R.~N.,  Slepian Z.,  2020, Isotropic N-Point Basis Functions and Their
  Properties (\mn@eprint {arXiv} {2010.14418})

\bibitem[\protect\citeauthoryear{Carrasco, Hertzberg  \& Senatore}{Carrasco
  et~al.}{2012}]{Carrasco201209}
Carrasco J. J.~M.,  Hertzberg M.~P.,   Senatore L.,  2012, \mn@doi [Journal of
  High Energy Physics] {10.1007/jhep09(2012)082}, 2012

\bibitem[\protect\citeauthoryear{Chen}{Chen}{2010}]{Chen201002}
Chen X.,  2010, \mn@doi [Advances in Astronomy] {10.1155/2010/638979}, 2010,
  1–43

\bibitem[\protect\citeauthoryear{Chen, Huang, Kachru  \& Shiu}{Chen
  et~al.}{2007}]{Chen200701}
Chen X.,  Huang M.-x.,  Kachru S.,   Shiu G.,  2007, \mn@doi [Journal of
  Cosmology and Astroparticle Physics] {10.1088/1475-7516/2007/01/002}, 2007,
  002–002

\bibitem[\protect\citeauthoryear{Chiang, Wagner, Schmidt  \& Komatsu}{Chiang
  et~al.}{2014}]{Chiang201405}
Chiang C.-T.,  Wagner C.,  Schmidt F.,   Komatsu E.,  2014, \mn@doi [Journal of
  Cosmology and Astroparticle Physics] {10.1088/1475-7516/2014/05/048}, 2014,
  048–048

\bibitem[\protect\citeauthoryear{{Chudaykin} \& {Ivanov}}{{Chudaykin} \&
  {Ivanov}}{2019}]{euclid_forecast}
{Chudaykin} A.,  {Ivanov} M.~M.,  2019, \mn@doi [\jcap]
  {10.1088/1475-7516/2019/11/034}, \href
  {https://ui.adsabs.harvard.edu/abs/2019JCAP...11..034C} {2019, 034}

\bibitem[\protect\citeauthoryear{Chudaykin, Ivanov, Philcox  \&
  Simonović}{Chudaykin et~al.}{2020}]{Chudaykin202009}
Chudaykin A.,  Ivanov M.~M.,  Philcox O.~H.,   Simonović M.,  2020, \mn@doi
  [Physical Review D] {10.1103/physrevd.102.063533}, 102

\bibitem[\protect\citeauthoryear{{DESI Collaboration} et~al.,}{{DESI
  Collaboration} et~al.}{2016}]{DesiCollaboration2016}
{DESI Collaboration} et~al., 2016, arXiv e-prints, \href
  {https://ui.adsabs.harvard.edu/abs/2016arXiv161100036D} {p. arXiv:1611.00036}

\bibitem[\protect\citeauthoryear{DLMF}{DLMF}{}]{NIST:DLMF}
DLMF, {\it NIST Digital Library of Mathematical Functions}.
\url {http://dlmf.nist.gov/}

\bibitem[\protect\citeauthoryear{{Dawson} et~al.,}{{Dawson}
  et~al.}{2013}]{Dawson2013}
{Dawson} K.~S.,  et~al., 2013, \mn@doi [\aj] {10.1088/0004-6256/145/1/10},
  \href {https://ui.adsabs.harvard.edu/abs/2013AJ....145...10D} {145, 10}

\bibitem[\protect\citeauthoryear{Deadman, Higham  \& Ralha}{Deadman
  et~al.}{2013}]{Deadman2013}
Deadman E.,  Higham N.~J.,   Ralha R.,  2013, in Manninen P.,  {\"O}ster P.,
  eds, Applied Parallel and Scientific Computing. Springer Berlin Heidelberg,
  Berlin, Heidelberg, pp 171--182

\bibitem[\protect\citeauthoryear{Dizgah, Lee, Schmittfull  \& Dvorkin}{Dizgah
  et~al.}{2020}]{Dizgah202004}
Dizgah A.~M.,  Lee H.,  Schmittfull M.,   Dvorkin C.,  2020, \mn@doi [Journal
  of Cosmology and Astroparticle Physics] {10.1088/1475-7516/2020/04/011},
  2020, 011–011

\bibitem[\protect\citeauthoryear{Dodelson \& Schneider}{Dodelson \&
  Schneider}{2013}]{Dodelson201309}
Dodelson S.,  Schneider M.~D.,  2013, \mn@doi [Physical Review D]
  {10.1103/physrevd.88.063537}, 88

\bibitem[\protect\citeauthoryear{{Eisenstein}, {Seo}, {Sirko}  \&
  {Spergel}}{{Eisenstein} et~al.}{2007}]{Eisenstein200708}
{Eisenstein} D.~J.,  {Seo} H.-J.,  {Sirko} E.,   {Spergel} D.~N.,  2007,
  \mn@doi [\apj] {10.1086/518712}, \href
  {https://ui.adsabs.harvard.edu/abs/2007ApJ...664..675E} {664, 675}

\bibitem[\protect\citeauthoryear{Fabrikant}{Fabrikant}{2013}]{Fabrikant2013}
Fabrikant V.~I.,  2013, Quarterly of Applied Mathematics, 71, 573

\bibitem[\protect\citeauthoryear{{Feldman}, {Kaiser}  \& {Peacock}}{{Feldman}
  et~al.}{1994}]{Feldman1994}
{Feldman} H.~A.,  {Kaiser} N.,   {Peacock} J.~A.,  1994, \mn@doi [\apj]
  {10.1086/174036}, \href {http://adsabs.harvard.edu/abs/1994ApJ...426...23F}
  {426, 23}

\bibitem[\protect\citeauthoryear{{Fry} \& {Gaztanaga}}{{Fry} \&
  {Gaztanaga}}{1993}]{Fry1993}
{Fry} J.~N.,  {Gaztanaga} E.,  1993, \mn@doi [\apj] {10.1086/173015}, \href
  {https://ui.adsabs.harvard.edu/abs/1993ApJ...413..447F} {413, 447}

\bibitem[\protect\citeauthoryear{{Gil-Mar{\'\i}n}, {Percival}, {Verde},
  {Brownstein}, {Chuang}, {Kitaura}, {Rodr{\'\i}guez-Torres}  \&
  {Olmstead}}{{Gil-Mar{\'\i}n} et~al.}{2017}]{Gil-Marin201702}
{Gil-Mar{\'\i}n} H.,  {Percival} W.~J.,  {Verde} L.,  {Brownstein} J.~R.,
  {Chuang} C.-H.,  {Kitaura} F.-S.,  {Rodr{\'\i}guez-Torres} S.~A.,
  {Olmstead} M.~D.,  2017, \mn@doi [\mnras] {10.1093/mnras/stw2679}, \href
  {https://ui.adsabs.harvard.edu/abs/2017MNRAS.465.1757G} {465, 1757}

\bibitem[\protect\citeauthoryear{Grieb, Sánchez, Salazar-Albornoz  \&
  Dalla Vecchia}{Grieb et~al.}{2016}]{Grieb2016}
Grieb J.~N.,  Sánchez A.~G.,  Salazar-Albornoz S.,   Dalla Vecchia C.,  2016,
  \mn@doi [Monthly Notices of the Royal Astronomical Society]
  {10.1093/mnras/stw065}, 457, 1577–1592

\bibitem[\protect\citeauthoryear{{Gualdi}, {Gil-Marin}  \& {Verde}}{{Gualdi}
  et~al.}{2021}]{Gualdi202104}
{Gualdi} D.,  {Gil-Marin} H.,   {Verde} L.,  2021, arXiv e-prints, \href
  {https://ui.adsabs.harvard.edu/abs/2021arXiv210403976G} {p. arXiv:2104.03976}

\bibitem[\protect\citeauthoryear{{Hahn}, {Villaescusa-Navarro}, {Castorina}  \&
  {Scoccimarro}}{{Hahn} et~al.}{2020}]{Hahn202003}
{Hahn} C.,  {Villaescusa-Navarro} F.,  {Castorina} E.,   {Scoccimarro} R.,
  2020, \mn@doi [\jcap] {10.1088/1475-7516/2020/03/040}, \href
  {https://ui.adsabs.harvard.edu/abs/2020JCAP...03..040H} {2020, 040}

\bibitem[\protect\citeauthoryear{Hand, Feng, Beutler, Li, Modi, Seljak  \&
  Slepian}{Hand et~al.}{2018}]{Hand2018}
Hand N.,  Feng Y.,  Beutler F.,  Li Y.,  Modi C.,  Seljak U.,   Slepian Z.,
  2018, \mn@doi [The Astronomical Journal] {10.3847/1538-3881/aadae0}, 156, 160

\bibitem[\protect\citeauthoryear{Ivanov, Simonović  \& Zaldarriaga}{Ivanov
  et~al.}{2020}]{Ivanov202005}
Ivanov M.~M.,  Simonović M.,   Zaldarriaga M.,  2020, \mn@doi [Journal of
  Cosmology and Astroparticle Physics] {10.1088/1475-7516/2020/05/042}, 2020,
  042–042

\bibitem[\protect\citeauthoryear{{Jasche} \& {Lavaux}}{{Jasche} \&
  {Lavaux}}{2019}]{Jasche201905}
{Jasche} J.,  {Lavaux} G.,  2019, \mn@doi [\aap] {10.1051/0004-6361/201833710},
  \href {https://ui.adsabs.harvard.edu/abs/2019A&A...625A..64J} {625, A64}

\bibitem[\protect\citeauthoryear{Jasche \& Wandelt}{Jasche \&
  Wandelt}{2013}]{Jasche201304}
Jasche J.,  Wandelt B.~D.,  2013, \mn@doi [Monthly Notices of the Royal
  Astronomical Society] {10.1093/mnras/stt449}, 432, 894–913

\bibitem[\protect\citeauthoryear{{Kaiser}}{{Kaiser}}{1987}]{Kaiser1987}
{Kaiser} N.,  1987, \mn@doi [\mnras] {10.1093/mnras/227.1.1}, \href
  {https://ui.adsabs.harvard.edu/abs/1987MNRAS.227....1K} {227, 1}

\bibitem[\protect\citeauthoryear{Kamalinejad \& Slepian}{Kamalinejad \&
  Slepian}{2020}]{Kamalinejad202011}
Kamalinejad F.,  Slepian Z.,  2020, A Non-Degenerate Neutrino Mass Signature in
  the Galaxy Bispectrum (\mn@eprint {arXiv} {2011.00899})

\bibitem[\protect\citeauthoryear{{Kitaura} et~al.,}{{Kitaura}
  et~al.}{2016}]{Kitaura201603}
{Kitaura} F.-S.,  et~al., 2016, \mn@doi [\mnras] {10.1093/mnras/stv2826}, \href
  {https://ui.adsabs.harvard.edu/abs/2016MNRAS.456.4156K} {456, 4156}

\bibitem[\protect\citeauthoryear{{Kofman}}{{Kofman}}{1991}]{Kofman199101}
{Kofman} L.,  1991, \mn@doi [Physica Scripta Volume T]
  {10.1088/0031-8949/1991/T36/012}, \href
  {https://ui.adsabs.harvard.edu/abs/1991PhST...36..108K} {36, 108}

\bibitem[\protect\citeauthoryear{Komatsu et~al.,}{Komatsu
  et~al.}{2003}]{Komatsu200309}
Komatsu E.,  et~al., 2003, \mn@doi [The Astrophysical Journal Supplement
  Series] {10.1086/377220}, 148, 119–134

\bibitem[\protect\citeauthoryear{Kullback \& Leibler}{Kullback \&
  Leibler}{1951}]{Kullback1951}
Kullback S.,  Leibler R.~A.,  1951, \mn@doi [The Annals of Mathematical
  Statistics] {10.1214/aoms/1177729694}, 22, 79

\bibitem[\protect\citeauthoryear{{LSST Science Collaboration} et~al.,}{{LSST
  Science Collaboration} et~al.}{2009}]{collaboration2009lsst}
{LSST Science Collaboration} et~al., 2009, LSST Science Book, Version 2.0
  (\mn@eprint {arXiv} {0912.0201})

\bibitem[\protect\citeauthoryear{{Laureijs} et~al.,}{{Laureijs}
  et~al.}{2011}]{Laureijs2011}
{Laureijs} R.,  et~al., 2011, arXiv e-prints, \href
  {https://ui.adsabs.harvard.edu/abs/2011arXiv1110.3193L} {p. arXiv:1110.3193}

\bibitem[\protect\citeauthoryear{{Li}, {Singh}, {Yu}, {Feng}  \& {Seljak}}{{Li}
  et~al.}{2019}]{Li201901}
{Li} Y.,  {Singh} S.,  {Yu} B.,  {Feng} Y.,   {Seljak} U.,  2019, \mn@doi
  [\jcap] {10.1088/1475-7516/2019/01/016}, \href
  {https://ui.adsabs.harvard.edu/abs/2019JCAP...01..016L} {2019, 016}

\bibitem[\protect\citeauthoryear{Linde \& Mukhanov}{Linde \&
  Mukhanov}{1997}]{Linde199707}
Linde A.,  Mukhanov V.,  1997, \mn@doi [Physical Review D]
  {10.1103/physrevd.56.r535}, 56, R535–R539

\bibitem[\protect\citeauthoryear{{Massara}, {Villaescusa-Navarro}, {Ho},
  {Dalal}  \& {Spergel}}{{Massara} et~al.}{2021}]{marked1}
{Massara} E.,  {Villaescusa-Navarro} F.,  {Ho} S.,  {Dalal} N.,   {Spergel}
  D.~N.,  2021, \mn@doi [\prl] {10.1103/PhysRevLett.126.011301}, \href
  {https://ui.adsabs.harvard.edu/abs/2021PhRvL.126a1301M} {126, 011301}

\bibitem[\protect\citeauthoryear{{Mehrem}, {Londergan}  \&
  {Macfarlane}}{{Mehrem} et~al.}{1991}]{Mehrem1991}
{Mehrem} R.,  {Londergan} J.~T.,   {Macfarlane} M.~H.,  1991, \mn@doi [Journal
  of Physics A Mathematical General] {10.1088/0305-4470/24/7/018}, \href
  {https://ui.adsabs.harvard.edu/abs/1991JPhA...24.1435M} {24, 1435}

\bibitem[\protect\citeauthoryear{{O'Connell} \& {Eisenstein}}{{O'Connell} \&
  {Eisenstein}}{2019}]{rascal2}
{O'Connell} R.,  {Eisenstein} D.~J.,  2019, \mn@doi [\mnras]
  {10.1093/mnras/stz1359}, \href
  {https://ui.adsabs.harvard.edu/abs/2019MNRAS.487.2701O} {487, 2701}

\bibitem[\protect\citeauthoryear{{O'Connell}, {Eisenstein}, {Vargas}, {Ho}  \&
  {Padmanabhan}}{{O'Connell} et~al.}{2016}]{rascal1}
{O'Connell} R.,  {Eisenstein} D.,  {Vargas} M.,  {Ho} S.,   {Padmanabhan} N.,
  2016, \mn@doi [\mnras] {10.1093/mnras/stw1821}, \href
  {https://ui.adsabs.harvard.edu/abs/2016MNRAS.462.2681O} {462, 2681}

\bibitem[\protect\citeauthoryear{O’Connell, Eisenstein, Vargas, Ho  \&
  Padmanabhan}{O’Connell et~al.}{2016}]{OConnell201607}
O’Connell R.,  Eisenstein D.,  Vargas M.,  Ho S.,   Padmanabhan N.,  2016,
  \mn@doi [Monthly Notices of the Royal Astronomical Society]
  {10.1093/mnras/stw1821}, 462, 2681–2694

\bibitem[\protect\citeauthoryear{{Padmanabhan}, {White}  \&
  {Cohn}}{{Padmanabhan} et~al.}{2009}]{Padmanabhan200903}
{Padmanabhan} N.,  {White} M.,   {Cohn} J.~D.,  2009, \mn@doi [\prd]
  {10.1103/PhysRevD.79.063523}, \href
  {https://ui.adsabs.harvard.edu/abs/2009PhRvD..79f3523P} {79, 063523}

\bibitem[\protect\citeauthoryear{{Pearson} \& {Samushia}}{{Pearson} \&
  {Samushia}}{2018}]{Pearson201808}
{Pearson} D.~W.,  {Samushia} L.,  2018, \mn@doi [\mnras]
  {10.1093/mnras/sty1266}, \href
  {https://ui.adsabs.harvard.edu/abs/2018MNRAS.478.4500P} {478, 4500}

\bibitem[\protect\citeauthoryear{{Peebles}}{{Peebles}}{1978}]{Peebles1978}
{Peebles} P.~J.~E.,  1978, in {Longair} M.~S.,  {Einasto} J.,  eds,  Vol. 79,
  Large Scale Structures in the Universe. p.~217

\bibitem[\protect\citeauthoryear{{Percival} et~al.,}{{Percival}
  et~al.}{2014}]{Percival201404}
{Percival} W.~J.,  et~al., 2014, \mn@doi [\mnras] {10.1093/mnras/stu112}, \href
  {https://ui.adsabs.harvard.edu/abs/2014MNRAS.439.2531P} {439, 2531}

\bibitem[\protect\citeauthoryear{Philcox \& Eisenstein}{Philcox \&
  Eisenstein}{2019}]{Philcox201910}
Philcox O. H.~E.,  Eisenstein D.~J.,  2019, \mn@doi [Monthly Notices of the
  Royal Astronomical Society] {10.1093/mnras/stz2896}, 490, 5931–5951

\bibitem[\protect\citeauthoryear{Philcox \& Slepian}{Philcox \&
  Slepian}{2021}]{Philcox202106ND}
Philcox O. H.~E.,  Slepian Z.,  2021, Efficient Computation of N-point
  Correlation Functions in D Dimensions (\mn@eprint {arXiv} {2106.10278})

\bibitem[\protect\citeauthoryear{Philcox, Eisenstein, O’Connell  \&
  Wiegand}{Philcox et~al.}{2019}]{Philcox201911}
Philcox O. H.~E.,  Eisenstein D.~J.,  O’Connell R.,   Wiegand A.,  2019,
  \mn@doi [Monthly Notices of the Royal Astronomical Society]
  {10.1093/mnras/stz3218}, 491, 3290–3317

\bibitem[\protect\citeauthoryear{{Philcox}, {Massara}  \& {Spergel}}{{Philcox}
  et~al.}{2020a}]{marked2}
{Philcox} O. H.~E.,  {Massara} E.,   {Spergel} D.~N.,  2020a, \mn@doi [\prd]
  {10.1103/PhysRevD.102.043516}, \href
  {https://ui.adsabs.harvard.edu/abs/2020PhRvD.102d3516P} {102, 043516}

\bibitem[\protect\citeauthoryear{{Philcox}, {Eisenstein}, {O'Connell}  \&
  {Wiegand}}{{Philcox} et~al.}{2020b}]{rascal3}
{Philcox} O. H.~E.,  {Eisenstein} D.~J.,  {O'Connell} R.,   {Wiegand} A.,
  2020b, \mn@doi [\mnras] {10.1093/mnras/stz3218}, \href
  {https://ui.adsabs.harvard.edu/abs/2020MNRAS.491.3290P} {491, 3290}

\bibitem[\protect\citeauthoryear{Philcox, Hou, Slepian, Cahn  \&
  Eisenstein}{Philcox et~al.}{2021b}]{Philcox2021boss4pcf}
Philcox O. H.~E.,  Hou J.,  Slepian Z.,  Cahn R.~N.,   Eisenstein D.~J.,
  2021b, 4PCF BOSS

\bibitem[\protect\citeauthoryear{Philcox, Slepian, Hou, Warner, Cahn  \&
  Eisenstein}{Philcox et~al.}{2021a}]{Philcox2021encore}
Philcox O. H.~E.,  Slepian Z.,  Hou J.,  Warner C.,  Cahn R.~N.,   Eisenstein
  D.~J.,  2021a, encore: Estimating Galaxy N-point Correlation Functions in
  O(N2) Time

\bibitem[\protect\citeauthoryear{Philcox, Ivanov, Zaldarriaga,
  Simonovi\ifmmode~\acute{c}\else \'{c}\fi{}  \& Schmittfull}{Philcox
  et~al.}{2021c}]{Philcox202002}
Philcox O. H.~E.,  Ivanov M.~M.,  Zaldarriaga M.,
  Simonovi\ifmmode~\acute{c}\else \'{c}\fi{} M.,   Schmittfull M.,  2021c,
  \mn@doi [Phys. Rev. D] {10.1103/PhysRevD.103.043508}, 103, 043508

\bibitem[\protect\citeauthoryear{{Portillo}, {Slepian}, {Burkhart}, {Kahraman}
  \& {Finkbeiner}}{{Portillo} et~al.}{2018}]{Portillo201808}
{Portillo} S. K.~N.,  {Slepian} Z.,  {Burkhart} B.,  {Kahraman} S.,
  {Finkbeiner} D.~P.,  2018, \mn@doi [\apj] {10.3847/1538-4357/aacb80}, \href
  {https://ui.adsabs.harvard.edu/abs/2018ApJ...862..119P} {862, 119}

\bibitem[\protect\citeauthoryear{Putter, Wagner, Mena, Verde  \&
  Percival}{Putter et~al.}{2012}]{Putter201204}
Putter R.~d.,  Wagner C.,  Mena O.,  Verde L.,   Percival W.~J.,  2012, \mn@doi
  [Journal of Cosmology and Astroparticle Physics]
  {10.1088/1475-7516/2012/04/019}, 2012, 019–019

\bibitem[\protect\citeauthoryear{Rodríguez-Torres et~al.,}{Rodríguez-Torres
  et~al.}{2016}]{RodriguezTorres201604}
Rodríguez-Torres S.~A.,  et~al., 2016, \mn@doi [Monthly Notices of the Royal
  Astronomical Society] {10.1093/mnras/stw1014}, 460, 1173–1187

\bibitem[\protect\citeauthoryear{Ruggeri, Castorina, Carbone  \&
  Sefusatti}{Ruggeri et~al.}{2018}]{Ruggeri201803}
Ruggeri R.,  Castorina E.,  Carbone C.,   Sefusatti E.,  2018, \mn@doi [Journal
  of Cosmology and Astroparticle Physics] {10.1088/1475-7516/2018/03/003},
  2018, 003–003

\bibitem[\protect\citeauthoryear{{Samushia}, {Slepian}  \&
  {Villaescusa-Navarro}}{{Samushia} et~al.}{2021}]{Samushia202102}
{Samushia} L.,  {Slepian} Z.,   {Villaescusa-Navarro} F.,  2021, arXiv
  e-prints, \href {https://ui.adsabs.harvard.edu/abs/2021arXiv210201696S} {p.
  arXiv:2102.01696}

\bibitem[\protect\citeauthoryear{Schmidt, Elsner, Jasche, Nguyen  \&
  Lavaux}{Schmidt et~al.}{2019}]{Schmidt201901}
Schmidt F.,  Elsner F.,  Jasche J.,  Nguyen N.~M.,   Lavaux G.,  2019, \mn@doi
  [Journal of Cosmology and Astroparticle Physics]
  {10.1088/1475-7516/2019/01/042}, 2019, 042–042

\bibitem[\protect\citeauthoryear{Schmittfull \& Dizgah}{Schmittfull \&
  Dizgah}{2021}]{Schmittfull202103}
Schmittfull M.,  Dizgah A.~M.,  2021, \mn@doi [Journal of Cosmology and
  Astroparticle Physics] {10.1088/1475-7516/2021/03/020}, 2021, 020

\bibitem[\protect\citeauthoryear{{Schmittfull}, {Feng}, {Beutler}, {Sherwin}
  \& {Chu}}{{Schmittfull} et~al.}{2015}]{Schmittfull201512}
{Schmittfull} M.,  {Feng} Y.,  {Beutler} F.,  {Sherwin} B.,   {Chu} M.~Y.,
  2015, \mn@doi [\prd] {10.1103/PhysRevD.92.123522}, \href
  {https://ui.adsabs.harvard.edu/abs/2015PhRvD..92l3522S} {92, 123522}

\bibitem[\protect\citeauthoryear{{Schmittfull}, {Baldauf}  \&
  {Zaldarriaga}}{{Schmittfull} et~al.}{2017}]{Schmittfull201707}
{Schmittfull} M.,  {Baldauf} T.,   {Zaldarriaga} M.,  2017, \mn@doi [\prd]
  {10.1103/PhysRevD.96.023505}, \href
  {https://ui.adsabs.harvard.edu/abs/2017PhRvD..96b3505S} {96, 023505}

\bibitem[\protect\citeauthoryear{{Schneider} \& {Bartelmann}}{{Schneider} \&
  {Bartelmann}}{1995}]{Schneider1995}
{Schneider} P.,  {Bartelmann} M.,  1995, \mn@doi [\mnras]
  {10.1093/mnras/273.2.475}, \href
  {https://ui.adsabs.harvard.edu/abs/1995MNRAS.273..475S} {273, 475}

\bibitem[\protect\citeauthoryear{{Scoccimarro}}{{Scoccimarro}}{2000}]{Scoccimarro200012}
{Scoccimarro} R.,  2000, \mn@doi [\apj] {10.1086/317248}, \href
  {https://ui.adsabs.harvard.edu/abs/2000ApJ...544..597S} {544, 597}

\bibitem[\protect\citeauthoryear{{Scoccimarro}, {Colombi}, {Fry}, {Frieman},
  {Hivon}  \& {Melott}}{{Scoccimarro} et~al.}{1998}]{Scoccimarro199803}
{Scoccimarro} R.,  {Colombi} S.,  {Fry} J.~N.,  {Frieman} J.~A.,  {Hivon} E.,
  {Melott} A.,  1998, \mn@doi [\apj] {10.1086/305399}, \href
  {https://ui.adsabs.harvard.edu/abs/1998ApJ...496..586S} {496, 586}

\bibitem[\protect\citeauthoryear{Scoccimarro, Zaldarriaga  \& Hui}{Scoccimarro
  et~al.}{1999}]{Scoccimarro199912}
Scoccimarro R.,  Zaldarriaga M.,   Hui L.,  1999, \mn@doi [The Astrophysical
  Journal] {10.1086/308059}, 527, 1–15

\bibitem[\protect\citeauthoryear{{Sefusatti} \& {Scoccimarro}}{{Sefusatti} \&
  {Scoccimarro}}{2005}]{Sefusatti200503}
{Sefusatti} E.,  {Scoccimarro} R.,  2005, \mn@doi [\prd]
  {10.1103/PhysRevD.71.063001}, \href
  {https://ui.adsabs.harvard.edu/abs/2005PhRvD..71f3001S} {71, 063001}

\bibitem[\protect\citeauthoryear{{Seljak}, {Aslanyan}, {Feng}  \&
  {Modi}}{{Seljak} et~al.}{2017}]{Seljak201712}
{Seljak} U.,  {Aslanyan} G.,  {Feng} Y.,   {Modi} C.,  2017, \mn@doi [\jcap]
  {10.1088/1475-7516/2017/12/009}, \href
  {https://ui.adsabs.harvard.edu/abs/2017JCAP...12..009S} {2017, 009}

\bibitem[\protect\citeauthoryear{{Sellentin} \& {Heavens}}{{Sellentin} \&
  {Heavens}}{2016}]{Sellentin201602}
{Sellentin} E.,  {Heavens} A.~F.,  2016, \mn@doi [\mnras]
  {10.1093/mnrasl/slv190}, \href
  {https://ui.adsabs.harvard.edu/abs/2016MNRAS.456L.132S} {456, L132}

\bibitem[\protect\citeauthoryear{Senatore}{Senatore}{2015}]{Senatore201511}
Senatore L.,  2015, \mn@doi [Journal of Cosmology and Astroparticle Physics]
  {10.1088/1475-7516/2015/11/007}, 2015, 007–007

\bibitem[\protect\citeauthoryear{Senatore \& Zaldarriaga}{Senatore \&
  Zaldarriaga}{2014}]{Senatore201409}
Senatore L.,  Zaldarriaga M.,  2014, Redshift Space Distortions in the
  Effective Field Theory of Large Scale Structures (\mn@eprint {arXiv}
  {1409.1225})

\bibitem[\protect\citeauthoryear{Slepian \& Eisenstein}{Slepian \&
  Eisenstein}{2015a}]{Slepian201506}
Slepian Z.,  Eisenstein D.~J.,  2015a, \mn@doi [Monthly Notices of the Royal
  Astronomical Society] {10.1093/mnras/stv2119}, 454, 4142–4158

\bibitem[\protect\citeauthoryear{Slepian \& Eisenstein}{Slepian \&
  Eisenstein}{2015b}]{Slepian201510}
Slepian Z.,  Eisenstein D.~J.,  2015b, \mn@doi [Monthly Notices of the Royal
  Astronomical Society: Letters] {10.1093/mnrasl/slv133}, 455, L31–L35

\bibitem[\protect\citeauthoryear{Slepian \& Eisenstein}{Slepian \&
  Eisenstein}{2018}]{Slepian201804}
Slepian Z.,  Eisenstein D.~J.,  2018, \mn@doi [Monthly Notices of the Royal
  Astronomical Society] {10.1093/mnras/sty1063}, 478, 1468–1483

\bibitem[\protect\citeauthoryear{Slepian et~al.,}{Slepian
  et~al.}{2017}]{Slepian201702}
Slepian Z.,  et~al., 2017, \mn@doi [Monthly Notices of the Royal Astronomical
  Society] {10.1093/mnras/stw3234}, 468, 1070–1083

\bibitem[\protect\citeauthoryear{{Sugiyama}, {Saito}, {Beutler}  \&
  {Seo}}{{Sugiyama} et~al.}{2020a}]{Sugiyama202009}
{Sugiyama} N.~S.,  {Saito} S.,  {Beutler} F.,   {Seo} H.-J.,  2020a, \mn@doi
  [\mnras] {10.1093/mnras/staa1940}, \href
  {https://ui.adsabs.harvard.edu/abs/2020MNRAS.497.1684S} {497, 1684}

\bibitem[\protect\citeauthoryear{Sugiyama, Saito, Beutler  \& Seo}{Sugiyama
  et~al.}{2020b}]{Sugiyama202012}
Sugiyama N.~S.,  Saito S.,  Beutler F.,   Seo H.-J.,  2020b, \mn@doi [Monthly
  Notices of the Royal Astronomical Society] {10.1093/mnras/staa3725}, 501,
  2862–2896

\bibitem[\protect\citeauthoryear{Taylor \& Joachimi}{Taylor \&
  Joachimi}{2014}]{Taylor201406}
Taylor A.,  Joachimi B.,  2014, \mn@doi [Monthly Notices of the Royal
  Astronomical Society] {10.1093/mnras/stu996}, 442, 2728–2738

\bibitem[\protect\citeauthoryear{Taylor, Joachimi  \& Kitching}{Taylor
  et~al.}{2013}]{Taylor201305}
Taylor A.,  Joachimi B.,   Kitching T.,  2013, \mn@doi [Monthly Notices of the
  Royal Astronomical Society] {10.1093/mnras/stt270}, 432, 1928–1946

\bibitem[\protect\citeauthoryear{{Villaescusa-Navarro}
  et~al.,}{{Villaescusa-Navarro} et~al.}{2020}]{Quijote_sims}
{Villaescusa-Navarro} F.,  et~al., 2020, \mn@doi [\apjs]
  {10.3847/1538-4365/ab9d82}, \href
  {https://ui.adsabs.harvard.edu/abs/2020ApJS..250....2V} {250, 2}

\bibitem[\protect\citeauthoryear{Wadekar, Ivanov  \& Scoccimarro}{Wadekar
  et~al.}{2020}]{Wadekar202012}
Wadekar D.,  Ivanov M.~M.,   Scoccimarro R.,  2020, \mn@doi [Physical Review D]
  {10.1103/physrevd.102.123521}, 102

\bibitem[\protect\citeauthoryear{{White}}{{White}}{2015}]{White201504}
{White} M.,  2015, \mn@doi [\mnras] {10.1093/mnras/stv842}, \href
  {https://ui.adsabs.harvard.edu/abs/2015MNRAS.450.3822W} {450, 3822}

\bibitem[\protect\citeauthoryear{Wishart}{Wishart}{1928}]{wishart28}
Wishart J.,  1928, Biometrika, 20A, 32

\bibitem[\protect\citeauthoryear{de Belsunce \& Senatore}{de~Belsunce \&
  Senatore}{2019}]{deBelsunce201902}
de Belsunce R.,  Senatore L.,  2019, \mn@doi [Journal of Cosmology and
  Astroparticle Physics] {10.1088/1475-7516/2019/02/038}, 2019, 038–038

\bibitem[\protect\citeauthoryear{{eBOSS Collaboration} et~al.,}{{eBOSS
  Collaboration} et~al.}{2021}]{eBOSSCollaboration2021}
{eBOSS Collaboration} et~al., 2021, \mn@doi [\prd]
  {10.1103/PhysRevD.103.083533}, \href
  {https://ui.adsabs.harvard.edu/abs/2021PhRvD.103h3533A} {103, 083533}

\makeatother
\end{thebibliography}

\label{lastpage}
\end{document}